\documentclass[11pt]{article}
\usepackage{amssymb,latexsym,amsmath,amsthm,graphicx,amscd}

\makeatletter \@addtoreset{figure}{section}
 \def\thefigure{\thesection.\@arabic\c@figure}
\def\fps@figure{h, t}
\@addtoreset{table}{bsection}
\def\thetable{\thesection.\@arabic\c@table}
\def\fps@table{h, t}
\newcommand{\todo}[1]{\vspace{5 mm}\par \noindent
\marginpar{\textsc{ToDo}}
\framebox{\begin{minipage}[c]{0.86\textwidth}
\tt #1 \end{minipage}}\vspace{5 mm}\par}

\newtheorem{theorem}{Theorem}[section]
\newtheorem{lemma}[theorem]{Lemma}
\newtheorem{proposition}[theorem]{Proposition}

\newfont{\tenbi}{cmbxti10}
\newcommand{\la}{\lambda}

\DeclareMathOperator{\Prym}{Prym}
\DeclareMathOperator{\Jac}{Jac}
\DeclareMathOperator{\Sp}{Sp}
\DeclareMathOperator{\Cov}{Cov}
\begin{document}

\title{Algebraic description of Jacobians isogeneous to certain Prym varieties with polarization 
(1,2)\footnote{{\small AMS Subject Classification 14K12, 14H40, 14H70, 70H06} }}

\author{V.Z. Enolski \\
School of Mathematics, University of Edinburgh\footnote{
 On leave from Institute of Magnetism, National Academy of Sciences of
Ukraine, Kiev} \\ {\tt venolski@googlemail.com} \\
 Yu.N.Fedorov \\
Department of Mathematics I, Polytechnic university of Catalonia, Barcelona \\
 {\tt Yuri.Fedorov@upc.edu} }
\maketitle

\begin{abstract}
For a class of non-hyperelliptic genus 3 curves $C$ which are 2-fold coverings of elliptic curves $E$,
we give an explicit
algebraic description of all birationally non-equivalent genus 2 curves whose Jacobians are degree 2 isogeneous
to the Prym varieties associated to such coverings. Our description is based on previous studies of Prym varieties with
polarization (1,2) in connection with 
separation of variables in a series of classical and new algebraic integrable systems linearized on such varieties.

We also consider some special cases of the covering $C\to E$, in particular, when the corresponding Prym varieties contain
pairs of elliptic curves and the Jacobian of $C$ is isogeneous (but not isomorphic) to the product of 3 different 
elliptic curves.

Our description is accompanied with explicit numerical examples.
 \end{abstract}

\section{Introduction}

Algebraic curves with involutions and related Abelian varieties have been objects of throughout study
during many decades, mostly in connection with integrable systems in mathematical physics and geometry.

In particular, a non-hyperelliptic genus 3 curve $C$ having an involution $\sigma$ and covering an elliptic curve
$E=C/\sigma$ appears as
a spectral curve of Lax representations of many algebraic integrable systems, such as
the Clebsch integrable case of the Kirchoff equations, the Frahm--Manakov top on the Lie algebra $so(4)$,
and the Kovalevskaya top, as well as their generalizations.

As was shown in \cite{hai83, hvm89, avm88, brs89}, generic complexified invariant tori of all these systems are
open subsets\footnote{More precisely, according to \cite{brs89},
complexified invariant tori of the Kovalevskaya top are given by 2 copies of such open subsets.} of
2-dimensional Abelian subvarieties of the Jacobian of $C$, namely, Prym
varieties $\Prym(C,\sigma)$ associated with the covering $C \to E$.
In this case $\Prym(C,\sigma)$ has polarization
(1,2), so it is not the Jacobian of a genus 2 curve.

On the other hand, in her celebrated paper \cite{kow}, S. Kovalevskaya presented a separation of variables
of the corresponding integrable system on a genus 2 (and, therefore, a hyperelliptic) curve $\Gamma$,
whereas the complex flow of the system was linearized on the Jacobian of
$\Gamma$. For the Clebsch integrable case and the Frahm--Manakov top on $so(4)$, similar complicated
linearizations were made by K\"otter \cite{kot892} and, respectively, Schottky \cite{scho891, scho925}.

Apparently, the papers \cite{hvm89, avm88} were the first ones where this seeming contradiction had been explicitly explained:
the subvariety $\Prym(C,\sigma)$ can be viewed as 2-fold covering of the Jacobians of 3 different, i.e., birationally
non-equivalent, genus 2 curves. In addition, there exist three other different genus 2 curves whose Jacobians are 2-fold coverings of
$\Prym(C,\sigma)$. In \cite{hvm89} it was shown that, for the case of the Kovalevskaya top,
one of these hyperelliptic curves is equivalent to the original curve of separation
found in \cite{kow}. Next, following a remarkable geometric construction of \cite{bw85} inspired by \cite{hai83},
the papers \cite{hvm89, avm88} presented a specific family $D_\lambda$, $\lambda \in {\mathbb P}$
of plane genus 3 curves covering elliptic curves
${\cal E}_\la$, which gives rise to the same Prym variety {\it dual} to $\Prym(C,\sigma)$ arising in the Kovalevskaya problem.
Then the three genus 2 curves whose Jacobians are isogeneous to $\Prym(C,\sigma)$ appear as regulazations of singular
curves in the family $D_\lambda$. In \cite{avm88} the coefficients of $D_\lambda$
have been expressed as complicated algebraic functions of parameters (the Weierstrass points) of the above genus 2 curves.

The covering $C \to E$ and the related Prym varieties have also been sources of several beautiful
geometric constructions related to intersection of quadrics in ${\mathbb P}^6$, ${\mathbb P}^7$
(see \cite{bw85, hai83, avm87}), which, in turn, has different connections with integrable systems.
\medskip

In this paper, using the modern results of \cite{hai83, hvm89, avm88} and, on the other hand, calculations of the classical paper
\cite{kot892}, we present a complete solution to the ''direct'' problem:
given an {\it almost} generic genus 3 curve $C$ covering
an elliptic curve $E$ (equation \eqref{mastercurve} below),
to find explicitly Weierstrass equations of all the six non-equivalent genus 2 curves $\Gamma$ whose Jacobians
are {\it degree} 2 isogeneous to the Prym variety $\Prym(C,\sigma)$ (see Theorem \ref{main_th}). Moreover, it appears that
the Weierstrass points of these curves are calculated via one and same universal formula \eqref{d_s},
whereas the passage from one curve to another is made by a mere permutation of roots of certain quartic equations associated
to $C$.

Thus, if such a curve $C$ appears as the spectral curve of a Lax pair of an algebraic integrable system, one can perform
explicitly a separation of variables on any of the six curves.

We also consider interesting special cases of the covering $C \to E$: first,
when the genus 3 curve $C$ is itself hyperelliptic
and has 2 commuting involutions (as occurs in the above mentioned integrable systems when the area integral or
its analog equals zero); second, when $C$ is not hyperelliptic, but $\Prym(C,\sigma)$ contains 2 elliptic curves and,
as a result, the Jacobian of $C$ is isogeneous (but not isomorphic) to a product of 3 different elliptic curves.
(In the latter case the solutions
of the corresponding integrable systems can be written in terms of elliptic functions.)

The algebraic description of the Jacobians is accompanied by giving the structure of
their Riemann period matrices $\tau$ and the relations between them. We observe that some of the
period matrices are obtained one from the other one by multiplication by 2, thus the equations of
corresponding genus 2 curves are connected
via the algebraic Richelot transformation, as described in \cite{bm88}, see also \cite{lm99}.

In Appendices we present a detailed algorithm of explicit calculation of the period matrix
of $\Prym(C,\sigma)$ and apply it to a first numerical example in this field:
given a covering $C\to E$, we derive the $2\times 2$ Riemann matrices of all the six Jacobians, degree 2 isogeneous
to $\Prym(C,\sigma)$. Then we calculate the Weierstrass equations of the corresponding genus 2 curves
by using the main Theorem \ref{main_th} and, finally, see that the period matrices of the Jacobians
and the curves share the same sets of absolute invariants. This provides an experimental
confirmation of the main results of the paper.

The concrete formulas of the absolute invariants of genus 2 curves, both in an algebraic and theta-function form,
suitable for calculations, are also presented for the first time.

\medskip

We now specify the class of curves (over the field $\mathbb C$ of complex numbers)
and Prym varieties we consider in this paper, and start with the following properties.

\begin{theorem} \label{gen_cover} Let $E$ be a generic elliptic curve in ${\mathbb P}^2$ whose affine part in
${\mathbb C}^2(x,y)$ is given by equation
$$
y^2= \Phi(x): = (x-c_1)(x-c_2)(x-c_3).
$$
1) Any 2-fold covering
$C\to E$ ramified at 4 arbitrary chosen finite points \\ $Q_1=(x_1,y_1), \dots, Q_4=(x_4,y_4)$ on
$E$ can be written in form
\begin{equation} \label{gen_gen_3}
C \; : \quad  w^2 = g_3(x) + (\alpha x+ \beta) y, \quad  y^2=\Phi(x),
\end{equation}
where $g_3(x)$ is a polynomial of degree at most 3 and $\alpha,\beta$ are constants such that
$$
 g_3^2(x)- (\alpha x+ \beta)^2 \Phi (x) = (x-x_1)\cdots (x-x_4) \rho^2(x)
$$
for a certain linear function $\rho(x)$.

2) For any divisor $Q=\{Q_1,\dots,Q_4\}$ on $E$, the set $\Cov_E^2(Q)$ of 2-fold covers of $E$ ramified
exactly at $Q$ consists of $2^2=4$ birationally non-equivalent covering curves $C$.
\end{theorem}

The idea of the proof of the theorem is due to A. Levin \cite{LevA}. For the reasons of brevity we do not give it here.
\medskip

The involution $\sigma : (x,w)\to (x,-w)$ on the curve $C$ extends to its Jacobian variety,
$\Jac (C)$. Thus the latter contains two Abelian subvarieties: the elliptic curve $E$ itself and the 2-dimensional
Prym variety denoted as $ \Prym(C,\sigma)$, which is anti-symmetric with respect to the extended involution, whereas
$E$ is invariant. Equivalently,
$\Prym(C,\sigma)=\text{ker}\, (1+\sigma)$, see e.g., \cite{Mum}.

$\Prym(C,\sigma)$ is not a Jacobian variety, in particular just because it does not carry a principal polarization.
Namely, if we regard this Abelian variety as a complex torus ${\mathbb C}^2/{\Lambda}$,
with ${\Lambda}$ being a lattice generated
by 4 period vectors, then, in appropriate coordinates $z_1,z_2$ in ${\mathbb C}^2$, its period matrix takes the form
\begin{equation} \label{Lambda_periods}
\Lambda =  \left(
{\begin{array}{rrcc}
1 & 0 & 2\,P & 2\,p \\
0 & 2 & 2\,p & 2\,\pi
\end{array}}
 \right), \quad \text{equivalent to} \quad
\widehat \Lambda =\left(
{\begin{array}{rrcc}
2 & 0 & 2\,\pi & 2\,p \\
0 & 1 & 2\,p & 2\,P
\end{array}}
 \right),
\end{equation}
with some constants $p,P,\pi$ which are related to periods of holomorphic differentials on $C$, which are 
anti-symmetric with respect to $\sigma$. 
(A brief derivation of the period matrix can be found in Appendix A). The right half of $\Lambda$ is a Riemann matrix, and
the diagonal of the left half indicates the polarization of $\Prym(C,\sigma)$, namely (1,2).

According to \cite{bw85}, any 2-dimensional Abelian variety $\cal A$ with polarization (1,2) can be
realized as the Prym variety of a covering $C\to E$ described in Theorem \ref{gen_cover}.
Moreover, for a generic $\cal A$ there is a one-parametric family of such coverings, in which both $C$ and $E$ depend
on the parameter.

It should be mentioned that, for a generic divisor $Q$ on $E$, the four coverings $C\to E$ of the set
$\Cov_E^2(Q)$ give rise to non-isogeneous Prym varieties.
\medskip

In this paper we will concentrate of an important particular case of the covering $C\to E$ given by equation
\begin{equation}
C: \;\;  w^2 = g(x) + 2 h_3 \sqrt{\Phi(x)}, \qquad g(x) = h_2 x^2+ h_1 x+ h_0, \quad h_2\ne 0, \label{mastercurve}
\end{equation}
where either $\Phi =(x-c_1)(x-c_2)(x-c_3)$ (odd order case) or \\ $\Phi =(x-a_1)(x-a_2)(x-a_3)(x-a_4)$ (even order case).
An equivalent algebraic form of $C$ is
$$
(w^2- g(x))^2= 4 h_3^2 \Phi (x) .
$$
In the sequel, without loss of generality, we will assume $h_2=1$.

As above, the covering is ramified at 4 zeros of $w$, i.e., at four points $Q_{i}=(s_{i},\sqrt{\Phi (s_i)})\in E$,
where $s_i$ are the roots of the polynomial
\begin{equation}
\psi (x) =\varkappa \left(g^{2}(x)-4 h^{2}_{3}\Phi(x) \right)  = (x-s_{1}) (x-s_{2})(x-s_{3})(x-s_{4}) . \label{Psi0}
\end{equation}
Here $\varkappa=1$ in the odd order case and $\varkappa=1/(h_2^2-4h_3^2)$ in the even order case.
 
We stress that, given a generic elliptic curve $E$ and a divisor $Q=\{Q_1,\dots, Q_4 \}$ on it,
not any covering $C\to E$ ramified
at $Q$ can be written in the form \eqref{mastercurve}: this imposes one algebraic condition on the
coefficients of $\Phi(x)$ and $\psi(x)$. More precisely, the coefficients must satisfy a quartic equation which
is too long to show here.

If this condition holds, then only one curve $C$ of the set $\Cov_E^2(Q)$ is given by equation
\eqref{mastercurve}, whereas the other three curves have the form
\footnote{The proof is omited for brevity reasons.}
\begin{equation} \label{alt_curves}
  w^2 = (x-c_\alpha)\,g(x) + 2 h_3 (x-c_\alpha)\,\sqrt{\Phi(x)}, \qquad \alpha =1,2,3
\end{equation}
i.e., they belong to the general family \eqref{gen_gen_3}.

The latter implies that for the covering $C\to E$ given by \eqref{mastercurve},
the corresponding variety $\Prym (C,\sigma)$ can be uniquely reconstructed from the data
$\{E\, ; Q_1,\dots, Q_4 \in E\}$, whereas in the general case described by Theorem \ref{gen_cover}, the same data
are not sufficient for this purpose, as they lead to four different covering $C\to E$ and 4 non-isogeneous Pryms.

The curves of the form \eqref{mastercurve} appear as spectral curves in the Frahm--Manakov top on $so(4)$, the Clebsch
integrable case, the elliptic Gaudin model, and several other integrable problems, for example the generalized Hen\'on--Heiles system with a
quartic potential (\cite{bef95}).
Notice, however, that the spectral curve of the Lax pair for the Kovalevskaya top found in
\cite{brs89} (more precisely, its factor by an extra involution) is equivalent to the genus 3 curve
\begin{equation} \label{sp_kow}
   y^2 = (x^2-2H x+2) x + x \sqrt{-4L^2 x^3 +(4H^2-K+4)x^2 - 8 Hx+4}\, ,
\end{equation}
where $H,K$ are non-trivial constants of motion and $L$ is the area integral.
Hence, the curve \eqref{sp_kow} fits only into the general form 
\eqref{gen_gen_3}\footnote{Although the curve \eqref{sp_kow} is also rather
special: two of the corresponding branch points $Q_1,\dots, Q_4$ are in the hyperelliptic involution on the
elliptic curve $E$}.

In the next Section we consider in detail some classical systems leading to the curves \eqref{mastercurve} and the related
Prym varieties.

\section{Relevant integrable systems: Lax pairs, separation of variables, and algebraic geometric description}
\label{systems}

\paragraph{The Clebsch case.}
Recall that the Kirchhoff equations on $e(3)=\{K,p\}={\mathbb R}^6$ with a Hamilton function $H=H(K,p)$
\begin{align*}
 \dot K & = K \times \frac{\partial H}{\partial K} + p \times \frac{\partial H}{\partial p} , \\
 \dot p & = p \times \frac{\partial H}{\partial K}
\end{align*}
describing the motion of a rigid body in an ideal fluid, always have 2 integrals
$\langle p,p\rangle, \langle K,p \rangle$, the Casimir functions
of the Lie--Poisson bracket on $e(3)$. For the quadratic Hamiltonian proposed by Clebsch
$$
H = \frac 12 (c_{1}K^{2}_{1}+c_{2}K^{2}_{2}+c_{3}K^{2}_{3})
+\frac 12 (b_{1}p^{2}_{1}+b_{2}p^{2}_{2}+b_{3}p^{2}_{3}),
$$
with the parameters $b_i, c_i$ satisfying the condition
$$
{b_1-b_2\over c_3}+{b_2-b_3\over c_1}+{b_3-b_1\over c_2}=0 ,
$$
the Kirchhoff equations are completely integrable, and their generic real invariant manifolds are 2-dimensional tori. The
system admits a $3\times 3$ matrix Lax representation with an {\it elliptic} spectral
parameter which can be written in the form (see e.g., \cite{Bel,fed95})
\begin{gather}
\dot L(\la) =[L(\la),A(\la)], \quad \la\in {\mathbb C}, \label{Lax1} \\
L_{ij}(\la) = \varepsilon_{ijk} \biggl( \sqrt{\la-c_k} \, K_{k}+ \sqrt{(\la-c_i)(\la-c_j)} \, p_k\biggr)\, , \quad
(i,j,k)=(1,2,3) , \notag
\end{gather}
$\varepsilon_{ijk}$ being the Levi-Civita tensor.  
The matrix $A_{ij}(\la)$ has a similar structure and is not shown here.
The roots $\sqrt{\la-c_k}$, $\sqrt{(\la-c_i)(\la-c_j)}$ are meromorphic functions on
the elliptic curve isomorphic to $E\, : \; y^2 =(x-c_1)(x-c_2)(x-c_3)$.

The spectral curve $|L(\la)- {\bf I} y|=0$ is reducible, and its non-trivial component reads
\begin{equation} \label{sp_C}
C : \left\{ y^2 =h_{2} \la^{2}+h_{1}\la + h_{0}+ 2h_{3} \sqrt{(\la-c_1)(\la-c_2)(\la-c_3)} \right \} ,
\end{equation}
where $h_{0},\ldots,h_{3}$ are constants of the integrals
\begin{equation}
\begin{aligned}
p_1^2+p_2^2+p_3^2 & =h_{2}\, ,  \\
K^{2}_{1}+K^{2}_{2}+K^{2}_{3}
-(c_{2}+c_{3})p^{2}_{1}
-(c_{3}+c_{1})p^{2}_{2}
-(c_{1}+c_{2})p^{2}_{3}
 & =h_{1}\, ,  \\
-c_{1}K^{2}_{1}
-c_{2}K^{2}_{2}
-c_{3}K^{2}_{3}
+c_{2}c_{3}p^{2}_{4}
+c_{1}c_{3}p^{2}_{5}
+c_{1}c_{2}p^{2}_{6}
& =h_{0}\, ,  \\
 K_{1}p_{1}+K_{2}p_{2}+K_{3}p_{3}& =h_{3}\, .
\end{aligned}
\label{integrals}
\end{equation}
Thus the curve \eqref{sp_C} has the same structure as our model curve \eqref{mastercurve}.

As was noticed in many publications, in the case $h_3=0$  equations can be reduced to an integrable system on the
cotangent bundle of the sphere $S^2 =\{\langle p, p\rangle=h_2\}$,
which has the same first integrals as the classical Neumann system describing a motion of a point on $S^2$ in a
force field with a quadratic potential. In this case the explicit theta-function solution of the Clebsch system
was obtained in \cite{web878}.  We will return to details of this case in Section \ref{h30}.

\paragraph{A scheme of F. K\"otter's integration of the Clebsch system.}
In the general case $h_3 \ne 0$ the Clebsch system was first integrated in quadratures and theta-functions by F. K\"otter
\cite{kot892} (1891), who, of course, did not use the Lax pair \eqref{Lax1}, however
noticed that the above quadratic integrals \eqref{integrals}
can be included to the following family of integrals with a parameter $x\in {\mathbb C}$
\begin{gather} \label{3.20}
f(x) = \sum^{3}_{\alpha =1}
\biggl(\sqrt{x-c_\alpha}\,  K_{\alpha}
+ {\sqrt{\Phi(x)} \over \sqrt{x-c_\alpha}}\,  p_\alpha \biggr)^2
= h_{2}x^{2}+h_{1}x+h_{0}+2h_{3} \sqrt{\Phi (x)} \, ,   \\
\Phi(x) =(x-c_{1})(x-c_{2})(x-c_{3}) \, .   \notag
\end{gather}
The family also represents a family of quadrics of rank $\le 4$ in ${\mathbb C}^6(K,p)$, which appeared in the later papers
\cite{hai83, avm84, avm87}.

Let $s_1,\dots, s_4$ be the zeroes of $f(x)$ in \eqref{3.20}, then
$Q_1=(s_1,\sqrt{\Phi (s_1)})$, $\dots$, $Q_4=(s_4,\sqrt{\Phi (s_4)}) \in E$ are the branch points of the covering $C\to E$,
where now $C$ is the spectral curve \eqref{sp_C} identified with \eqref{mastercurve}. As before, we assume $h_2=1$.

In \cite{kot892} (page 68), K\"otter derived expressions for the phase variables
$K_\alpha, p_\alpha$ in terms of two points on the genus 2 hyperelliptic curve
\begin{equation}
{\mathcal G} : \; w^2 = z(z-d_1^2) (z-d_2^2)(z-d_3^2)(z-d_1^2 d_2^2 d_3^2)   \label{GammaK}
\end{equation}
whose branch points are determined by the following complicated expressions
\begin{gather}
d_{\alpha}={A_{\alpha 3}+iA_{\alpha 4} \over A_{\alpha 1}+iA_{\alpha 2}}
 = {B_{\alpha 3}+iB_{\alpha 4}\over B_{\alpha 1}+iB_{\alpha 2}}, \quad \alpha=1,2,3, \label{6.9}\\
A_{\alpha i} = {\sqrt{\Phi (s_i)}
\over \sqrt{\psi '(s_i)} \sqrt{s_i -c_\alpha}} \, ,  \qquad
B_{\alpha i}
= {\sqrt{s_i -c_\alpha} \over \sqrt{\psi '(s_i)}}.  \nonumber
\end{gather}
Here, as above, $\psi(s)=(s-s_1)\dots (s-s_4)$, $\psi'(s)=\frac{d}{ds}(s-s_1)\dots (s-s_4)$, and
the signs of the quotients $B_{\alpha i}$  
 are chosen in such a way that for any $\alpha=1,2,3$
\begin{gather} \label{signs}
 \prod\limits_{j=1}^4 \dfrac{\sqrt{s_j -c_\alpha}}{\sqrt{\psi '(s_j)}}
= - \frac{g(c_{\alpha}) }{(s_1-s_2)(s_2-s_4)(s_3-s_1)(s_3-s_2)(s_4-s_1)(s_4-s_2) } ,  \\
g(x)= x^2+ h_1 x+h_0 . \notag
\end{gather}
Note that, in view of \eqref{Psi0}, $g^2(c_\alpha)= (c_\alpha-s_1) \cdots (c_\alpha-s_4)$ giving
$g(c_\alpha)=\pm \sqrt{\psi(c_\alpha)}$.
Then the condition \eqref{signs} implies
\begin{equation} \label{imp_cond}
   \sqrt{\psi(c_\alpha)} = -g(c_\alpha), \qquad \alpha=1,2,3 .
\end{equation}
If these conditions are fulfilled, one also has
\begin{equation}
{A_{\alpha 3}-iA_{\alpha 4} \over A_{\alpha 1}-iA_{\alpha 2}}
={B_{\alpha 3}-iB_{\alpha 4} \over B_{\alpha 1}-iB_{\alpha 2}}
= - {1\over d_{\alpha}}\, .    \label{6.10}
\end{equation}

As one can notice, the condition \eqref{signs} is still insufficient to fix the signs of
$B_{\alpha\,i}$ and, therefore, the constants $d^2_\alpha$ in \eqref{6.9} uniquely. Below, in Lemma \ref{reciprocal}, we
will see that, in fact, all the signs of $B_{\alpha \, i}$ allowed by \eqref{signs} lead to birationally
equivalent genus 2 curves $\cal G$.

Following K\"otter, introduce new variables
\begin{equation}
\begin{aligned}
\xi_{\alpha} & = (A_{\alpha 1}+iA_{\alpha 2})p_{\alpha}
+(B_{\alpha 1}+iB_{\alpha 2})K_{\alpha}\,  ,               \\
\eta_{\alpha}& = (A_{\alpha 1}-iA_{\alpha2})p_{\alpha}
+(B_{\alpha 1}-iB_{\alpha 2})K_{\alpha}\, ,
\end{aligned}
\qquad \alpha =1,2,3 \,    \label{6.11}
\end{equation}
and observe that the family of integrals \eqref{3.20} gives rise to the following three independent relations
\begin{equation}
\sum^{3}_{\alpha =1}
\left( \xi^{2}_{\alpha}+\eta^{2}_{\alpha}\right)=0\, ,\quad
\sum^{3}_{\alpha =1}
\left(d^{2}_{\alpha} \xi^{2}_{\alpha}+\frac{\eta^2_\alpha}{d^2_\alpha} \right)=0\, ,
\quad \sum^{3}_{\alpha=1}\xi_{\alpha}\eta_{\alpha}=0 \, ,
\label{6.13}
\end{equation}
which define a two-dimensional manifold in the projective space
$\mathbb{P}^{5}=(\xi_{1}:\xi_{2}:\xi_{3}:\eta_{1}:\eta_{2}:\eta_{3})$.

Remarkably, this manifold is
isomorphic to the set $\mathcal T$ of common tangent lines $\ell $ of two confocal
quadrics in $\mathbb{C}^{3}=(X_{1},X_{2},X_{3})$ given by
\begin{gather}
{\mathcal Q}_j = \biggl\{
{X^{2}_{1}\over d_{1}^2-\nu_{j}} +
{X^{2}_{2}\over d_{2}^2-\nu_{j}} +
{X^{2}_{3}\over d_{3}^2-\nu_{j}}
=1\biggr\}\, ,\quad j=1,2\, , \label{6.14} \\ 
\nu_{1}=0 \, , \quad \nu_{2}=d_{1}^2 d_{2}^2 d_{3}^2. \notag
\end{gather}
Indeed, let $\ell\subset {\mathbb C}^3$ be a line tangent to ${\mathcal Q}_1$,
let $\eta=(\eta_1 : \eta_2 : \eta_3)^T\in {\mathbb P}^2$ be the direction of $\ell$, and
$\imath \xi=\imath (\xi_1 :\xi_2 : \xi_3)^T$, $\imath=\sqrt{-1}$
be a vector normal to ${\mathcal Q}_1$ at the contact point
$\ell \cap {\mathcal Q}_1$. Assume $\left| \xi \right|=\left| \eta \right|$, which is described by
the first relation in (\ref{6.13}). The third relation expresses the orthogonality of $\eta, \xi$, whereas
the second one is the condition of tangency of $\ell$ and ${\mathcal Q}_2$ (this is a non-trivial observation).

On the other hand, as was shown in \cite{Knorr}, the set $\mathcal T$ is isomorphic to an 8-fold unramified covering
of the Jacobian variety of the curve
$$
  w^2 = (z-d_1^2) (z-d_2^2)(z-d_3^2)(z-\nu_1)(z-\nu_2)\, ,
$$
obtained by doubling three of four independent period vectors of the Jacobian. The above curve
is precisely the K\"otter curve ${\mathcal G}$.

In particular, up to a common factor $L$, the components of $\eta, \xi$ can ve written in terms of
2 ellipsoidal coordinates
on the quadric ${\mathcal Q}_1$, as was done in \cite{Knorr} following K.Jacobi's integration of the problem on geodesics
on the ellipsoid (see also \cite{Moser_var}),
and $\eta, \xi$  are known meromorphic functions on $\cal T$.
 The factor $L$ is also a meromorphic function on $\cal T$, which
can be calculated by using the last integral of the Clebsch system, independent of \eqref{6.13}.

Then the relations \eqref{6.11} imply that the original variables
$K_\alpha, p_\alpha$ of the Clebsch system are meromorphic functions on $\cal T$,
that is, the generic complex invariant manifold of the system is an open subset of $\cal T$, which, in turn,
is the 8-fold covering of $\Jac({\cal G})$.

The same observation also follows directly from the
explicit expressions for $K_\alpha, p_\alpha$ in terms of two points on ${\mathcal G}$ given in \cite{kot892}.

\paragraph{The Frahm--Manakov top on $so(4)$.} Recall that the Euler equations describing the free motion of
$n$-dimensional rigid body (top) have the form
\begin{equation} \label{eu}
   \dot M = [M,\Omega], \qquad M, \Omega \in so(n),
\end{equation}
where $\Omega, M$ are the matrices of the angular velocity and the angular momentum.
These equations were first written in \cite{FRahm} (in a scalar form), where it was also shown that they
are completely integrable at least for $n=4$ provided that
$$
M_{ij}=(a_i-a_j)/(b_i-b_j)\, \Omega_{ij}, \qquad 1 \le i<j \le 4 ,
$$
where $a_i, b_i$ are arbitrary distinct parameters.
In this case the system on the 6-dimensional space $so(4)$ has 4 independent quadratic integrals
\begin{gather} \label{ints_M}
\begin{aligned}
(a_{1}+a_{4})M^{2}_{1}+(a_{2}+a_{4})M^{2}_{2}+(a_{3}+a_{4})M^{2}_{3} \quad & \\
+(a_{2}+a_{3})M^{2}_{4} +(a_{1}+a_{3})M^{2}_{5}+(a_{1}+a_{2})M^{2}_{6} & =h_{0}\, , \\
a_{1}a_{4}M^{2}_{1} +a_{2}a_{4}M^{2}_{2} +a_{3}a_{4}M^{2}_{3}
+a_{2}a_{3}M^{2}_{4} +a_{1}a_{3}M^{2}_{5} +a_{1}a_{2}M^{2}_{6} & =h_{1}\, , \\
 M^{2}_{1}+M^{2}_{2}+M^{2}_{3}+M^{2}_{4}+M^{2}_{5}+M^{2}_{6}& =h_{2}\, , \\
M_{1}M_{4}+M_{2}M_{5}+M_{3}M_{6} & =h_{3}\, ,
\end{aligned} \\
\quad M_{\alpha}=\varepsilon_{\alpha \beta \gamma}M_{\beta \gamma}\, , \quad M_{\alpha+3}=M_{\alpha 4} \, ,
\qquad h_{1}, h_{2}, h_{3}, h_{4}=\hbox{const.} \notag
\end{gather}

Following Schottky \cite{scho891} (1891), see also \cite{hai83},
 the integrals can be included to the family of integrals
\begin{gather}
F(s) = \sum^{3}_{\alpha =1} \biggl( \sqrt{(s-a_\alpha )(s-a_4)}\,
M_{\beta \gamma} + \sqrt{(s-a_\beta)(s-a_\gamma)}\,  M_{\alpha 4} \biggr)^2,        \\
(\alpha ,\beta ,\gamma)=(1,2,3)\, , \qquad s\in {\mathbb C}. \notag
\end{gather}
Moreover, in \cite{scho891} it was shown that the integrals of the Frahm-Manakov top on $so(4)$
are obtained from those of the Clebsch system by a {\it linear} change of variables.
Hence, generic complex invariant manifolds of both systems are isomorphic as complex algebraic varieties,
we will use this fact several times below.

Like the Clebsch system, the Euler equations \eqref{eu} for $n=4$
admit $3\times 3$ matrix Lax representation with an elliptic spectral parameter (\cite{fed95})
\begin{gather}
\dot L(s) =[L(s),A(s)], \quad s\in {\mathbb C}, \qquad L(s), A(s) \in so(3), \label{Lax2} \\
L_{\alpha \beta} = \sqrt{(s-a_\alpha )(s-a_4)}\,M_{\beta \gamma} + \sqrt{(s-a_\beta)(s-a_\gamma)}\,  M_{\alpha 4},
\qquad (\alpha,\beta,\gamma)=(1,2,3). \notag
\end{gather}
The corresponding spectral curve $|L(s)- {\bf I} w|=0$ is reducible and its nontrivial component has the form
$$
C : \left\{ w^2 =h_{2} s^{2}+h_{1}s + h_{0}+ 2 h_{3} \sqrt{(s-a_1)(s-a_2)(s-a_3)(s-a_4)} \right \} ,
$$
$h_0,\dots h_3$ being constants of motion in \eqref{ints_M}.
Thus $C$ is an even order version of the genus 3 curve \eqref{mastercurve}.

\paragraph{Algebraic geometric structure of the complex invariant manifolds.}
As was shown in \cite{hai83}, the generic complex invariant manifold ${\mathcal I}_h$ of the Fram--Manakov top on $so(4)$,
given by the intersection of quadrics \eqref{ints_M} in ${\mathbb C}^6$ is an open subset of the Prym
variety $\Prym(C,\sigma)$, on which the flow of the system becomes linear.

Combining our previous observations and using the isomorphism between the invariant tori of
Clebsch system and the Fram--Manakov top on $so(4)$, we arrive at

\begin{proposition} \label{alg_prop_C}
A generic complex invariant manifold ${\mathcal I}_h$ of the Clebsch system is an open subset of 
the Prym variety $\Prym(C,\sigma)$, where $C$ is the spectral curve \eqref{sp_C}.
The variety $\Prym(C,\sigma)$ is isomorphic to an 8-fold unramified covering $\cal T$ of the
Jacobian of K\"otter's genus 2 curve ${\mathcal G}$ in \eqref{GammaK} and is obtained by doubling 3 of four independent
period vectors of $\Jac({\mathcal G})$.
\end{proposition}

As a result, $\Prym(C,\sigma)$ and $\Jac({\mathcal G})$ are isogeneous varieties.
\medskip

It is then natural to ask how the curve ${\mathcal G}$, on which a separation of variables is made,
can be found from the equation of the spectral curve $C$. The major contribution to solution of this problem
was made in papers \cite{avm88, hvm89}, which described how isogeneous principally polarized 2-dimensional
Abelian varieties
(and, therefore, Jacobians of genus 2 curves) can be constructed
from $\Prym(C,\sigma)$ in different ways (see also Section 3).
For the specific Abelian variety arising in the Kovalevskaya top problem, \cite{hvm89} also considered a pencil
$D_\lambda$, $\lambda \in {\mathbb P}$ of genus 3 curves coverings elliptic curves ${\cal E}_\la$, 
such that the corresponding Prym variety is dual to the same
$\Prym(C,\sigma)$ for any $\la$. Then it was shown that $D_\lambda$ contains
3 different hyperelliptic curves of genus 3, which are, respectively,
2-fold unramified coverings of three different genus 2 curves. As follows from the construction of \cite{bw85},
the Jacobians of the latter curves are all isogeneous to $\Prym(C,\sigma)$ (see also Section \ref{h30}).
It was proven that one of these genus 2 curves is birationally equivalent to the famous Kovalevskaya curve,
which appeared in the quadratures for the Kovalevskaya top (\cite{kow}).

Similar aspects of the problem  were studied in \cite{avm88},
where, in particular, the above variety $\Prym(C,\sigma)$
was described as a complete intersection of
six specific quadrics in ${\mathbb P}^7$ or the intersection of 4 specific quadrics in
${\mathbb P}^6$, the quadrics being depending on three parameters $b_1, b_2, b_3 \in {\mathbb C}$.
The paper \cite{avm88} also presented explicit algebraic equations of the three genus 2 hyperelliptic curves,
whose Jacobians are degree 2
isogeneous to $\Prym(C,\sigma)$, in terms of $b_1, b_2, b_3$. Lastly, the intersection
of the quadrics has been related (in a rather complicated way)
with the invariant tori of the Frahm--Manakov and Kovalevskaya tops, as well as of 
the case (ii) of the Hen\'on--Heiles system. 

\paragraph{Remark.} One can observe that K\"otter's expressions \eqref{6.9} for $d_\alpha$ depend on
an ordering of the roots $s_1,\dots, s_4$ of $f(r)$ in
\eqref{3.20}. Changing the ordering may transform the curve \eqref{GammaK} to another, birationally non-equivalent one.
This observation is closely related to the fact that $\Prym(C,\sigma)$ is degree 2 isogeneous to
different hyperelliptic Jacobians. Below (in Theorem \ref{main_th}) we will see that all the permutations of
$s_1,\dots, s_4$ lead to genus 2 curves giving all such Jacobians.

\section{Dual (1,2)-$\Prym$ varieties, genus 3 curves, and isogeneous hyperelliptic Jacobians} \label{dual_pryms}

Apart from the variety $\Prym(C,\sigma)$ with the period matrix $\Lambda$ in \eqref{Lambda_periods},
it is natural to consider also the (1,2)-polarized variety $\Prym^*(C,\sigma)$ {\it dual} to
$\Prym (C,\sigma)$, its period matrix can be written
as\footnote{A coordinate-free definition of the dual Prym variety can be found, e.g., in \cite{hai83} p. 466}
\begin{equation}
\label{Lambda*}
\Lambda^*=\left(
{\begin{array}{rrcc}
1 & 0 & \pi & 2\,p \\
0 & 2 & 2\,p & 4\,P
\end{array}}
 \right).\end{equation}

This matrix is obtained by dividing by 2 the 1st and 3rd periods vectors of  $\widehat\Lambda$
(equivalent to $\Lambda$) and the corresponding transformations of the coordinates in ${\mathbb C}^2$ to get a Riemann matrix. Namely,
$$
\widehat\Lambda= \left(
{\begin{array}{rrcc}
2 & 0 & 2\pi & 2p \\
0 & 1 & 2p & 2P
\end{array}}
 \right) \to \;
\left(
{\begin{array}{rrcc}
1 & 0 & \pi & 2\,p \\
0 & 1 & p & 2\,P
\end{array}}
 \right) \to \;
\left(
{\begin{array}{rrcc}
1 & 0 & \pi & 2\,p \\
0 & 2 & 2\,p & 4\,P
\end{array}}
 \right).
$$
 Note that the dual to $\Prym^*(C,\sigma)$ is just $\Prym (C,\sigma)$, and both varieties
can be seen as 4-fold coverings of each other.

The appearance of the dual Prym variety becomes more natural, if we recall the notion of dual genus 3 curves.
Namely, consider again the genus 3 curve \eqref{mastercurve} in the even order form
\begin{align*}
C : \; w^2 = g(x) +2h_{3} \sqrt{ (x-a_1)(x-a_2)(x-a_3)(x-a_4) } , \\
g(x)= h_{2} x^{2}+h_{1} x +h_{0},
\end{align*}
Again, the covering $C\to E$ is ramified at 4 zeros $Q_1,\dots,Q_4\in E$ of $w^2$ calculated
via the roots of the polynomial
\begin{align}
\begin{split}
\Psi (x) & =(h_{2}x^{2}+h_{1} x +h_{0})^{2}-4 h^{2}_{3}\Phi(x)  \\
         & = (h_2^2- 4h_3^2) (x-s_{1}) (x-s_{2})(x-s_{3})(x-s_{4})=(h_2^2- 4h_3^2) \psi(x), \\
         & \psi(x)= (x-s_{1}) (x-s_{2})(x-s_{3})(x-s_{4}) .
\end{split}\label{Psi}
\end{align}
Thus, the algebraic form of $C$ is $y^4-2 g(x) y^2 + \Psi (x)=0$.

Next, consider another genus 3 curve
\begin{gather} \label{dual_K}
K : \{ W^2 = g(x) + \sqrt{\Psi(x)} \}
\end{gather}
having polynomial form
\begin{equation} \label{dual_KK}
W^4-2 g(x) W^2 + 4 h_3^2 \Phi (x)=0  \quad  \Longleftrightarrow \quad h_3^2 y^4- g(x) y^2 + \Phi (x)=0 \quad
(W=\sqrt{2} h_3 y) .
\end{equation}
Under the involution $\rho\; :\, (x,W)\to (x, -W)$, the curve
a 2-fold covering of the elliptic curve ${\mathcal E}: \; \{Y^2=\Psi(x) \}$
ramified at 4 points ${\cal P}_1,\dots,{\cal P}_4$ on ${\mathcal E}$ whose $x$-coordinates are the roots of
$$
g^2(x)-\Psi(x) = 4 h_3^2 \Phi(x)= 4 h_3^2 (x-a_1)(x-a_2)(x-a_3)(x-a_4).
$$
Thus $x(Q_i)=s_i$ become $x$-coordinates of the branch points of the covering ${\mathcal E}\to {\mathbb P}=\{x\}$,
and, vice versa,
$x({\cal P}_i)=a_i$ are $x$-coordinates of the branch points of $E\to {\mathbb P}=\{x\}$. The branch loci of
$C \to {\mathbb P}$ and of $K\to {\mathbb P}$ have the same projection on $\mathbb P$.
In this sense, the curves $C, K$ can be called {\it dual}, or following \cite{pan86},
{\it bigonally related}\footnote{One should stress that $C$ and $K$ are not
birationally equivalent.}. As was shown in \cite{hai83, bw85}, see also \cite{au96},
$$
 K = \Prym (C,\sigma) \cap \Theta_C, \quad C = \Prym (K, \rho)\cap \Theta_K,
$$
where $\Theta_C\subset \Jac(C)$ and $\Theta_K\subset \Jac(K)$ are appropriate translations of the corresponding
theta-divisors.  Moreover, the following remarkable property holds.

\begin{proposition} \label{duality}
\textup{(\cite{hai83})} The Prym varieties $\mathrm{Prym}(C,\sigma)$ and $\mathrm{Prym}(K,\rho)$
are dual in the sense of the definition.
\end{proposition}

Curiously, in the case of the integrable Frahm--Manakov top on $so(4)$, the curves $K$ and $C$ arise as spectral curves
of the Manakov rational Lax pair (see \cite{hai83}) and, respectively, of the elliptic Lax pair \eqref{Lax2}.

We just mention that the above construction involving the dual genus 3 curves and the corresponding Prym varieties extends
to the case of the general coverings $C\to E$ described
in Theorem 1.1, as well as to two-fold coverings $S\to \Gamma$ of genus 2 curves
$\Gamma$ ramified at 6 points (see \cite{pan86} and the last section Conclusion).

\paragraph{Remark.} In this connection, it worths to notice that the curve {\it dual} to the factorized spectral curve
\eqref{sp_kow} for the Kovalevskaya top has the form
$$
 y^2 =(x^2-2Hx+2)x+ x^2 \sqrt{x^2+4(L^2-H) x + K} \, .
$$
It is singular, and its regularization is birationally equivalent to the genus 2 curve
$$
 {\cal K}\, :\;  Y^2 = (X^2-K) [ (X^2-K+4)(X-2H)+8 L^2],
$$
which, in turn, is birationally equivalent to the original Kovalevskaya curve derived in \cite{kow} for general constants of motion.
This, apparently new, observation has a transparent geometric explanation which perfectly fits into the framework of
\cite{avm88,hvm89}. The latter implies that the Jacobian of $\cal K$ is a 2-fold covering of the Prym variety associated
with the spectral curve \eqref{sp_kow}. The details will be presented elsewhere.

For the special case, when the area constant $L=0$, the curve \eqref{sp_kow} itself becomes hyperelliptic
of genus 2, and the relation between it and the Kovalevskaya curve was already described in \cite{lm99}:
these curves are connected via a Richelot transform (see also below). 

\paragraph{Relations between period matrices.} 
As was mentioned in Section 6 of \cite{hvm89}, given an arbitrary (1,2)-polarized 2-dimensional Abelian variety
${\mathcal A}_{2}$ having the period matrix $\Lambda$ as in \eqref{Lambda_periods}, namely,
$$
\Lambda =  \left(
{\begin{array}{rrcc}
1 & 0 & 2\,P & 2\,p \\
0 & 2 & 2\,p & 2\,\pi
\end{array}}
 \right),
$$
there are several ways to obtain a principally polarized 2-dimensional
Abelian variety (and therefore, the Jacobian of a genus 2 hyperelliptic curve) isogeneous to ${\mathcal A}_{2}$.
An obvious way is to divide the second period vector of $\Lambda$ by 2,
to get the period matrix
$$
\Omega_3 : = \left(
{\begin{array}{rrcc}
1 & 0 & 2\,P & 2\,p \\
0 & 1 & 2\,p & 2\,\pi
\end{array}}
 \right).
$$
On the other hand, one can double the first period vector of $\Lambda$ to get the period matrix
$$
\widetilde\Omega_3 : = \left(
{\begin{array}{rrcc}
2 & 0 & 2\,P & 2\,p \\
0 & 2 & 2\,p & 2\,\pi
\end{array}}
 \right) .
$$
This gives a variety with polarization (2,2), which is conformally equivalent to a principally polarized
variety with the period matrix
$$
\left(
{\begin{array}{rrcc}
1 & 0 & \pi & p \\
0 & 1 & p & P
\end{array}}
 \right) .
$$
In the sequel we will identify conformally equivalent Abelian tori.

Generic transformations which generate period matrices of principally polarized varieties
by {\it doubling} periods of $\Lambda$ are $\Lambda \to \Lambda {\mathcal T}$,
where $\mathcal T$ is a $4\times 4$ integer matrix, $\det {\mathcal T}= 2$, which admits the decomposition
$$
{\mathcal T}=S \,D_1 \, T,
$$
where $D_1=\mathrm{diag}(2,1,1,1)$, $T\in \Sp(4,{\mathbb Z})$,
and $S \in \Sp(4,{\mathbb Z}, (1,2))$,
the group of integer $4\times 4$ matrices preserving the symplectic matrix $J_{(1,2)}$ with polarization (1,2). That is,
$$
S J_{(1,2)} S^T = J_{(1,2)}, $$
where we denoted
$$
J_{(\alpha,\beta)}= \left(\begin{array}{rrrr}
 0 & 0 & \alpha  & 0 \\
 0 & 0 & 0 & \beta \\
 -\alpha & 0 & 0 & 0 \\
  0 & -\beta & 0& 0
\end{array}\right), \alpha,\beta\in \mathbb{N}, \qquad J_{(1,1)}=J\equiv  \left(\begin{array}{rrrr}
 0 & 0 & 1  & 0 \\
 0 & 0 & 0 &1 \\
 -1 & 0 & 0 & 0 \\
  0 & -1 & 0& 0
\end{array}\right).
$$
Thus, the action $\Lambda \to \Lambda S$ transforms ${\mathcal A}_2$ to an equivalent
(1,2)-polarized variety with permutated
period vectors, then $D_1$ multiplies the first period vector of $\Lambda S$ by 2,
producing the matrix of a principally polarized variety.

We will call
$A,B\in \Sp(4,{\mathbb Z}, (1,2))$ {\it $D_1$-equivalent} ($A \cong_{D_1} B$),
when the principally polarized period matrices
$\Lambda A D_1$, $\Lambda B D_1$ are symplectically equivalent, i.e., when there exists $T\in \Sp(4,{\mathbb Z})$
such that
\begin{equation} \label{equivABD1}
 \Lambda A D_1=\Lambda B D_1\, T, \qquad \text{that is,} \quad
T=D_1^{-1} B^{-1}A D_1 \in \Sp(4,{\mathbb Z}).
\end{equation}
The $D_1$-equivalence class of $A\in \Sp(4,{\mathbb Z}, (1,2))$ will be denoted as $\{A\}_{D_1}$.

Note that it is quite easy to check such a equivalence in practice:
since $J_{(1,2)} =D_1 J D_1$, from \eqref{equivABD1} we conclude that $D_1^{-1} B^{-1}A D_1$ is {\it always} symplectic:
\begin{gather*}
D_1^{-1} B^{-1}A D_1 \, J\, (D_1^{-1} B^{-1}A D_1)^T =
D_1^{-1} B^{-1}A \, J_{(1,2)}\, A^T (B^{-1})^T D_1^{-1}=  J,
\end{gather*}
but not always {\it integer} symplectic. So, $A,B\in \Sp(4,{\mathbb Z}, (1,2))$ are $D_1$-equivalent, if and only if
$D_1^{-1} B^{-1}A D_1$ is just integer.
\medskip

We are interested in transformations which produce all possible non-equivalent principally polarized tori.

\begin{proposition} \label{3_Ts} There are at least 3 $D_1$-equivalence classes in $\Sp(4,{\mathbb Z}, (1,2))$
represented by the matrices
\begin{equation} \label{SS}
S_1=\left(
{\begin{array}{rrrr}
0 & 0 & -1 & -2 \\
1 & -1 & -1 & -2 \\
1 & 0 & 0 & 2 \\
0 & 0 & 0 & -1
\end{array}}
 \right) , \quad
S_2= \left(
{\begin{array}{rrrr}
0 & 0 & -1 & -2 \\
1 & -1 & -1 & -2 \\
1 & 0 & -1 & 0 \\
0 & 0 & 0 & -1
\end{array}}
 \right), \quad S_3=1_4,
\end{equation}
which, under the action $\Lambda \to \Lambda S_i D_1$, give 3 non-equivalent principally polarized Abelian varieties.

\end{proposition}


The proof of Proposition \ref{3_Ts} is a direct calculation, it uses the above criterion of $D_1$-equivalence.
\medskip

Obviously, the action of $S_3$ combined with $D_1$
produces the above period matrix $\widetilde\Omega_3$, whereas $\Lambda S_1 D_1 T_1, \Lambda S_2 D_1 T_2$
are period matrices equivalent to, respectively,
$$
\widetilde \Omega_1  \sim  \left(
{\begin{array}{rrcc}
1 & 0 & 4\,P & 2\,p \\
0 & 1 & 2\,p & \pi
\end{array}}
 \right), \quad
\widetilde \Omega_2 \sim \left(
{\begin{array}{rrcc}
1 & 0 & P+ \frac 12 & p \\
0 & 1 & p & \pi
\end{array}} \right).
$$
Explicitly, here the corresponding transformations $T_i\in\mathrm{Sp}(4,\mathbb{Z})$ are
\[
T_1=\left(  \begin{array}{rrrr}0&0&1&1\\  1&-1&2&2\\  -1&0&0&2\\  0&0&0&-1 \end{array} \right),\quad
T_2=\left(  \begin{array}{rrrr}-1&0&0&1\\  0&-1&1&2\\  -2&0&-1&2\\  0&0&0&-1     \end{array} \right), \quad T_3=1_4
\]
Abelian varieties with the period matrices $\widetilde\Omega_1, \widetilde\Omega_2, \widetilde\Omega_3$ are Jacobian varieties and will be denoted as
$\Jac( \widetilde\Gamma_{1} ), \Jac(\widetilde\Gamma_{2}),\Jac(\widetilde\Gamma_{3})$ for appropriate genus 2 curves
$ \widetilde\Gamma_{1}, \widetilde\Gamma_{2}, \widetilde\Gamma_{3}$.
By construction, each of the Jacobians is a 2-fold covering of the variety ${\mathcal A}_2$.
\medskip

In a similar way, other three non-equivalent principally polarized varieties are obtained by the action
\begin{equation}
\label{low}
\Lambda \to \Lambda \widehat S D_2^{-1} T, \quad D_2 =\text{diag}(1,2,1,1), \quad
\widehat S \in \Sp(4,{\mathbb Z}, (1,2)),
\quad T \in \Sp(4,{\mathbb Z}) .
\end{equation}
Now $A,B\in \Sp(4,{\mathbb Z}, (1,2))$ will be called {\it $D_2$-equivalent} ($A \cong_{D_2} B$),
when the principally polarized period matrices
$\Lambda A D_2^{-1}$, $\Lambda B D_2^{-1}$ are symplectically equivalent, i.e.,
when there exists $T\in \Sp(4,{\mathbb Z})$ such that
\begin{equation} \label{equivAB}
 \Lambda A D_2^{-1} = \Lambda B D_2^{-1}\, T, \qquad \text{i.e., when } \quad
T=D_2 B^{-1}A D_2^{-1} \in \Sp(4,{\mathbb Z}).
\end{equation}
The $D_2$-equivalence class of ${\mathcal A}\in \Sp(4,{\mathbb Z}, (1,2))$ will be denoted as $\{ {\mathcal A} \}_{D_2}$.

Note that the classes $\{{\mathcal A} \}_{D_2}$ and $\{ {\mathcal A} \}_{D_1}$ are distinct. In particular,
$S_1, S_2, S_3$ in \eqref{SS} are not $D_1$-equivalent, but are $D_2$-equivalent, that is, the period matrices
$\Lambda S_i D_2$ are all equivalent.

Similarly to Proposition \ref{3_Ts}, one can prove that there are at least 3 $D_2$-equivalence classes represented by matrices
\begin{equation}
\widehat S_1 = J= \left(\begin{array}{rrrr}
 0 & 0 & 1  & 0 \\
 0 & 0 & 0 & 1 \\
 -1 & 0 & 0 & 0 \\
  0 & -1 & 0& 0
\end{array}\right), \quad
\widehat S_2 = \left(\begin{array}{rrrr}
 -2 & -2 & 1  & -2 \\
 -1 & -1 & 0 & 0 \\
 -1 & 0 & 0 & 0 \\
  0 & -1 & 0 & -1
\end{array}\right), \quad  \widehat {S}_3 =1_4 .
\end{equation}
Clearly, the period matrix $\Lambda \widehat{ S}_3 D_2^{-1}$ is just the matrix $\Omega_3$ indicated above,
and the other period matrices are equivalent to
$$
\Omega_1 = \Lambda \widehat S_1 D_2^{-1}T_1 \sim
\left(\begin{array}{cccc}
1 & 0 & 2P & p \\
0 & 1 & p & \pi/2
\end{array}\right), \quad
\Omega_2 = \Lambda \widehat S_2 D_2^{-1}T_2 \sim
\left(\begin{array}{rrrr}1&0&2P&p\\ 0&1&p&\frac{1+\pi}{2}\end{array}\right)
 $$
The transformations $T_1, T_2$ that take $\Lambda$ to the above form are
\[
 T_1=J^{-1}, \quad T_2=\left(  \begin{array}{rrrr}0&0&-1&0\\0&-2&2&-1\\1&0&-2&-1\\0&1&-1&0
\end{array} \right), \quad  T_3=1_4 .
\]
Abelian varieties with the period matrices $\Omega_1, \Omega_2, \Omega_3$  will be denoted
as $\Jac(\Gamma_1), \Jac(\Gamma_{2}),\Jac(\Gamma_{3})$ for appropriate genus 2 curves $\Gamma_1, \Gamma_2, \Gamma_3$.
Then ${\mathcal A}_2$ can be viewed as a 2-fold covering of each of the above Jacobians.

Note that the transformations of $\Lambda$ equivalent to those given by
$\widehat {S}_1 , \widehat {S}_2, \widehat {S}_3$ were indicated in \cite{hvm89}.
\medskip

Now let our abstract Abelian variety ${\mathcal A}_2$ be the Prym variety $\Prym(C,\sigma)$.
Take the dual variety $\mathrm{Prym}^*(C,\sigma)$ with the period matrix $\Lambda^*$ given in \eqref{Lambda*}
and apply to it the same algorithm generating three principally polarized tori, i.e., Jacobians ''above''
$\mathrm{Prym}^*(C,\sigma)$, and three Jacobians ''below'' it,
by applying transformations of the same equivalence classes $\{S_1\}, \{S_2\}, \{S_3\}$ and
$\{\hat S_1\}, \{\hat S_3\}, \{\hat S_3\}$ respectively.

\begin{proposition} \label{updown}
The 3 Jacobians above $\mathrm{Prym}^*(C,\sigma) $ are equivalent to the 3 Jacobians below $\mathrm{Prym}(C,\sigma)$
and vice versa: the 3 Jacobians below $\mathrm{Prym}^*(C,\sigma)$ are equivalent to
the 3 Jacobians above $\mathrm{Prym}(C,\sigma)$.
\end{proposition}

Indeed, the right action of $T_i = S_i D_1$, $i=1,2,3$ on $\Lambda^*$ give, respectively, the
period matrices equivalent to
\begin{equation} \label{dual_upper}
\begin{gathered}
\Omega_1^* \sim \left( \begin{array}{cccc} 1 & 0 & 2\pi & 2p \\ 0 & 1 & 2p & 2P  \end{array} \right), \quad
\Omega_2^*\sim \left(\begin{array}{cccc} 1 & 0 & (\pi+1)/2 & p \\ 0 & 1 & p & 2P \end{array} \right), \\
\Omega_3^*\sim \left( \begin{array}{cccc} 1 & 0 & \pi/2 & p \\ 0 & 1 & p & 2P \end{array} \right).
\end{gathered}
\end{equation}
One observes that $\Omega_1^*, \Omega_2^*, \Omega_3^*$ are  equivalent to, respectively,
the period matrices $\Omega_3, \Omega_2, \Omega_1$ given above. The second statement of the proposition is proved
in a similar way. $\square$

These observations can be depicted in the following diagram, where the arrows denote the
corresponding 2:1 coverings. 

\vskip 0.5cm
\begin{figure} \hskip2.5cm
\unitlength=1mm
\begin{picture}(60,20)(0,0)
\put(10,20){\makebox(0,0){$\left( \begin{array}{cc} 4P&2p\\ 2p&\pi  \end{array}\right)$}}
\put(50,20){\makebox(0,0){$\left( \begin{array}{cc} P+1/2&p\\ p&\pi  \end{array}\right)$}}
\put(90,20){\makebox(0,0){$\left( \begin{array}{cc} \pi&p\\ p&P  \end{array}\right)$}}
\put(52,-5){\makebox(0,0){$\left( \begin{array}{cccc} 1&0&2P&2p\\ 0&2&2p&2\pi  \end{array}\right)$}}
\put(52,2){\vector(0,1){14}}
\put(33,-3){\vector(-1,1){16}}
\put(72,-3){\vector(1,1){16}}
\put(52,-12){\vector(0,-1){14}}
\put(33,-9){\vector(-1,-1){16}}
\put(72,-9){\vector(1,-1){16}}
\put(10,-35){\makebox(0,0){$\left( \begin{array}{cc} 2P&p\\ p&\pi/2  \end{array}\right)$}}
\put(50,-35){\makebox(0,0){$\left( \begin{array}{cc} 2P&p\\ p&(\pi+1)/2  \end{array}\right)$}}
\put(90,-35){\makebox(0,0){$\left( \begin{array}{cc} 2\pi&2p\\ 2p&2P  \end{array}\right)$}}
\put(52,-65){\makebox(0,0){$\left( \begin{array}{cccc} 1&0&\pi&2p\\ 0&2&2p&4P  \end{array}\right)$}}
\put(52,-58){\vector(0,1){14}}
\put(33,-60){\vector(-1,1){16}}
\put(33,-69){\vector(-1,-1){16}}
\put(72,-60){\vector(1,1){16}}
\put(72,-69){\vector(1,-1){16}}
\put(52,-70){\vector(0,-1){14}}
\put(10,-90){\makebox(0,0){$\left( \begin{array}{cc} \pi&2p\\ 2p&4P  \end{array}\right)$}}
\put(50,-90){\makebox(0,0){$\left( \begin{array}{cc} \pi&p\\ p&P+1/2  \end{array}\right)$}}
\put(90,-90){\makebox(0,0){$\left( \begin{array}{cc} \pi&p\\ p&P \end{array}\right)$}}
\put(16,4){\makebox(0,0){$D_1\circ S_1$}}
\put(44,8){\makebox(0,0){$D_1\circ S_2$}}
\put(88,4){\makebox(0,0){$D_1\circ S_3$}}
\put(16,-15){\makebox(0,0){$ D_2^{-1}\circ \hat{S}_1$}}
\put(43,-19){\makebox(0,0){$D_2^{-1}\circ \hat{S}_2$}}
\put(88,-15){\makebox(0,0){$D_2^{-1}\circ \hat{S}_3$}}
\put(17,-55){\makebox(0,0){$D_1\circ S_3$}}
\put(60,-51){\makebox(0,0){$D_1\circ S_2$}}
\put(85,-55){\makebox(0,0){$D_1\circ S_1$}}
\put(16,-75){\makebox(0,0){$D_2^{-1}\circ \hat{S}_3$}}
\put(60,-78){\makebox(0,0){$D_2^{-1}\circ \hat{S}_2$}}
\put(88,-75){\makebox(0,0){$D_2^{-1}\circ \hat{S}_1$}}
\end{picture}
\end{figure}
\vskip 10cm
Diagram 1. Relation between period matrices of the six Jacobians and two Prym varieties.
(For the Jacobians, only the Riemann matrices are shown.)
\vskip1cm

Notice that each of the 3 Jacobians ''above'' $\Prym(C,\sigma)$ is a 16-fold covering of the one of the
3 Jacobians ''below'' $\Prym^*(C,\sigma)$, obtained by doubling each of the 4 period vectors.
Hence they are not only isogeneous,
but {\it isomorphic}, and, as shown on the diagram, can be identified with each other.

Moreover, since the dual to  $\Prym^*(C,\sigma)$ is $\Prym(C,\sigma)$ again,
the above diagram is closed in the sense that one cannot obtain other non-equivalent Jacobians
by the actions of the above equivalence classes.

\paragraph{The Richelot correspondences.}
One can observe that some of the period matrices below and above $\Prym(C,\sigma)$ are obtained from each other
by multiplying the 3rd ad 4th columns (i.e., the Riemann matrix $\tau$) by 2. Namely, the above diagram can be added with
the following one
\begin{equation} \label{tau->2tau}
\begin{array}{ccccc}
\tilde \Omega _{1} &  & \tilde \Omega _{2} &  & \tilde\Omega _{3} \\
\uparrow  & \nwarrow  &  & \searrow  & \downarrow  \\
\Omega _{1} &  & \Omega _{2} &  & \Omega _{3}
\end{array}
\end{equation}
where arrows denote the duplication $(1_2\, \tau) \to (1_2\, 2\tau)$). So, the
corresponding curves $\Gamma_\alpha, \widetilde\Gamma_\beta$ are related algebraically via the
{\it Richelot transformation} (see, e.g., \cite{bm88}), which
generalizes the Landen transformation for elliptic curves to the genus 2 case. 
One should stress that a Richelot transformation is, in fact, only a correspondence:
for example, $\tilde \Omega _{1}$ is obtained by the duplication of symplectically non-equivalent period matrices
$\Omega_1$ and $\Omega_2$.

\section{The six genus 2 curves in an algebraic form} \label{alg6}

Before we proceed to the main theorem of the section, we will need the following

\begin{lemma} \label{reciprocal}
Given a genus 2 curve $G$ written in form
\begin{equation} \label{G_canon}
w^2 = z(z-e_1) (z-e_2)(z-e_3)(z-e_1 e_2 e_3)
\end{equation}
with distinct $e_1,e_2,e_3$, any of the reciprocal transformations of the type
\begin{equation} \label{e-e}
\begin{gathered}
(e_1,e_2,e_3) \to  (1/e_1,1/e_2,1/e_3) , \\
(e_1,e_2,e_3 ) \to  (e_1,e_2,1/e_3)\quad \text{or}
\quad(e_1,e_2,e_3 ) \to  (e_1,1/e_2,1/e_3)
\end{gathered}
\end{equation}
and similar transformations obtained by permutations of indices,
takes $G$ to a birationally equivalent curve. The same holds for the sign flip
$(e_1,e_2,e_3) \to (-e_1,-e_2,-e_3)$.
\end{lemma}

Indeed, for the substitutions \eqref{e-e}, the corresponding M\"obius transformations $z\to \zeta$ that take the curve back to
the original form \eqref{G_canon} are, respectively,
\[
z=\frac{1}{\zeta},\quad  z=\frac{e_1 e_2}{\zeta},\quad z=\frac{a\zeta+b}{c\zeta+1} ,
\]
where
\begin{gather*}
a=\frac{e_3^2-1}{e_3\, \Delta},\quad
b=- \frac{e_1 (e_2 e_3 -1)+e_3-e_2}{e_3\, \Delta},
\quad c=\frac{e_1 e_2 (e_3^2-1)}{e_3\, \Delta},\\
\Delta=e_1 (e_3-e_2)+ e_2 e_3-1 \, .
\end{gather*}
The change $(e_1,e_2,e_3) \to (-e_1,-e_2,-e_3)$ is compensated by $z \to -z$. $\square$
\medskip

We now give an algebraic description of the 6 Jacobians already presented in Diagram 1.

\begin{theorem}[Main] \label{main_th}
1) For a generic genus 3 curve \eqref{mastercurve}
there are exactly 6 birationally non-equivalent genus 2 curves $\Gamma_1, \Gamma_2, \Gamma_3$ and
$\tilde\Gamma_1, \tilde\Gamma_2, \tilde\Gamma_3$ such that
$\Prym(C,\sigma)$ is a 2-fold unramified covering of $\Jac(\Gamma_{1}), \Jac(\Gamma_{2}), \Jac(\Gamma_{3})$, and
$\Jac(\widetilde\Gamma_{1}), \Jac(\widetilde\Gamma_{2}), \Jac(\widetilde\Gamma_{3})$ are
2-fold unramified coverings of $\Prym(C,\sigma)$.  \\
Moreover, $\Jac(\Gamma_{1}), \Jac(\Gamma_{2}), \Jac(\Gamma_{3})$ are also 2 fold coverings of the dual variety
$\Prym^*(C,\sigma) \cong \Prym(K,\rho)$,
whereas the latter is a 2-fold covering of $\Jac(\widetilde\Gamma_{1}), \Jac(\widetilde\Gamma_{2}),
\Jac(\widetilde\Gamma_{3})$\footnote{
Here we identify 2-dimensional Abelian varieties obtained one from another by duplication of all of the 4 periods.},
as depicted in the following diagram, where the arrows denote the corresponding 2:1 coverings and double
arrows indicate 4-fold coverings:
\begin{equation} \label{diagram}
\begin{array}{ccccc}
\Jac(\widetilde\Gamma_{1}) &  & \Jac(\widetilde\Gamma_{2}) &  & \Jac(\widetilde\Gamma_{3}) \\
\Vert & \searrow  & \downarrow  & \swarrow  & \Vert \\
\Vert &  &  \Prym(C,\sigma ) &  & \Vert \\
\Downarrow  & \swarrow  & \downarrow & \searrow  & \Downarrow  \\
\Jac(\Gamma_1) &  & \Jac(\Gamma_{2}) &  & \Jac(\Gamma_3 ) \\
\Vert & \searrow  & \downarrow  & \swarrow  &  \Vert \\
\Vert &  & \Prym^* (C,\sigma ) &  &  \Vert \\
\Downarrow & \swarrow  & \downarrow  & \searrow  &  \Downarrow \\
\Jac(\widetilde\Gamma_{1}) &  & \Jac(\widetilde\Gamma_{2}) &  & \Jac (\widetilde\Gamma_{3})
\end{array}
\end{equation}

2) Assume that $h_3\ne 0$ and all the roots $s_1,\dots,s_4$ of $\psi(x)$ in \eqref{Psi0} are distinct.
The curves $\tilde\Gamma_1, \tilde\Gamma_2, \tilde\Gamma_3$ can be written in the hyperelliptic form
\begin{equation} \label{Gamma}
\widetilde\Gamma \, :\;  w^2 = z(z-d_1^2) (z-d_2^2)(z-d_3^2)(z-d_1^2 d_2^2 d_3^2) \, ,
\end{equation}
where, for the odd order case described by \eqref{mastercurve},
the branch points of the curves are determined as any solutions of the following three equations,
respectively
\begin{align}
\widetilde\Gamma_1 \, : \;
d_{\alpha }^{2}+\frac{1}{d_{\alpha }^{2}}
& =2 \frac{ S_{\left( 14\right) \left( 23\right) }^{(\alpha) } - S_{\left( 42\right) \left(13\right) }^{(\alpha) } }
{S_{\left(12\right) \left( 34\right) }^{(\alpha) }} \; \Longleftrightarrow \;
d_\alpha = \frac{ \sqrt{S_{(14)\, (23)}^{(\alpha)}} \pm  \sqrt{S_{(24)\,(13)}^{(\alpha)} }  }
{\sqrt{ S_{(12)\, (34)}^{(\alpha)} }}, \notag \\
\widetilde\Gamma_2 \, : \;
d_{\alpha }^{2}+\frac{1}{d_{\alpha }^{2}}
& =2 \frac{ S_{\left( 42\right) \left( 13\right) }^{(\alpha) } - S_{\left( 12\right) \left(24\right) }^{(\alpha) } }
{S_{\left(14\right) \left( 23\right) }^{(\alpha) }} \; \Longleftrightarrow \;
d_\alpha = i \frac{ \sqrt{S_{(12)\, (34)}^{(\alpha)}} \pm  \sqrt{S_{(24)\,(13)}^{(\alpha)} }  }
{\sqrt{ S_{(14)\, (23)}^{(\alpha)} }},   \label{d_s} \\
\widetilde\Gamma_3 \, : \;
d_{\alpha }^{2}+\frac{1}{d_{\alpha }^{2}}
& =2 \frac{ S_{\left(12\right) \left( 34\right) }^{(\alpha) } - S_{\left( 14\right) \left(23\right) }^{(\alpha)} }
{S_{\left(42\right) \left( 13\right) }^{(\alpha) }} \; \Longleftrightarrow \;
d_\alpha = \frac{ \sqrt{S_{(14)\, (23)}^{(\alpha)}} \pm  \sqrt{S_{(12)\,(34)}^{(\alpha)} }  }
{\sqrt{ S_{(42)\, (13)}^{(\alpha)} }}, \notag \\
\quad  \alpha=1,2,3 .\notag
\end{align}
The equations are obtained from each other by permutations of the terms
$S_{(12)(34)}^{(\alpha)}$, $S_{(14)(23)}^{(\alpha)}$, $S_{(42) (13)}^{(\alpha)}$ given by
\begin{gather}
S_{(ij)(kl)}^{(\alpha) }
=(s_{i}-s_{j})(s_{k}-s_{l}) \left[ (s_i - c_\alpha)(s_j -c_\alpha) +(s_{k}-c_\alpha)(s_{l}-c_\alpha)
+2 \sqrt{\psi (c_\alpha)}\; \right ] \label{d_ss}
\end{gather}
and having the properties
\begin{gather}
 S_{(ij) (kl) }^{(\alpha) } = - S_{(ji)(kl) }^{(\alpha) } = - S_{(ij)(lk) }^{(\alpha) }=S_{(ji)(lk)}^{(\alpha) }  , \quad
S_{(ij)(kl) }^{(\alpha) } = S_{(kl)(ij) }^{(\alpha) } . \label{symm_rel}
\end{gather}

Equivalently, in view of the relations
$\sqrt{\psi (c_\alpha)}= - g(c_\alpha)$ in \eqref{imp_cond}, for $h_2=1$ we have
\begin{gather}
S_{(ij)(kl) }^{(\alpha) }
 =(s_{i}-s_{j})(s_{k}-s_{l}) ( (s_{i}s_{j}+s_{k}s_{l}) - 4 c_\alpha h_3^2-2h_0 ).  \label{d_ss_a}
 \end{gather}
Note that
\begin{equation} \label{sum_0}
S_{(12)(34)}^{(\alpha)} + S_{(14)(23)}^{(\alpha)}+ S_{(42)(13)}^{(\alpha)}=0, \qquad \alpha=1,2,3,
\end{equation} 
which justifies the equivalences in \eqref{d_s}. 
Flipping the sign $\pm$ in \eqref{d_s} implies the change $d_\alpha \to -1/d_\alpha$,
which, in view of Lemma \ref{reciprocal}, leads to an equivalent curve \eqref{Gamma}.
\medskip

3) In the even order case in \eqref{mastercurve}, under the assumption of item (2),
the curves $\tilde\Gamma_1, \tilde\Gamma_2, \tilde\Gamma_3$ have the form \eqref{Gamma}, \eqref{d_s} with
$S_{(ij)(kl)}^{(\alpha)}$ replaced by
\begin{align}
S_{(ij)(kl)}^{(\alpha,4) }
 & =(s_{i}-s_{j})(s_{k}-s_{l}) \left[
(s_i - a_\alpha)(s_j -a_\alpha) (s_k - a_4)(s_j -a_4) \right. \notag \\
&\quad \left. + (s_i - a_4)(s_j -a_4) (s_k - a_\alpha)(s_j -a_\alpha)
- 2 \sqrt{ \psi(a_\alpha)} \sqrt{\psi(a_4) } \right], \label{S_even_roots}  \\
 & \qquad \alpha=1,2,3 . \notag
\end{align}
In view of relations
\begin{equation} \label{imp_even}
\sqrt{\psi (a_j)}= - \dfrac{ g(a_j) }{\sqrt{h_2^2-4 h_3^2}}, \quad j =1,2,3,4,
\end{equation}
the above rereads in the rational form
\begin{align}
S_{(ij)(kl)}^{(\alpha,4) }
& =(s_{i}-s_{j})(s_{k}-s_{l}) \bigg [
(s_i - a_\alpha)(s_j -a_\alpha) (s_k - a_4)(s_j -a_4)  \notag \\
 & \quad + \left. (s_i - a_4)(s_j -a_4) (s_k - a_\alpha)(s_j -a_\alpha)
  -2 \frac{g(a_\alpha) g(a_4)}{h_2^2-4 h_3^2} \right].  \label{S_even}
\end{align}
Note that, apart from the analogs of \eqref{symm_rel}, one has
$$
   S_{(ij)(kl)}^{(\alpha,4) } =S_{(ij)(kl)}^{(4, \alpha) }, \quad
\frac{ S_{\left( 14\right) \left( 23\right) }^{(\alpha,4) } - S_{\left( 42\right) \left(13\right) }^{(\alpha,4) } }
{S_{\left(12\right) \left( 34\right) }^{(\alpha,4)}} =
\frac{ S_{\left( 14\right) \left( 23\right) }^{(\beta,\gamma) } - S_{\left( 42\right) \left(13\right) }^{(\beta,\gamma) } }
{S_{\left(12\right) \left( 34\right) }^{(\beta,\gamma)}} \, ,
$$
for $(\alpha,\beta,\gamma)=(1,2,3)$ and similar relations generated by permutations of $(1,2,3,4)$.
The expressions \eqref{d_s}, \eqref{S_even} and, therefore, the curves \eqref{Gamma} are invariant
with respect to simultaneous projective transformations of $s_1,\dots, s_4, a_1,\dots, a_4$ induced by such transformations
of the $x$-coordinate in the equation of $C$.
\medskip

4) In the same even order case the curves $\Gamma_1, \Gamma_2, \Gamma_3$ can be
written in the form
\begin{equation} \label{Gamma0}
\Gamma \, :\;  w^2= z(z-k_1^2) (z-k_2^2)(z-k_3^2)(z-k_1^2 k_2^2 k_3^2) \, ,
\end{equation}
where $k_1, k_2, k_3$ are solutions of \eqref{d_s}, \eqref{d_ss}, \eqref{d_ss_a} after the permutation
$$
\{s_1,s_2,s_3,s_4\} \longleftrightarrow \{a_1,a_2,a_3,a_4\}.
$$
Namely they are solutions of
\begin{gather}
k_{\alpha }^{2}+\frac{1}{k_{\alpha }^{2}}
=2 \frac{ {\bar S}_{\left( 14\right) \left( 23\right) }^{(\alpha,4) } - {\bar S}_{\left( 42\right) \left(13\right) }^{(\alpha,4) } }
{{\bar S}_{\left(12\right) \left( 34\right) }^{(\alpha) }} \,, \notag \\
k_{\alpha }^{2}+\frac{1}{d_{\alpha }^{2}}
=2 \frac{ {\bar S}_{\left( 42\right) \left( 13\right) }^{(\alpha,4) } - {\bar S}_{\left( 12\right) \left(24\right) }^{(\alpha,4) } }
{{\bar S}_{\left(14\right) \left( 23\right) }^{(\alpha,4) }} \, , \label{d_k} \\
k_{\alpha }^{2}+\frac{1}{d_{\alpha }^{2}}
=2 \frac{ {\bar S}_{\left(12\right) \left( 34\right) }^{(\alpha,4) } - {\bar S}_{\left( 14\right) \left(23\right) }^{(\alpha,4)} }
{{\bar S}_{\left(42\right) \left( 13\right) }^{(\alpha,4) }} \, , \notag \\
\quad  \alpha=1,2,3 ,\notag
\end{gather}
with\footnote{Note the difference in the last terms in \eqref{S_even} and \eqref{S_even_dual}. }
\begin{align}
{\bar S}_{(ij)(kl)}^{(\alpha,4) }
& = (a_{i}-a_{j})(a_{k}-a_{l}) \notag \\
& \quad \cdot \left[
\sqrt{ (a_i - s_\alpha)(a_j -s_\alpha) (a_k - s_4)(a_j -s_4)} -
 \sqrt{ (a_i - s_4)(a_j -s_4) (a_k - s_\alpha)(a_j -s_\alpha) } \right]^2 \notag \\
& = (a_{i}-a_{j})(a_{k}-a_{l})  \bigg [(a_i - s_\alpha)(a_j -s_\alpha) (a_k - s_4)(a_j -s_4) \notag \\
& \qquad \qquad +(a_i - s_4)(a_j -s_4) (a_k - s_\alpha)(a_j -s_\alpha)- \frac{g(s_\alpha) g(s_4) }{2 \, h_3^2} \bigg]. \label{S_even_dual}
\end{align}

Thus, various permutations of $\{s_1,s_2,s_3,s_4\}, \{a_1,a_2,a_3,a_4\}$ in the formulas
\eqref{d_s}, \eqref{S_even} produce
the branch points of all the six genus 2 curves whose Jacobians appear in the diagram \eqref{diagram}.
 \medskip

5) In the odd order case described by equation \eqref{mastercurve}, the corresponding
 curves $\Gamma_1, \Gamma_2, \Gamma_3$ have
the form \eqref{Gamma0}, where $k_\alpha^2$ are the solutions of the following equations, respectively
\begin{gather}
k_{\alpha }^{2}+\frac{1}{k_{\alpha }^{2}}
=2 \frac{ {\bar S}_{(23)}^{(\alpha) } - {\bar S}_{(31) }^{(\alpha) } }
{{\bar S}_{(12)}^{(\alpha)}} \,, \quad
k_{\alpha }^{2}+\frac{1}{k_{\alpha }^{2}}
=2 \frac{ {\bar S}_{(31) }^{(\alpha) } - {\bar S}_{(12 ) }^{(\alpha) } }
{{\bar S}_{(23) }^{(\alpha,4) }} \, , \label{k_k} \\
k_{\alpha }^{2}+\frac{1}{k_{\alpha }^{2}}
=2 \frac{ {\bar S}_{(12) }^{(\alpha) } - {\bar S}_{(23) }^{(\alpha)} }
{{\bar S}_{(31) }^{(\alpha) }} \, , \qquad  \alpha=1,2,3 ,\notag
\end{gather}
with
\begin{align}
{\bar S}_{(ij)}^{(\alpha)} & 
 = (c_{i}-c_{j}) \left[ \sqrt{ (c_i - s_\alpha)(c_j -s_\alpha)(c_k - s_4)}
+ \sqrt{ (c_i - s_4)(c_j -s_4)(c_k - s_\alpha)} \right] ^2  \notag \\
 & = (c_{i}-c_{j})  \left[ (c_i - s_\alpha)(c_j -s_\alpha)(c_k - s_4) + (c_i - s_4)(c_j -s_4)(c_k - s_\alpha)
  + \frac{g(s_\alpha) g(s_4)}{2 \, h_3^2} \right] ,  \label{S_odd_dual} \\
& \qquad (i,j,k) =(1,2,3). \notag
\end{align}
Note that again ${\bar S}_{(23)}^{(\alpha) } + {\bar S}_{(31)}^{(\alpha)} + {\bar S}_{(12)}^{(\alpha)}=0$.

\end{theorem}


We stress that, in view of Lemma \ref{reciprocal}, two different solutions $d_\alpha^2$
of \eqref{d_s} lead to birationally equivalent curves $\tilde\Gamma, \Gamma$. The same holds for
the two solutions $k_\alpha^2$ of \eqref{d_k}.

\paragraph{Remark.} Theorem \ref{main_th} also holds when one of the roots $s_1,\dots,s_4$ coincides
with a root $c_i$ ($a_i$) of $\Phi(x)$, as happens, for example, in the case of the spectral curve $C$ of
the generalized H\'enon--Heiles system with a quartic potential considered in \cite{bef95}).
In this case the expressions for $d_\alpha$ can be considerably simplified.
\medskip

The theorem can be accompanied with the following observation, which was first made in \cite{avm88} in another context.

\begin{proposition} Let the triples
$$
\{d_1^2, d_2^2, d_3^2 \}, \quad \{(d_1')^2, (d_2')^2, (d_3')^2 \}, \quad
\{(d_1'')^2, (d_2'')^2, (d_3'')^2 \}
$$
be any solutions of the three equations \eqref{d_s} respectively. Then, for an appropriate choice of signs,
\begin{equation} \label{ddd}
{d_\alpha}' = i \frac{ 1 \pm d_\alpha }{1 \mp d_\alpha}, \quad
{d_\alpha}'' = \frac{i  \pm d_\alpha }{i \mp d_\alpha}, \qquad \alpha=1,2,3 .
\end{equation}
\end{proposition}

\noindent{\it Proof.} Using the first and second lines in \eqref{d_s},
we rewrite the first relation in \eqref{ddd} as
$$
d_\alpha'= \frac{ \sqrt{S_{(42)\, (13)}^{(\alpha)}} + i\sqrt{S_{(12)\,(34)}^{(\alpha)} }  }
{\sqrt{ S_{(14)\, (23)}^{(\alpha)} }}
= \frac {\sqrt{S_{(12)\, (34)}^{(\alpha)}}- \sqrt{S_{(14)\, (23)}^{(\alpha)}}- i \sqrt{S_{(42)\, (13)}^{(\alpha)}} }
{\sqrt{S_{(12)\, (34)}^{(\alpha)}}+ \sqrt{S_{(14)\, (23)}^{(\alpha)}} + i \sqrt{S_{(24)\, (13)}^{(\alpha)}} } .
$$
The latter indeed holds in view of \eqref{sum_0}. The second formula in \eqref{ddd} is proved in a similar way.
$\square$

\paragraph{Remark.} Theorem \ref{main_th} says that the choice of one of 3 genus 2 curves
$\tilde\Gamma_1, \tilde\Gamma_2, \tilde\Gamma_3$
depends on the choice of
partition of the roots $\{s_1,s_2,s_3,s_4\}$ of $\psi(x)$ into 2 unordered pairs (in total, three different partitions).
This, in particular, implies that the coefficients of any of $\tilde\Gamma_1, \tilde\Gamma_2, \tilde\Gamma_3$
cannot be expressed only in terms of the coefficients of the genus 3 curve $C$.
The same holds for the curves $\Gamma_1, \Gamma_2, \Gamma_3$.

If $C$ is the spectral curve of a Lax pair of one of the systems described in Section 2,
the above observation also implies that
no one of the above curves $\tilde\Gamma_i, \Gamma_i$, on which a separation of variables can be constructed,
depends on the constants of motion $h_0,h_1, h_2,h_3$ in a rational way. Thus we arrive at

\begin{proposition} It is impossible to construct a $2\times 2$ matrix Lax representation of the Clebsch or Frahm--Manakov
systems with a rational spectral parameter, such that its spectral curve $\mathcal G$ be equivalent to one of the six genus 2
curves $\tilde\Gamma_\alpha, \Gamma_\alpha$, $\alpha=1,2,3$ and such that the coefficients of $\mathcal G$ produce first integrals which are
rational in the original variables of the problem.
\end{proposition}

Notice that the proposition does not exclude a (hypothetical) existence of other \\ $2\times 2$ matrix Lax representation with a rational spectral
parameter such that its genus 2 spectral curve ${\cal H}$ is not birationally equivalent to the 6 curves
$\tilde\Gamma, \Gamma$, but $\Jac({\cal H})$ is related to $\Prym^*(C,\sigma)$ via an isogeny of degree {\it higher} than 2.



This situation described above contrasts to the case when a curve $C$ covers a hyperelliptic curve with
just 2 branch points:
it is known that then the corresponding Prym variety is principally polarized and isomorphic to the Jacobian
of another hyperelliptic curve (\cite{Mum, Dal}). The latter is unique (up to birational equivalence), and its equation
can be written in terms of the coefficients of $C$, as was made in \cite{Levin}.
\medskip


\noindent{\it Proof of Theorem} \ref{main_th}.
1) The fact that there are {\it exactly} 3 non-isomorphic Jacobians which are
2-fold unramified coverings of $\Prym(C,\sigma)$ and that the latter covers exactly 3 others non-isomorphic Jacobians
was proved in \cite{hvm89} (Theorems 3 and 5). The rest of item (1) is a simplified reformulation of
Proposition \ref{updown} and Diagram 1.

It should be noticed that a part of Diagram \eqref{diagram} (including both Prym varieties and the Jacobians
of $\tilde\Gamma_1, \tilde\Gamma_2, \tilde\Gamma_3$) already appeared in Section 3 of \cite{avm88}.
\medskip

2) Expressions \eqref{d_s} for the odd order case
follow straightforwardly from the K\"otter formulas \eqref{6.9} for the branch points $d_\alpha^2$ of the
curve \eqref{GammaK}. 
Namely, if we identify $\Jac({\mathcal G})$ with its 16-fold covering obtained by doubling all 4 period vectors, Proposition
\ref{alg_prop_C} implies that $\Jac({\mathcal G})$ can also be regarded as a 2-fold unramified covering of
$\Prym (C,\sigma)$. Hence ${\mathcal G}$ must be equivalent to one of the curves $\tilde\Gamma_1, \tilde\Gamma_2, \tilde\Gamma_3$
mentioned in item 1). Fix an order of the roots $s_1,\dots,s_4$.
According to \eqref{6.9} and \cite{kot892} (page 93), one has
\begin{gather} \label{d-d}
d_\alpha- \frac{1}{d_\alpha} = -2 \sqrt{\frac {(s_1-s_4) (s_2-s_3) } {(s_1-s_2)(s_3-s_4) } }\,
\frac{ \sqrt{c_\alpha-s_1} \sqrt{c_\alpha-s_4} + \sqrt{c_\alpha-s_2} \sqrt{c_\alpha-s_3} }
{ \sqrt{c_\alpha-s_1} \sqrt{c_\alpha-s_2} + \sqrt{c_\alpha-s_3} \sqrt{c_\alpha-s_4}  }, \\
\alpha = 1,2,3, \notag
\end{gather}
which implies
\begin{align*}
d_{\alpha }^{2}+\frac{1}{d_{\alpha }^{2}}-2 &=
4\frac{ (s_{1}-s_{4})(s_{2}-s_{3}) }{ (s_{1}-s_{2})(s_{3}-s_{4}) } \,
\frac{ (c_\alpha-s_1) (c_\alpha-s_4) + (c_\alpha-s_2)(c_\alpha-s_3)+ 2\sqrt{\psi(c_\alpha)} }
{ (c_\alpha-s_1)(c_\alpha-s_2) + (c_\alpha-s_3)(c_\alpha-s_4) + 2\sqrt{\psi(c_\alpha)} } \\
 & = 4 \frac{ S_{(14)(23) }^{(\alpha) } }{ S_{(12)(34) }^{(\alpha) } }.
\end{align*}
In view of the identities \eqref{sum_0}, the above is equivalent to the first relation in \eqref{d_s}.

It remains to observe that, in view of symmetries \eqref{symm_rel}, any permutation of $s_1,\dots,s_4$ may lead only to one
of the three expressions \eqref{d_s} or to their modifications leading to a simultaneous change
$(d_1^2, d_2^2, d_3^2) \to (-d_1^2, -d_2^2, -d_3^2)$ and, therefore, to an equivalent curve $\widetilde\Gamma$.

The three permutations in \eqref{d_s} describe three curves \eqref{Gamma} which are birationally non-equivalent:
they have different absolute invariants as was checked in examples (see Appendix C.1).
As a result, all the permutations of $s_1,\dots,s_4$ give the curves
$\tilde\Gamma_1, \tilde\Gamma_2, \tilde\Gamma_3$ in the diagram \eqref{diagram}.
\medskip

3) The three curves $\tilde\Gamma_\alpha$ are obviously invariant with respect to M\"obius transformations
of the curve $C$ in \eqref{mastercurve} generated by substitutions $x \to (ax+b)/(cx+1)$. Applying this transformation to
$s_1,\dots, s_4, c_1, c_2, c_3$ in \eqref{d-d} and denoting the images of $c_1, c_2, c_3,\infty$ by $a_1,\dots, a_4$,
after simplifications we get
\begin{align}
d_\alpha- \frac{1}{d_\alpha} & = -2 \sqrt{\frac {(s_1-s_4) (s_2-s_3) } {(s_1-s_2)(s_3-s_4) } } \notag \\
& \cdot \frac{ \sqrt{ (s_1-a_\alpha) (s_4-a_\alpha)(s_2-a_4)(s_3-a_4) } -
\sqrt{ (s_1-a_4 ) (s_4-a_4) (s_2-a_\alpha) (s_3-a_\alpha ) } }
{ \sqrt{  (s_1-a_\alpha) (s_2-a_\alpha) (s_3-a_4)(s_4-a_4)} -
\sqrt{ (s_1-a_4) (s_2-a_4)(s_3-a_\alpha) (s_4-a_\alpha)  }} .
\end{align}
The latter gives rise to \eqref{d_s}, \eqref{S_even_roots}. Conditions \eqref{imp_even} are obtained from
\eqref{imp_cond} by the same M\"obius transformation. 
\medskip

4) This follows from Propositions \ref{updown}, \ref{duality}, and the observation that for the dual curves $K$ in
\eqref{dual_K} and $C$ the sets $\{s_1,\dots, s_4\}$ and $\{a_1,\dots, a_4\}$ are interchanged.
\medskip

5) Taking the limit $a_4\to \infty$ in
 \eqref{d_k}, \eqref{S_even_dual} and denoting the remaining finite $a_1,a_2,a_3$ as $c_1, c_2, c_3$, we get
the expressions \eqref{k_k}, \eqref{S_odd_dual}. $\square$




\section{The limit case $h_3\to 0$} \label{h30} In the case $h_3=0$ the genus 3 curve $C$ for the odd order case
can be written in spatial form
\begin{equation} \label{spatial}
  w^2 = g(x)=(x-\rho_1)(x-\rho_2), \quad y^2 = \Phi(x)\equiv (x-c_1)(x-c_2)(x-c_3).
\end{equation}
Hence the branch points $Q_1,\dots, Q_4$ of the covering $C\to {\mathcal E}$ are divided in pairs of
hyperelliptically involutive points:
$$
Q_{1,2} = (\rho_1, \pm \sqrt{\Phi(\rho_1)}), \quad Q_{3,4} = (\rho_2, \pm \sqrt{\Phi(\rho_2)}),
$$
for an appropriate choice of indices of $Q_i$.
It is known (\cite{hai83, au96, acgh84}) that in this case $C$ becomes hyperelliptic and admits 2 involutions
$$
\sigma \, : \; (x,y,w) \to (x,y,-w), \quad \iota \, : \; (x,y,w) \to (x,-y,w),
$$
such that $C/\sigma =E$ and $C/\iota$ is a genus 2 hyperelliptic curve
\begin{equation} \label{moser}
G\, : \, \mu^2=g(x) \Phi(x) \equiv (x-\rho_1)(x-\rho_2)(x-c_1)(x-c_2) (x-c_3) .
\end{equation}

The Jacobian of $G$ is a 2-fold covering of $\Prym (C,\sigma)$.
Equivalently, if we replace $\Jac(G)$ by its conformal image dividing all its 4 periods by 2,
$\Prym (C,\sigma)$ can be regarded as an 8-fold unramified covering of $\Jac(G)$.

Explicitly, this can be described as follows (see e.g., \cite{acgh84}): setting in \eqref{spatial} $w=W (x-\rho_1)$,
which gives $W^2=(x-\rho_2)/(x-\rho_1)$,
we eliminate $x$ from the second equation in \eqref{spatial} to get the equation of $C$ in the hyperelliptic form
\begin{equation} \label{G3}
   Y^2 = (W^2-1) \left( W^2- \frac{c_1-\rho_2}{c_1-\rho_1} \right)\left( W^2- \frac{c_2-\rho_2}{c_2-\rho_1} \right)
\left( W^2- \frac{c_3-\rho_2}{c_3-\rho_1} \right).
\end{equation}
Factorizing it by the involution $\sigma$ and, respectively, $\iota$, we obtain
$$
Z^2 = (W-1) \left( W- \frac{c_1-\rho_2}{c_1-\rho_1} \right)\left( W- \frac{c_2-\rho_2}{c_2-\rho_1} \right)
\left( W- \frac{c_3-\rho_2}{c_3-\rho_1} \right) ,
$$
respectively,
$$
V^2 = W(W-1) \left( W- \frac{c_1-\rho_2}{c_1-\rho_1} \right)\left( W- \frac{c_2-\rho_2}{c_2-\rho_1} \right)
\left( W- \frac{c_3-\rho_2}{c_3-\rho_1} \right).
$$
They define the curves birationally equivalent to $E$ and, respectively, $G$.

\paragraph{Remark.} For the case $h_3=0$ the Clebsch system with the integrals \eqref{integrals}
was solved by H. Weber \cite{web878} in terms of theta-functions of $G$.
A similar results was obtained by Moser for the Frahm--Manakov top on $so(4)$
(with $G$ replaced by the even order curve) in \cite{Moser_var}. This corresponds with the observation of \cite{hai83}
that for $h_3=0$ the compactification of the complex invariant manifold of the Frahm--Manakov top (i.e., Prym$(C,\sigma)$)
is an 8-fold covering of $\Jac(G)$.
\medskip

The curve \eqref{moser} must be equivalent to one of the 3 curves
$\widetilde\Gamma_{1}, \widetilde\Gamma_{2}, \widetilde\Gamma_{3}$ of Theorem \ref{main_th},
and its equation can be derived from the K\"otter formulas \eqref{d_s}, \eqref{d_ss_a} by an appropriate limit procedure.
Namely, setting in \eqref{d_ss_a} or \eqref{d_ss} directly $s_1=s_2$, $s_3=s_4$
gives $S_{\left( 12\right) \left( 34\right) }^{\alpha }=0$ and also
$S_{\left( 14\right) \left( 24\right) }^{\alpha }= S_{\left( 42\right) \left( 13\right) }^{\alpha }=0$, which leads to an indetermination in the right hand sides of \eqref{d_s}.
For this reason, in the relation
$$
(x-\rho_1)^2(x-\rho_2)^2 - 4h_{3}^2 (x-c_1)(x-c_2)(x-c_3) =(x-s_1)\cdots (x-s_4) .
$$
we let $h_3 << 1$ and get the following expansions
\begin{gather}
 s_{1,2}= \rho_1 \pm h_3 \frac{2 \sqrt{\Phi(\rho_1) } }{\rho_1-\rho_2}+ O(h_3^2), \quad
 s_{3,4}= \rho_2 \pm h_3 \frac{2\sqrt{\Phi(\rho_2)}}{\rho_1-\rho_2}+ O(h_3^2), \label{exp_s_rho} \\
\begin{aligned}
\sqrt{\psi (c_\alpha)} & = (\rho_1-c_\alpha) (\rho_2-c_\alpha) \\
 & \quad - \frac{2 h_2^2} {(\rho_1-\rho_2)^2}
\left( (\rho_{1}-c_\alpha ) \frac{\Phi(\rho_2)}{\rho_2-c_\alpha}
+\frac{\Phi(\rho_1)}{\rho_1-c_\alpha} (\rho_2-c_\alpha) \right)+ O(h_3^4).
\end{aligned}
\end{gather}
Substituting them into \eqref{d_ss}, up to terms of order $h_3^4$, we obtain
\begin{align}
S_{(12) (34) }^{(\alpha) }
& = 
 - 16 h_3^2  \sqrt{\Phi(\rho_1)} \sqrt{\Phi(\rho_2)}  , \notag \\
S_{(14) (23) }^{(\alpha) }
& = 8 h_3^2 \sqrt{\Phi(\rho_1)} \sqrt{\Phi(\rho_2)} -
4 h_3^2 \left( (\rho_2-c_\alpha) \frac{\Phi(\rho_1)}{\rho_1-c_\alpha}
+ (\rho_1-c_\alpha) \frac{\Phi(\rho_2)}{\rho_2-c_\alpha} \right) \notag \\
 & = -4 h_3^2 \left( \sqrt{(\rho_2-c_\alpha) (\rho_1-c_\beta)(\rho_1-c_\gamma)}
- \sqrt{ (\rho_1-c_\alpha) (\rho_2-c_\beta)(\rho_2-c_\gamma) } \right)^2  , \label{exp_S} \\
S_{(42) (13) }^{(\alpha) }
& = 8 h_3^2 \sqrt{\Phi(\rho_1)} \sqrt{\Phi(\rho_2)} + 4 h_3^2
\left( (\rho_2-c_\alpha) \frac{\Phi(\rho_1)}{\rho_1-c_\alpha}
+ (\rho_1-c_\alpha) \frac{\Phi(\rho_2)}{\rho_2-c_\alpha} \right) \notag \\
 & = 4 h_3^2 \left( \sqrt{(\rho_2-c_\alpha) (\rho_1-c_\beta)(\rho_1-c_\gamma)}
+ \sqrt{ (\rho_1-c_\alpha) (\rho_2-c_\beta)(\rho_2-c_\gamma)}  \right)^2 . \notag
\end{align}

This leads to the following
\begin{theorem} \label{h00}
1) In the case $h_3=0$ the curves $\widetilde\Gamma_{1}, \widetilde\Gamma_{2}, \widetilde\Gamma_{3}$ can be
written in the form
$$
\widetilde\Gamma \; :\; w^2 = z(z-d_1^2) (z-d_2^2)(z-d_3^2)(z-d_1^2 d_2^2 d_3^2) \, ,
$$ 
where, up to a sign,  for the first curve
\begin{gather}
d_{\alpha } =  \frac{  \sqrt[4]{P_{\alpha 2} }} {\sqrt[4]{P_{\alpha 1} } } \quad \text{or} \quad
 d_{\alpha } =  \frac{  \sqrt[4]{P_{\alpha 1} }} {\sqrt[4]{P_{\alpha 2} }}, \qquad \alpha=1,2,3, \label{dd_1} \\
 P_{\alpha 1} =(\rho_1-c_\alpha) (\rho_2-c_\beta)(\rho_2-c_\gamma) ,  \quad
 P_{\alpha 2} =(\rho_2-c_\alpha) (\rho_1-c_\beta)(\rho_1-c_\gamma) , \notag
\end{gather}
and for the other two curves, with $d_\alpha$ replaced by ${d_\alpha}', {d_\alpha}''$ respectively,
\begin{align}
{d_\alpha} ' & =
i \frac{  \sqrt[4]{P_{\alpha 1} } + \sqrt[4]{P_{\alpha 2} } } { \sqrt[4]{P_{\alpha 1}} - \sqrt[4]{P_{\alpha 2}} }
=  i \frac{ 1+ d_\alpha }{1- d_\alpha}\, , \label{dd_2} \\
{d_\alpha}'' & =
 \frac{  \sqrt[4]{P_{\alpha 1} } -i \sqrt[4]{P_{\alpha 2} }} { \sqrt[4]{P_{\alpha 1}} + i\sqrt[4]{P_{\alpha 2}} }
=  \frac{ i+ d_\alpha }{i-d_\alpha}\,  \label{dd_3}.
\end{align}

2) The first curve $\widetilde\Gamma_1$ defined by \eqref{dd_1} is birationally equivalent to the Weber curve $G$
in \eqref{moser}. The genus 3 curve $C$ itself is equivalent to
\begin{equation} \label{CZW}
   {\cal W}^2 = (Z^2- d_1^2) (Z^2- d_2^2) (Z^2- d_3^2) (Z^2- d_1^2 d_2^2 d_3^2) .
\end{equation}
 \end{theorem}

We note that the last relations in \eqref{dd_2}, \eqref{dd_3} are consistent with the relations \eqref{ddd} and
stress that now, among the 3 curves $\widetilde\Gamma_{1}, \widetilde\Gamma_{2}, \widetilde\Gamma_{3}$,
the curve $\widetilde\Gamma_{1} \cong G$ has a special feature:
it is covered by the genus 3 curve $C$, whereas, apparently, the other ones are not.
\medskip

\noindent{\it Proof of Theorem} \ref{h00}. 1) Substituting the expansions \eqref{exp_S} into the first permutation in \eqref{d_s} one gets
\begin{align}
d_{\alpha }^{2} +\frac{1}{d_{\alpha }^{2}} & = \lim_{h_3\to 0}
2\frac{ S_{\left( 14\right) \left( 23\right)}^{\alpha } + S_{\left( 24\right) \left(13\right)}^{\alpha }}
{S_{\left(12\right) \left( 34\right) }^{\alpha }} \notag \\
& = \frac{1}{ \sqrt{\Phi(\rho_1)} \sqrt{\Phi(\rho_2)}} \left( (\rho_2-c_\alpha) \frac{\Phi(\rho_1)}{\rho_1-c_\alpha}
+ (\rho_1-c_\alpha) \frac{\Phi(\rho_2)} {\rho_2-c_\alpha} \right) , \quad \alpha=1,2,3. 
\end{align}
Each of the above equations has 2 solutions, namely \eqref{dd_1}.

For the second and third permutations in \eqref{d_s} the expansions \eqref{exp_S} yield, up to a sign,
\begin{align*}
{d_\alpha} ' & = \lim_{h_3\to 0} i \frac{\sqrt{S_{(12)\, (34)}^{(\alpha)}} + \sqrt{S_{(24)\,(13)}^{(\alpha)} } }
{ \sqrt{ S_{(14)\, (23)}^{(\alpha)} }} \notag \\
& = i \frac{ \sqrt{P_{\alpha 1} } + \sqrt{P_{\alpha 2} } + 2 \sqrt[4]{\Phi(\rho_1) \Phi(\rho_2)} }
{ \sqrt{P_{\alpha 1}} - \sqrt{P_{\alpha 2}}  }
= i \frac{ \left( \sqrt[4]{P_{\alpha 1} } + \sqrt[4]{P_{\alpha 2} } \right)^2  }
{ \sqrt{P_{\alpha 1}} - \sqrt{P_{\alpha 2}}  } \\
& = i \frac{\left( \sqrt[4]{P_{\alpha 1} } + \sqrt[4]{P_{\alpha 2} } \right)^2  }
{ \left( \sqrt[4]{P_{\alpha 1}} - \sqrt[4]{P_{\alpha 2}} \right)
\left( \sqrt[4]{P_{\alpha 1}} + \sqrt[4]{P_{\alpha 2}} \right)} =
i \frac{  \sqrt[4]{P_{\alpha 1} } + \sqrt[4]{P_{\alpha 2} }} { \sqrt[4]{P_{\alpha 1}} - \sqrt[4]{P_{\alpha 2}}  },
\end{align*}
and
$$
  d_\alpha '' = \lim_{h_3\to 0} \frac{\sqrt{S_{(12)\, (34)}^{(\alpha)}} + \sqrt{-S_{(14)\,(23)}^{(\alpha)} } }
{ \sqrt{ S_{(42)\, (13)}^{(\alpha)} }} = \frac{ \sqrt[4]{P_{\alpha 1} } - i \sqrt[4]{P_{\alpha 2} }  }
{ \sqrt[4]{P_{\alpha 1}} + i \sqrt[4]{P_{\alpha 2}}  },
$$
that is, \eqref{dd_2}, \eqref{dd_3}.
\medskip

2) Choose for definiteness the first expression in \eqref{dd_1}, which gives
$$
d_1^2 d_2^2 d_3^2 =\frac{ \sqrt{\Phi(\rho_2)}}{\sqrt{ \Phi(\rho_1)} }.
$$
Under the rescaling
$$
z\to z \frac{ \sqrt{\Phi(\rho_1)}}{\sqrt{ \Phi(\rho_2) } }
$$
the branch points $0, d_1^2, d_2^2, d_3^2$, and $d_1^2 d_2^2 d_3^2, \infty$ of $\widetilde\Gamma_1$ transform
to, respectively,
$$
0,\;  \frac{\rho_1-c_1}{\rho_2-c_1}, \; \frac{\rho_1-c_2}{\rho_2-c_2},\; \frac{\rho_1-c_3}{\rho_2-c_3}, \; 1, \;\infty.
$$
These are the images of the branch points $x=\rho_1,  c_1, c_2, c_3, \infty, \rho_2$ of \eqref{moser}
under the M\"obius transformation $z=(x-\rho_1)/(x-\rho_2)$.

Next, by construction, $C$ is obtained as a regularization of the curve cut out by $W^2=(x-\rho_1)/(x-\rho_2)$ with $x$ being
a function on $G$. As $x=\rho_1, x=\rho_2$ correspond to $z=0, z=\infty$ on $\widetilde\Gamma$, the equation
\eqref{CZW} of $C$ is obtained from that of $\widetilde\Gamma$ by substitution $z=Z^2$ and subsequent regularization.
$\square$

\paragraph{The dual curve $K$ and the Jacobians.} In the case $h_3=0$ the dual curve $K$ described by
\eqref{dual_K} becomes singular (as the underlying elliptic curve
${\cal E}= \{y^2=(x-s_1)\cdots (x-s_4) \}$ is singular).
As follows from the second equation in \eqref{dual_KK}, the
regularization of $K$ takes the form $y^2 g(x)+ \Phi(x)=0$, equivalent to the genus 2 curve $G$ in \eqref{moser}.
Then the variety $\Prym(K,\sigma)$ can be recovered as an
Abelian subvariety of the generalized Jacobian of $K$, an extension of $\Jac(G)$ by ${\mathbb C}^*$, following the
procedure described in \cite{be77}. $\Prym(K,\sigma)$ appears as a smooth 2-fold covering of $\Jac(G)$.

The equations of the genus 2 curves $\Gamma_1, \Gamma_2, \Gamma_3$ can be calculated by
substituting the expansions \eqref{exp_s_rho} to the expressions \eqref{S_odd_dual}, \eqref{k_k}.

 \begin{theorem} \label{dual_h3} In the case $h_3=0$ the genus 2 curves $\Gamma_1, \Gamma_2, \Gamma_3$ are described by
equations \eqref{Gamma0}, \eqref{k_k} with
\begin{align}
{\bar S}_{(ij)}^{(1)} & = (c_i-c_j)\left( \sqrt{P_{k1} } + \sqrt{P_{k2}}  \right)^2, \notag \\
{\bar S}_{(ij)}^{(2)} & = (c_i-c_j)\left( \sqrt{P_{k1} } - \sqrt{P_{k2}}  \right)^2, \label{S_h3}  \\
{\bar S}_{(ij)}^{(3)} & = (c_i-c_j)\cdot 4 (\rho_2-c_1)(\rho_2-c_2)(\rho_2-c_3),  \quad (i,j,k)=(1,2,3), \notag
\end{align}
where, as above,
\begin{gather*}
P_{\alpha 1} =(\rho_1-c_\alpha) (\rho_2-c_\beta)(\rho_2-c_\gamma) ,  \quad
 P_{\alpha 2} =(\rho_2-c_\alpha) (\rho_1-c_\beta)(\rho_1-c_\gamma) , \\
(\alpha, \beta,\gamma) = (1,2,3),
\end{gather*}
and ${\bar S}_{(12)}^{(\alpha)}+{\bar S}_{(23)}^{(\alpha)}+{\bar S}_{(31)}^{(\alpha)}=0$. Up to a sign, thi implies
\begin{align}
   k_1 & = \frac{ \sqrt{c_2-c_3}\, ( \sqrt{P_{11}} + \sqrt{P_{12}} ) + \imath \sqrt{c_3-c_1}\, (\sqrt{P_{21}} + \sqrt{P_{22}} ) }
 {\sqrt{c_1-c_2} \, (\sqrt{P_{31}} + \sqrt{P_{32}} )}, \notag \\
k_2 & =  \frac{ \sqrt{c_2-c_3}\, ( \sqrt{P_{11}} - \sqrt{P_{12}} ) + \imath \sqrt{c_3-c_1}\, (\sqrt{P_{21}} - \sqrt{P_{22}} ) }
 {\sqrt{c_1-c_2} \, (\sqrt{P_{31}} - \sqrt{P_{32}} )}, \label{k_h3} \\
k_3 & = \frac{ \sqrt{c_2-c_3} + \imath \sqrt{c_3-c_1}} { \sqrt{c_1-c_2} },  \notag
\end{align}
and the triples $ k_1', k_2',k_3', \; k_1'', k_2'',k_3''$ are obtained from the above by cyclic permutations of
$c_1, c_2, c_3$.
\end{theorem}

The fact that the expression for ${\bar S}_{(ij)}^{(3)}$ in \eqref{S_h3} depends only on $\rho_2$, does not play any role,
as $\rho_1, \rho_2$ do not appear in the formula for $k_3$ in \eqref{k_h3}.
\medskip

\noindent{\it Proof of Theorem} \ref{dual_h3}. Setting in \eqref{S_odd_dual} $\alpha=3$ and applying the
limit $s_1,s_2 \to \rho_1$, $s_3,s_4 \to \rho_2$ described by the expansions \eqref{exp_s_rho}, we get
$$
{\bar S}_{(ij)}^{(3)} = (c_i-c_j)\cdot [2(c_1-\rho_2)(c_2-\rho_2)(c_3-\rho_2)- 2\Phi(\rho_2)] .
$$
Repeating this for $\alpha=1,2$ yields
$$
\frac { {\bar S}_{(ij)}^{(1,2)} }{c_i-c_j } =
(c_k-\rho_1)(c_i-\rho_2)(c_j-\rho_2)  + (c_k-\rho_2)(c_i-\rho_1)(c_j-\rho_1)
\pm 2 \sqrt{\Phi(\rho_1)} \sqrt{\Phi(\rho_2)} .
$$
These expressions are equivalent to \eqref{S_h3} up to factor $-1$. Expressions \eqref{k_h3} are then obtained by
using equivalences \eqref{d_s}.  $\square$

\section{The singular cases: $\Jac (C)$ contains 3 elliptic curves}\label{sec:sing_case}

Apart from the case $h_3=0$, the formulas of Theorem \ref{main_th} for $d_\alpha^2$ are not directly applicable
in the other cases when all $s_1,\dots, s_4$ are distinct, but,
for a certain $\alpha\in \{1,2,3\}$, one has
$$
S_{(12)\, (34)}^{(\alpha)} =0, \quad \text{or} \; S_{(14)\, (32)}^{(\alpha)}=0  \quad \text{or } \quad
S_{(24)\, (13)}^{(\alpha)}=0 .
$$
This gives $d_\alpha^2=0$ or $\infty$ for one of the curves
$\widetilde\Gamma_{1}, \widetilde\Gamma_{2}, \widetilde\Gamma_{3}$,
which means that at least two branch points of it coincide, so this curve is {\it singular}.
However, as Theorem \ref{3-curve} below says, the other 2 curves $\tilde\Gamma$ are not singular,
but their Jacobians contain 2 elliptic curves. (The curve $C$ itself also remains smooth and non-hyperelliptic).

\subsection{The conditions of singularity.}
Consider, for concreteness, $\alpha=3$ and the case $S_{(12)\, (34)}^{(3)}=0$.
As follows from the definition of $S_{(12)\, (34)}^{(\alpha)}$, when $s_1,\dots, s_4$ are distinct, this gives
\begin{gather}
(s_1-c_3)(s_2-c_3) + (s_3-c_3)(s_4-c_3) + 2\sqrt{\psi(c_3)} =0, \notag \\
 \text{or, in view of \eqref{imp_cond},} \notag \\
 (s_1-c_3)(s_2-c_3) + (s_3-c_3)(s_4-c_3) = 2 g(c_3), \label{ccond}
\end{gather}
where, as above, $g(x)= x^2 + h_1 x+h_0 =(x-\rho_1)(x-\rho_2)$ (without loss of generality we again assumed that $h_2=1$).

Since $S_{(12)\, (34)}^{(3)}=0$, the denominator of \eqref{d-d} must be zero as well, then
the condition \eqref{ccond} is equivalent to
\begin{equation} \label{sp_cond}
(s_1-c_3)(s_2-c_3) = (s_3-c_3)(s_4-c_3) =(c_3-\rho_1)(c_3-\rho_2) .
\end{equation}


\begin{proposition} \label{Sing_alg_cond}
The condition $(s_1-c_3)(s_2-c_3) = (s_3-c_3)(s_4-c_3)$ or its analog obtained by a permutation of $s_1,s_2,s_3,s_4$
hold if and only if one of the following two conditions on the
coefficients of the genus 3 curve $C$ is satisfied
\begin{gather}
 (c_3-\rho_1)(c_3-\rho_2) = (c_3-c_1)(c_3-c_2), \label{cond_s1} \\
 h_3^2 \,(c_3-c_1)(c_3-c_2)- (c_3-\rho_1)(c_3-\rho_2)\cdot (2c_3-\rho_1-\rho_2-h_3^2) =0 \,. \label{cond_s22}
\end{gather}
\end{proposition}

We note that these two conditions are independent. It follows that if \eqref{cond_s1} is satisfied then
\eqref{sp_cond} holds for any value of the constant $h_3$.
\medskip

\noindent{\it Proof of Proposition} \ref{Sing_alg_cond}.
The condition \eqref{sp_cond} or its permutation implies
\begin{gather*}
{\mathfrak P} :=[({s_{1}} - c)\,({s_{4}} - c) - ({s_{3}} - c)\,({s_{2}} - c)] \\
\cdot \,[({s_{1}} - c)\,({s_{2}} - c) - ({s_{3}} - c)\,({s_{4}} - c)]
[({s_{1}} - c)\,({s_{3}} - c) - ({s_{4}} - c)\,({s_{2}} - c)]=0
\end{gather*}
with $c$ being one of $c_1, c_2, c_3$.

The expression $\mathfrak P$ is invariant with respect to any permutation of $\{s_1,\dots,s_4\}$, hence one can rewrite
it in terms of elementary symmetric functions of $s_j$, that is, the coefficients $\Delta_j$ of the polynomial
$$
\psi=(x-s_1)\cdots(x-s_4)= x^4+ \Delta_1 x^3 + \Delta_2 x^2 + \Delta_3 x + \Delta_4 .
$$
Explicitly, one has
\begin{gather*}
{\mathfrak P}= ( - {\Delta _{1}}^{3} - 8\,{\Delta _{3}} + 4\,{\Delta _{1}}\,{
\Delta _{2}})\,c^{3} +
( - 16\,{\Delta _{4}} - 2\,{\Delta _{1}}\,{\Delta _{3}}-{\Delta _{1}}^{2}\,\Delta_2
+ 4\,{\Delta_2}^{2})\,c^{2}  \\
+ (- {\Delta _{1}}^{2}\,{\Delta _{3}} - 8\,{\Delta _{1}}
\,{\Delta _{4}} + 4\,{\Delta _{2}}\,{\Delta _{3}})\,c + {\Delta_{3}}^{2} - \Delta_1^2\, {\Delta_4} .
\end{gather*}
Now replacing $\Delta_j$ by the corresponding coefficients of the expansion
\begin{align*}
\psi=g^2(x)-4 h_3^2 \Phi(x) & = x^{4} + (- 4\,h_3^2 + 2\,h_1)\,x^{3}
+ (4\,h_3^2 \,(c_1 + c_2+c_3)+ 2\,h_0 + h_1^2)\,x^{2} \\
&\quad + ( - 4 h_3^2 (c_1c_2+c_2 c_3+c_1c_3) + 2\,h_0\,h_1)\,x + h_{0}^{2} + 4\,h_{3}^2\,c_1 c_2 c_3\,
\end{align*}
and setting $c=c_3$, one obtains the factorization
\begin{gather*}
\mathfrak P= 16 h_3^2 \, (c_1 c_2 - c_1 c_3-c_2 c_3 - c_3 h_1 -h_0) \\
\cdot (-2c_3^3+(2h_3^2-3h_1)c_3^2+(-h_1^2-h_3^2(c_1+c_2)+h_1 h_3^2-2 h_0)c_3
+c_1 c_2 h_3^2 + h_0 (h_3^2-h_1))\, .
\end{gather*}
Setting here $h_1=-\rho_1-\rho_2, h_0=\rho_1 \rho_2$, one finally arrives to
the conditions \eqref{cond_s1}, \eqref{cond_s22}. $\square$
\medskip

Now assume, without lost of generality, that $c_1+c_2+c_3=0$, i.e., the equation of the curve $E$ is canonical (this
can always be achieved by an appropriate shift of $x$ in the equation \eqref{mastercurve} of the curve $C$.)
Let $q_1,\dots, q_4$ be the images of the branch points $Q_1,\dots, Q_4$ in the parallelogram of periods of $E$, i.e.,
$$
s_i= \wp(q_i \,|2\varOmega_1, 2\varOmega_3), \quad
g(s_i)= - \wp'(q_i \,|2\varOmega_1, 2\varOmega_3), \qquad i=1,2,3,4, $$
where $\varOmega_1, \varOmega_3$ are half-periods of $E$, so that $c_\alpha=\wp(\varOmega_\alpha)$.


\begin{proposition} \label{Sing_test}
1) The conditions \eqref{sp_cond} and \eqref{cond_s1} hold if and only if,
apart from the known relation $q_1+q_2+q_3+q_4 \equiv 0$, one has
\begin{equation} \label{pairwise}
q_1+q_2 \equiv q_3+q_4 \equiv \omega_3 \equiv p_1+p_2,
\end{equation}
where $p_1, p_2$ are the Abel images of zeros of $g(x)$,
$$
\rho_k=\wp(p_k \,|2\varOmega_1, 2\varOmega_3), \quad h_3 \wp'(p_k \,|2\varOmega_1, 2\varOmega_3)= \psi(\rho_k), \qquad
k=1,2.
$$
2) When \eqref{pairwise} holds there are also the following relations
\begin{gather}
\begin{gathered}
 (c_1-s_1)(c_2-s_1) (c_3-s_2) = (c_1-s_2)(c_2-s_2) (c_3-s_1), \\
(c_1-s_3)(c_2-s_3) (c_3-s_4)= (c_1-s_4)(c_2-s_4) (c_3-s_3), \end{gathered} \label{eq_3_fac} \\
\begin{gathered}
 (\rho_1-s_1)(\rho_2-s_1) (c_3-s_2)= (\rho_1-s_2)(\rho_2-s_2) (c_3-s_1), \\
 (\rho_1-s_3)(\rho_2-s_3) (c_3-s_4)= (\rho_1-s_4)(\rho_2-s_4) (c_3-s_3).
\end{gathered} \label{eq_4}
\end{gather}
\end{proposition}

Notice that \eqref{pairwise} also says that the pair $(q_2, q_3)$ is obtained from $(q_4, q_1)$ by a translation on $E$.
\medskip

\paragraph{Remark.} Item 2 of Proposition \ref{Sing_test} has the following corollary:
Let $S_{(12)\, (34)}^{(3)}=0$ and the
curve $\widetilde\Gamma_1$ is singular. Then relations \eqref{eq_3_fac} together with item 5) of Theorem \ref{main_th}
imply that in the expressions \eqref{S_odd_dual} for the dual curves $\Gamma$
one has $\bar S_{(1,2,3)}^{(3)}=0$ and the corresponding branch point $k_\alpha^2=\infty$ or $0$.
That is, one of the curves $\Gamma_1, \Gamma_2, \Gamma_3$ is singular as well. This observation also appears in Theorem
\ref{3-curve} below in an independent way.
\medskip

\noindent{\it Proof of Proposition} \ref{Sing_test}.
1) We use the following classical addition formula (see, e.g., \cite{hc44}): given
an elliptic curve $y^2 = 4x^3 - g_2x-g_3=4(x-c_1)(x-c_2)(x-c_3)$ with $c_1+c_2+c_3=0$, the values
$$
  P_1=\wp(u_1), \quad P_2=\wp(u_2), \quad P_3=\wp(u_1+u_2)
$$
satisfy the algebraic relation
\begin{equation} \label{3_add}
(P_1+P_2+P_3)(4 P_1 P_2 P_3-g_3 )= (P_1 P_2 + P_2 P_3+ P_1 P_3-g_2/4 )^2 .
\end{equation}

Now let $P_1=\rho_1, P_2=\rho_2, P_3=c_3$. Then \eqref{3_add} takes the equivalent forms
\begin{gather} (\rho_1 \rho_2 -(\rho_1+\rho_2)c_3 +(c_1+c_2)c_3 -c_1c_2 )^2 =0 \; \Longleftrightarrow \notag \\
\left( (\rho_1-c_3)(\rho_2-c_3)- (c_1-c_3)(c_2-c_3) \right)^2 =0  .\label{3_rho}
\end{gather}
Hence, if $p_1+p_2=\varOmega_3$, the above condition says that \eqref{cond_s1} is satisfied. Vice versa, if
\eqref{cond_s1} (and, therefore, \eqref{3_rho}) holds, this implies $p_1+p_2=\varOmega_3$.

In a similar way, setting in \eqref{3_add} $P_1=s_1, P_2=s_2, P_3=c_3$, we get
$$
\left( (s_1-c_3)(s_2-c_3)- (c_1-c_3)(c_2-c_3) \right)^2 =0 ,
$$
which must be equivalent to $q_1+q_2=\varOmega_3$. The same argumentation holds for $P_1=s_3, P_2=s_4, P_3=c_3$.
As a result, we proved item 1) of the proposition.
\medskip

2) Probably, the fastest way to prove \eqref{eq_3_fac} is to use the classical formula
$$
 s_i-c_\alpha = \wp(q_i)-\wp(\varOmega_\alpha) = \left( \frac{e^{\eta_\alpha \, q_i}\sigma(q_i-\varOmega_\alpha) }
{\sigma (q_i) \sigma(\varOmega_\alpha) } \right)^2 ,
$$
where $\sigma(u)= \sigma(u |2\varOmega_1,2\varOmega_3)$ is the Weierstrass sigma function of $E$ and $\eta_\alpha$ are the
periods of the canonical differential of the second kind on $E$ satisfying $\eta_1+\eta_2+\eta_3=0$.
Then the condition \eqref{eq_3_fac} can be rewritten in the form
\begin{gather}
\left( \frac{e^{\eta_3 q_2+ (\eta_1+\eta_2)q_1 }
\sigma(q_2-\varOmega_3) \sigma(q_1-\varOmega_1) \sigma(q_1-\varOmega_2)}
{\sigma (q_1) } \right)^2 \notag \\
= \left( \frac{e^{\eta_3 q_1+ (\eta_1+\eta_2)q_2 }
\sigma(q_1-\varOmega_3) \sigma(q_2-\varOmega_1) \sigma(q_2-\varOmega_2)}
{\sigma (q_2) } \right)^2
\label{sigmasu}
\end{gather}
Using here $q_2=\varOmega_3-q_1$, the properties of $\sigma(u)$, and the Legendre relations, we see that
\eqref{sigmasu} holds. Relation \eqref{eq_4} is proved in the same way.
 $\square$

\subsection{Algebraic description of the Jacobians.}
In view of the identities \eqref{sum_0},
the condition $S_{(12),(34)}^{(3)}=0$ implies $S_{(14),(23)}^{(3)} = S_{(24),(13)}^{(3)}$. Hence, according
to \eqref{d_s}, for the remaining two curves $\tilde \Gamma_2, \tilde \Gamma_3$ one has $d_3=1$, and they take the form
\begin{gather}
\tilde \Gamma_2 \; : \; w^2=z(z-1) (z-\alpha_1)(z-\beta_1) (z-\alpha_1\beta_1), \notag \\
\alpha_1= \frac{\left( \sqrt{S_{(12)\, (34)}^{(1)}} \pm  \sqrt{S_{(24)\,(13)}^{(1)} } \right)^2 }{S_{(14)\, (23)}^{(1)}}, \quad
\beta_1=\frac{\left( \sqrt{S_{(12)\, (34)}^{(2)}} \pm  \sqrt{S_{(24)\,(13)}^{(2)} } \right)^2 }{S_{(14)\, (23)}^{(2)}},  \label{abs} \\
\tilde \Gamma_3 \; : \; w^2=z(z-1) (z-\alpha_2)(z-\beta_2) (z-\alpha_2\beta_2), \notag \\
\alpha_2=\frac{\left( \sqrt{S_{(12)\, (34) }^{(1)} } \pm  \sqrt{S_{(14)\,(23)}^{(1)} }\right)^2 }{S_{(24)\, (13)}^{(1)}}, \quad
\beta_2=\frac{\left(\sqrt{S_{(12)\, (34)}^{(2)}} \pm  \sqrt{S_{(14)\,(23)}^{(2)} }\right)^2 }{S_{(24)\, (13)}^{(2)}},
\label{abs2}
\end{gather}
where, under the condition \eqref{sp_cond},
\begin{align}
 S_{(12),(34)}^{(j)} & = (s_1-s_2)(s_3-s_4) \cdot 4h_3^2 (c_3-c_j) \notag \\
& = (s_1-s_2)(s_3-s_4) \cdot (c_3-c_j) [ (s_1 + s_2 + s_3+s_4-2(\rho_1+\rho_2)] , \notag \\
 S_{(14),(23)}^{(j)} & =(s_1-s_4)(s_2-s_3) \cdot[ (s_1-s_3)(s_4-s_2)+ 4h_3^2 (c_3-c_j) ],  \label{S_sp} \\
 S_{(24),(13)}^{(j)} & =(s_2-s_4)(s_1-s_3) \cdot[ (s_3-s_2)(s_1-s_4)+ 4h_3^2 (c_3-c_j)],\qquad j=1,2. \notag
\end{align}
Indeed, in view of the relations $\sqrt{\psi (c_\alpha)}= - g(c_\alpha)$ and expressions \eqref{symm_rel},
\begin{gather*}
\frac{ S_{(ij),(kl)}^{(\alpha)}}{ (s_{i}-s_{j})(s_{k}-s_{l})}
= (s_i - c_\alpha)(s_j -c_\alpha )+(s_{k}-c_\alpha)(s_{l}-c_\alpha)- 2 (c_\alpha-\rho_1)(c_\alpha-\rho_2)
  \end{gather*}
for $\alpha=1,2,3$. The right hand side can be written as
\begin{gather*}
  (c_3-c_\alpha) [ (s_1 + s_2 + s_3+s_4-2(\rho_1+\rho_2)] \hskip 4cm \\ +
(s_i - c_3)(s_j -c_3)+(s_{k}-c_3)(s_{l}-c_3)- 2 (c_3-\rho_1)(c_3-\rho_2) .
\end{gather*}
The first line in the above sum is $4 h_3^2 (c_3-c_\alpha)$.
By \eqref{sp_cond}, for $(i,j,k,l)=(1,2,3,4)$
the second line in the above sum is zero, which yields the first relation in \eqref{S_sp}.
Again by \eqref{sp_cond}, for $(i,j,k,l)=(2,4,1,3)$ the second line reads
\begin{gather*}
 (s_1-c_3)(s_3-c_3) + (s_2-c_3)(s_4-c_3) - (s_1-c_3)(s_2-c_3)- (s_3-c_3)(s_4-c_3) \\ =
(s_1 - s_{4})(s_3-s_2) .
\end{gather*}

According to the reduction case studied by Jacobi (see e.g. \cite{be01})  each of the curves
$\tilde \Gamma_2, \tilde \Gamma_3$ is a 2-fold covering of 2 elliptic curves,
which can be written in the following Legendre form
\begin{equation} \label{elliptic_W}
\begin{gathered}
\tilde \Gamma_2 \; \longrightarrow \; {\mathcal W}_\pm^{(1)} =\{ \mu^2 =\la (\la-1)(1- C_\pm^{(1)} \la) \}, \qquad
C_\pm^{(1)} = - \frac{(\sqrt{\alpha_1} \pm \sqrt{\beta_1} )^2 }{ (\alpha_1-1)(\beta_1-1)}, \\
\tilde \Gamma_3 \; \longrightarrow \; {\mathcal W}_\pm^{(2)} =\{ \mu^2 =\la (\la-1)(1- C_\pm^{(2)} \la) \}, \qquad
C_\pm^{(2)} = - \frac{(\sqrt{\alpha_2} \pm \sqrt{\beta_2} )^2 }{ (\alpha_2-1)(\beta_2-1)},
\end{gathered}
\end{equation}
and $\Jac(\tilde \Gamma_2)$ contains $ {\mathcal W}_+^{(1)},  {\mathcal W}_-^{(1)}$, whereas $\Jac(\tilde\Gamma_3)$ contains
${\mathcal W}_+^{(2)},  {\mathcal W}_-^{(2)}$\footnote{This is, the corresponding Jacobian is isogeneous
(but not isomorphic) to the direct product of the elliptic curves.}.
\medskip

{\bf Remark.} Notice that choosing different signs in \eqref{abs}, \eqref{abs2} may lead to transformations
$\alpha_i \to 1/\alpha_i$ and (or) $\beta_i \to 1/\beta_i$ and, therefore, to different moduli
$C_\pm^{(i)}$. However, according to Lemma \ref{reciprocal}, this cannot change the
invariants of $\tilde \Gamma_2, \tilde \Gamma_3$ and, therefore,
the absolute invariants of ${\mathcal W}_\pm^{(1)}, {\mathcal W}_\pm^{(2)}$. 

\begin{proposition} \label{W12} The elliptic curves
${\mathcal W}_+^{(1)}, {\mathcal W}_+^{(2)}$ and ${\mathcal W}_-^{(1)}, {\mathcal W}_-^{(2)}$ are pairwise isomorphic.
Their absolute $J$-invariants are
\begin{gather}
 J_{\pm} = 4 \frac{\left[ (s_1-s_3)(s_4-s_2)(s_1-s_4)(s_2-s_3) \Sigma_{\pm}
+ 64 h_3^4 (s_1-s_2)^2 (s_3-s_4)^2 (c_3-c_1)(c_3-c_2) \right]^3 }
{h_3^4 \, (c_3-c_1)(c_3-c_2) \, \Delta \, \Sigma_{\pm}^2} , \label{Jpm} \\
\Sigma_\pm = \left( \sqrt{ \bar S_{(14),(23)}^{(1)} \, \bar S_{(24),(13)}^{(2)} } \pm
\sqrt{\bar S_{(14),(23)}^{(2)}\, \bar S_{(24),(13)}^{(1)}  }  \right)^2, \notag \\
\begin{aligned}
\bar S_{(14),(23)}^{(j)} & = (s_1-s_3)(s_4-s_2) + 4h_3^2 (c_3-c_j) , \\
\bar S_{(24),(13)}^{(j)} & =(s_3-s_2)(s_1-s_4)+ 4h_3^2 (c_3-c_j) , \end{aligned} \notag
\end{gather}
and $\Delta$ is the discriminant of the polynomial $\psi(s)$:
$$
\Delta = (s_1-s_2)^2(s_3-s_4)^2 (s_1-s_3)^2(s_4-s_2)^2(s_3-s_2)^2(s_1-s_4)^2 \,.
$$
Thus the varieties $\Jac(\tilde \Gamma_2)$ and $\Jac(\tilde \Gamma_3)$ contain the same pair of elliptic curves, denoted as
${\mathcal W}_-, {\mathcal W}_+$, however $\tilde \Gamma_2, \tilde \Gamma_3$ are not birationally equivalent.
\end{proposition}

The expression \eqref{Jpm} can be further simplified by using one of the conditions of Proposition \ref{Sing_alg_cond}.
\medskip

\noindent{\it Proof.} The $J$-invariants of the curves ${\mathcal W}_+^{(1)}, {\mathcal W}_-^{(1)}$
given by \eqref{elliptic_W} read
$$
J_{\pm} = 256 \frac{(C_\pm^2- C_\pm +1)^3 }{C^2_\pm (C_\pm-1)^2} ,
$$
where $C_\pm$ are functions of $\alpha_1, \beta_1$. Here $J_{\pm}$ must be invariant with respect to
substitutions $\alpha_1 \to 1/\alpha_1$ or $\beta_1 \to 1/\beta_1$ or both, hence, they can be written in terms of
$A=\sqrt{\alpha_1} + 1/\sqrt{\alpha_1}, B= \sqrt{\beta_1} + 1/\sqrt{\beta}_1$ in the following compact form
\begin{equation} \label{J_symm}
J_{\pm} = 256 \frac{[(A \pm B)^2+ (A^4-4)(B^2-4)]^3}{(A^2-4)(B^2-4)(A\pm B)^4}.
\end{equation}
Taking into account \eqref{abs}, we get
\begin{gather*}
A^2 = (\sqrt{\alpha_1}+1/\sqrt{\alpha_1} )^2 = 4 \frac{ S_{(24),(13)}^{(1)}  }{S_{(14),(23)}^{(1)}}, \quad
 B^2 = (\sqrt{\beta_1}+1/\sqrt{\beta_1} )^2 = 4 \frac{S_{(24),(13)}^{(2)} }{S_{(14),(23)}^{(2)}}, \\
A^2-4 = (\sqrt{\alpha_1}- 1/\sqrt{\alpha_1} )^2 = 4 \frac{S_{(12),(34)}^{(1)} }{S_{(14),(23)}^{(1)}}, \quad
B^2-4 = (\sqrt{\beta_1}- 1/\sqrt{\beta_1} )^2 = 4 \frac{S_{(12),(34)}^{(2)} }{S_{(14),(23)}^{(2)}} ,
\end{gather*}
with $S_{(ij),(kl)}^{(\alpha)}$ described by \eqref{S_sp}. Substituting the above into \eqref{J_symm} and simplifying, we
get the expressions \eqref{Jpm}.

Repeating the calculations with $\alpha_1, \beta_1$ replaced by $\alpha_2, \beta_2$ we arrive at the same expressions.
As a result, the curves ${\mathcal W}_+^{(1)}, {\mathcal W}_+^{(2)}$ are birationally equivalent,
as well as ${\mathcal W}_-^{(1)}$ and ${\mathcal W}_-^{(2)}$.

This statement can also be proved independently, as follows:
In view of \eqref{ddd}, the relations \eqref{abs}, \eqref{abs2} imply
\begin{equation} \label{alphas12}
\sqrt{\alpha_2} = \left\{ i \frac{i +\sqrt{\alpha_1} }{ i -\sqrt{\alpha_1}} \;
-i \frac{i - \sqrt{\alpha_1} }{ i + \sqrt{\alpha_1}} \right\}\, , \quad
\sqrt{\beta_2} =\left\{ \frac{i +\sqrt{\beta_1} }{ i -\sqrt{\beta_1}}, \; \frac{i -\sqrt{\beta_1} }{ i +\sqrt{\beta_1}}  \right\}
\end{equation}
regardless to signs of the roots. Replacing $\sqrt{\alpha_2}, \sqrt{\beta_2}$ in $C_+^{(2)}$
by the first options in \eqref{alphas12}, we get the formula for $C_+^{(1)}$, and
replacing $\sqrt{\alpha_2}, \sqrt{\beta_2}$ in $C_-^{(2)}$
by the second options in \eqref{alphas12}, we get the expression for $C_-^{(1)}$. That is, for appropriate choices
of signs in \eqref{abs}, \eqref{abs2}, we get $C_-^{1}=C_-^{2}$, $C_+^{1}=C_+^{2}$.

Lastly, the claim that
$\tilde \Gamma_2, \tilde \Gamma_3$ are not birationally equivalent was checked comparing their absolute
invariants in numerical tests (see Appendix C). $\square$

\begin{theorem} \label{3-curve} \begin{description}
\item{1)} One can choose such normalized periods $\tau_1, \tau_2$ of the curves ${\mathcal W}_-, {\mathcal W}_+$
that in appropriate coordinates on the Abelian varieties
$$
\Jac(\tilde \Gamma_1), \; \Jac(\tilde \Gamma_2),\; \Jac(\tilde \Gamma_3), \; \Prym(C,\sigma), \quad
\Jac(\Gamma_1), \; \Jac(\Gamma_2),\; \Jac(\Gamma_3), \; \Prym^*(C,\sigma)
$$
their period matrices read, respectively, as follows
\begin{gather*}
\tilde\Omega_1 =
\begin{pmatrix}
1 & 0 & 2\tau_2 & 1 \\
0 & 1 & 1 & \tau_1/2 \end{pmatrix}, \quad
\tilde\Omega_2 =
\begin{pmatrix}
1 & 0 & \dfrac{\tau_2+1}{2} & 1/2 \\
0 & 1 & 1/2 & \tau_1/2 \end{pmatrix}, \quad
\tilde\Omega_3 = \begin{pmatrix}
1 & 0 & \tau_1/2 & 1/2 \\
0 & 1 & 1/2 & \tau_2/2
\end{pmatrix}, \\
\Lambda=\begin{pmatrix}
1 & 0 & \tau_2 & 1 \\
0 & 2 & 1 & \tau_1
\end{pmatrix} \, ,  \\
\Omega_1= \begin{pmatrix}
1 & 0 & \tau_2 & 1/2 \\
0 & 1 & 1/2 & \tau_1/4
\end{pmatrix}, \:
\Omega_2= \begin{pmatrix} 1 & 0 & \tau_2 & 1/2 \\
0 & 1 & 1/2 & \dfrac{1+\tau_1/2}{2}\end{pmatrix}, \;
\Omega_3 = \begin{pmatrix}
1& 0 & \tau_1 & 1 \\
0 & 1 & 1 & \tau_2 \end{pmatrix} , \\
\Lambda^* = \begin{pmatrix} 1 & 0 & \tau_1/2 & 1 \\
0 & 2 & 1 & 2\tau_2 \end{pmatrix} \, .
\end{gather*}

\item{2)} $\Jac(\tilde \Gamma_2)$ and $\Jac(\tilde \Gamma_3)$ both contain the elliptic curves
${\mathcal W}_1, {\mathcal W}_2$, whereas $\Jac(\Gamma_1), \Jac(\Gamma_2)$ both contain other
elliptic curves ${\mathcal U}_1, {\mathcal U}_2$ with the normalized periods $\tau_1/2, 2\tau_2$ respectively.
That is, ${\mathcal U}_1, {\mathcal U}_2$ can be regarded as 2-fold coverings of ${\mathcal W}_1, {\mathcal W}_2$
(and vice versa).


\item{3)} The variety $\Prym(C,\sigma)$ contains the elliptic curves ${\mathcal U}_1, {\mathcal U}_2$.
As a result, the Jacobian of the genus 3 non-hyperelliptic curve ${C}$ itself
contains 3 different elliptic curves ${\mathcal U}_1, {\mathcal U}_2$,
and the ''base'' elliptic curve $E=\{y^2 =\Phi(x)\equiv (x-c_1)(x-c_2)(x-c_3) \}$.

The dual variety $\Prym^*({C},\sigma) =\Prym(K, \rho)$ described in Section \ref{dual_pryms}
contains the elliptic curves ${\mathcal W}_1, {\mathcal W}_2$.
The Jacobian of the genus 3 non-hyperelliptic curve $K$ contains 3 different elliptic curves
${\mathcal W}_1, {\mathcal W}_2, {\mathcal E}=\{y^2 =\Psi(x)\equiv (x-s_1)\cdots (x-s_4) \}$.
\end{description}
\end{theorem}

The above can be summarized in the following diagram, where the vertical arrows denote coverings
and the horizontal ones denote inclusions. The periods of ${\cal U}_{1,2}, {\cal W}_{1,2}$ are also indicated.
$$
\begin{CD}
 (1,\tau_1/2): @. {\cal U}_1 @>>> \Jac(\Gamma_{1,2}) @<<<  {\cal U}_2 @. :(1, 2\tau_2) \\
       @.           @V 2:1 VV        @V 2:1 VV             @VV 2:1 V      @.  \\
(1/2,\tau_1/2): @. {\cal W}_1 @>>> \Prym^* (C,\sigma ) @<<<  {\cal W}_2 @. :(1, \tau_2) \\
        @\vert           @\vert             @V 2:1 VV             @\vert        \\
(1,\tau_1): @. {\cal W}_1 @ >>> \Jac(\tilde\Gamma_{2,3}) @<<<  {\cal W}_2 @. :(1, \tau_2)  \\
      @.        @V 2:1 VV            @V 2:1 VV                  @VV 2:1 V      @.  \\
(1,\tau_1/2): @. {\cal U}_1 @>>> \Prym(C,\sigma ) @<<<  {\cal U}_2  @. :(1/2, \tau_2) \\
       @.           @\vert             @V 2:1 VV           @\vert    @\vert   \\
(1,\tau_1/2): @. {\cal U}_1 @>>> \Jac(\Gamma_{1,2})  @<<<  {\cal U}_2  @. :(1, 2\tau_2)
\end{CD}
$$
The diagram is cyclic in the sense that the first and the last rows should be identified.

As follows from item 1) of Theorem \ref{3-curve}, now the singular curve $\Gamma_3$ is the Richelot image
of the regular curves $\tilde\Gamma_2, \tilde\Gamma_3$, whereas the singular curve $\tilde\Gamma_1$
is the Richelot image of $\Gamma_1,\Gamma_2$.
\medskip

\noindent{\it Proof of Theorem} \ref{3-curve}.
1) If a genus 2 curve $G$ is a 2-fold covering of two elliptic curves with normalized periods
$\tau_1, \tau_2$, then in appropriate homology basis on $G$ its Riemann matrix reads
$$
\begin{pmatrix}
\tau_1/2 & 1/2 \\
 1/2 & \tau_2/2 \end{pmatrix} ,
$$
and vice versa (see e.g. \cite{be01}). The Jacobian of $G$ contains both elliptic curves and is isogeneous (but not
isomorphic) to their direct product.

Thus, if $\tau_1, \tau_2$ are the periods of ${\mathcal W}_1$,
respectively ${\mathcal W}_2$,
and one of the curves $\tilde\Gamma$, say $\tilde\Gamma_3$, is a 2-fold cover of
${\mathcal W}_1, {\mathcal W}_2$, then, in an appropriate basis, the period matrix of $\tilde\Gamma_3$
has the above form, and the structure of period matrices of the rest of the curves immediately follows from Diagram 1.

2) Using the Thomae formulae for the branch points of a hyperelliptic genus 2 curve $G$ (\cite{be01, fay73}), one can see
that if its Riemann matrix has the structure
$$
\begin{pmatrix}
* & 1 \\
 1 & * \end{pmatrix} ,
$$
then two or several branch points of $G$ coincide, i.e., the curve is singular. Thus, as seen from their period
matrices, $\tilde\Gamma_1$ and also $\Gamma_3$ are singular, as was already shown in an algebraic way
the beginning of the section.

Next, $\Jac(\tilde\Gamma_2)$, having the period matrix $\tilde\Omega_2$, contains elliptic curves with the normalized
periods $\tau_1$ and $\tau_2+1$ (unimodularly equivalent to $\tau_2$), i.e.,
the same curves ${\mathcal W}_1, {\mathcal W}_2$, as already shown in Proposition \ref{W12}.

In a similar way, the structure of $\Omega_1, \Omega_2$ says that $\Jac(\Gamma_1), \Jac(\Gamma_2)$ both contain other
elliptic curves with the periods $(1,\tau_1/2)\cong (1,1+\tau_1/2)$ and $(1,2\tau_2)$,
namely ${\mathcal U}_1, {\mathcal U}_2$.
\medskip

3) Taking linear combinations of the periods of $\Prym(C,\sigma )$, we transform $\Lambda$ to fully diagonal form:
$$
 \Lambda \rightarrow \begin{pmatrix}
1 & 0 & 2\tau_2 & 1 \\
0 & 2 & 2 & \tau_1 \end{pmatrix} \rightarrow
\begin{pmatrix}
1 & 0 & 2\tau_2 & 0 \\
0 & 2 & 0 & \tau_1 \end{pmatrix} .
$$
Thus a sublattice of periods of $\Prym(C,\sigma )$ splits into a direct product of 2 lattices in $\mathbb C$ generated
by the periods $(1,2\tau_2)$ and $(2,\tau_1)\cong (1,\tau_1/2)$ of the curves ${\mathcal U}_2$ and ${\mathcal U}_1$
respectively. Similarly, the period matrix of $\Prym(K,\rho)$ can be transformed as follows
  $$
 \Lambda^* \rightarrow \begin{pmatrix}
1 & 0 & \tau_1 & 1 \\
0 & 2 & 2 &  2\tau_2 \end{pmatrix} \rightarrow
\begin{pmatrix}
1 & 0 & \tau_1 & 1 \\
0 & 2 & 0 & 2\tau_2 \end{pmatrix} ,
$$
hence $\Prym(K,\rho)$ contains elliptic curves with the periods $(1,\tau_1)$ and $(2,2\tau_2)\cong (1,\tau_2)$.

Finally, recalling that $\Jac(C)$ contains $\Prym(C,\sigma )$ and $E$, whereas $\Jac(K)$
contains $\Prym(K,\rho)$ and $\cal E$, we obtain the last statements of item 3).  $\square$

\paragraph{Remark.} We stress that in the case we just considered,
although $\Jac (C)$ contains 3 elliptic curves,
the curve $C$ admits, in general, only one involution $\sigma$ and cannot be regarded as a 2-fold covering of
the elliptic curves ${\mathcal W}_1, {\mathcal W}_2$.
\medskip

A numerical example for this section is given in Appendix C.2.

\section{Conclusion and open problems.} In given paper, by using previous results of \cite{kot892,hai83,avm88,hvm89},
we gave explicit description of all the genus 2 curves
whose Jacobians are related, via isogenies of degree 2, to
the Prym variety $\Prym(C,\sigma)\subset \Jac(C)$, the latter being associated to the double covering
$C\to E$ for the class of genus 3 curves $C$ of the form \eqref{mastercurve}.

This description may be useful in constructing a separation of variables for the algebraic integrable systems whose
complex invariant manifolds are open subsets of $\Prym(C,\sigma)$ or whose Lax representations yield spectral curves
of type \eqref{mastercurve}.
A concrete implementation of the algorithm of the separation lies beyond the scope of this paper and will be
describes elsewhere.

In Sections \ref{h30}, \ref{sec:sing_case} we also treated important special cases of the covering $C\to E$,
although, for brevity, our description did not cover another special case
when only two of the four branch points $Q_1,\dots,Q_4$ of the covering $C\to E$ are
in the hyperelliptic involution on $E$ and the other two are not.

\medskip

Theorem \ref{main_th} gives algebraic expressions for the parameters $d_\alpha^2$ of
the genus 2 curves $\widetilde\Gamma_i, \Gamma_i$ in terms of permutations of the roots of the quartic equation
$\psi(x)=0$. On the other hand,
it is known that a cross ratio of the roots of such an equation can be written as a quotient
of theta-constants of an appropriate elliptic curve (see, e.g., a classical monograph \cite{hc44}).
This property can be used to derive alternative compact expressions for $d_\alpha^2$ in terms of such theta-constants.
The explicit formulae will be given in a separate publication.
\medskip

It is natural to ask how the above algebraic description of hyperelliptic Jacobians isogeneous to
$\Prym(C,\sigma)\subset \Jac(C)$ can be extended to the case of the other 3 curves \eqref{alt_curves}
of the set $\Cov^2_E(Q)$, as well as to the general case of the covering $C\to E$ outlined in
Theorem \ref{gen_cover}. Recall that, since the data $\{E\, ; Q_1,\dots, Q_4 \in E\}$ do not define the covering
uniquely, the formulas of Theorem \ref{main_th}, which use only these data, are no more applicable in these cases.

Next, the model integrable systems described in Section \ref{systems} admit multidimensional generalizations, which
lead to spectral curves of higher genera. In particular, as was shown in \cite{pan86}, the generic complex invariant
varieties of the Frahm--Manakov top on $so(5)$ are open subsets of 4-dimensional Prym varieties
$\Prym(S,\sigma) \subset \Jac(S)$, where $S$ is the spectral curve of genus 6 of the form
$$
  w^2 = x^3+ h_2 x^2+h_1 x+h_0 + h_3 \sqrt{ (x-a_1)\cdots (x-a_5)} \,,
$$
$h_0,\dots, h_3$ being constants of motion, and $a_1,\dots, a_5$ being parameters of the Frahm--Manakov top.
Thus $S$ has involution $(x,y)\to (x,-y)$ and is
 a 2-fold covering of  a genus 2 curve $\Gamma \, : \, y^2=(x-a_1)\cdots (x-a_5)$.
The covering $S\to \Gamma$ has 6 ordinary branch points, the fixed points of the involution
$\sigma$. Thus the variety $\Prym(S,\sigma)$ has polarization $(1,1,2,2)$, and
is isogeneous to various 4-dimensional principally polarized Abelian varieties.
Then two natural questions arise: is it true that these varieties (or some of them) are Jacobians ?
And if yes, how to give an algebraic description of the corresponding genus 4 curves ? The questions are closely
related to the problem of separation of variables of the integrable top on $so(5)$ and other algebraic integrable systems.

We just mention that the recent results of \cite{pir12} give strong reasons to conjecture
that the answer to the first question is positive.
\newpage

\appendix

\section{Period matrix of the Prym variety}
Here we recall the algorithm of extracting the period matrix of the 2-dimensional Prym subvariety of the Jacobian
of a genus 3 curve $C$ admitting an involution
$\sigma: \, C \to C$ with 4 fixed points $Q_1,\dots, Q_4$ and covering
an elliptic curve $E=C/\sigma$. Our description mostly follows Section V of \cite{fay73}, although a
similar one can be found in \cite{hai83}.

Let
\begin{equation}
({\mathfrak a}_1, {\mathfrak a}_2, {\mathfrak a}_3; {\mathfrak b}_1, {\mathfrak b}_2, {\mathfrak b}_3 )=
(\mathfrak{a},\mathfrak{A},\bar{\mathfrak{a}};\mathfrak{b},\mathfrak{B},\bar{\mathfrak{b}}) \in \mathrm{H}_1( \widehat{\mathcal{C}},\mathbb{Z})
\label{cycles}
\end{equation}
 be a canonical basis of cycles on ${\mathcal{C}}$ such that
\begin{equation} \label{s-cycles}
\sigma (\mathfrak{a}) =-\bar{\mathfrak{a}},\quad \sigma (\mathfrak{A})=-\mathfrak{A},\quad
\sigma (\mathfrak{b})=-\bar{\mathfrak{b}},\quad \sigma (\mathfrak{B})=-\mathfrak{B},
\end{equation}
as depicted in Figure A.1,
and $u,w,\bar{u}$ be the corresponding \textit{normalized} holomorphic differentials, such that
$$
\left(
\begin{array}{ccc}
\oint\limits_{\mathfrak{a}}u & \oint\limits_{\mathfrak{A}}u & \oint\limits_{\bar{\mathfrak{a}}}u \\
\oint\limits_{\mathfrak{a}}w & \oint\limits_{\mathfrak{A}}w & \oint\limits_{\bar{\mathfrak{a}}}w \\
\oint\limits_{\mathfrak{a}}\bar{u} & \oint\limits_{\mathfrak{A}}\bar{u} & \oint\limits_{\bar{\mathfrak{a}}}\bar{u}
\end{array} \right)= 1_3 .
$$
Then, as one can deduce, $\sigma^*(u)=-\bar u$, $\sigma^*(w)=-w$.

\begin{figure}[h,t]
\begin{center}
\includegraphics[width=0.75\textwidth]{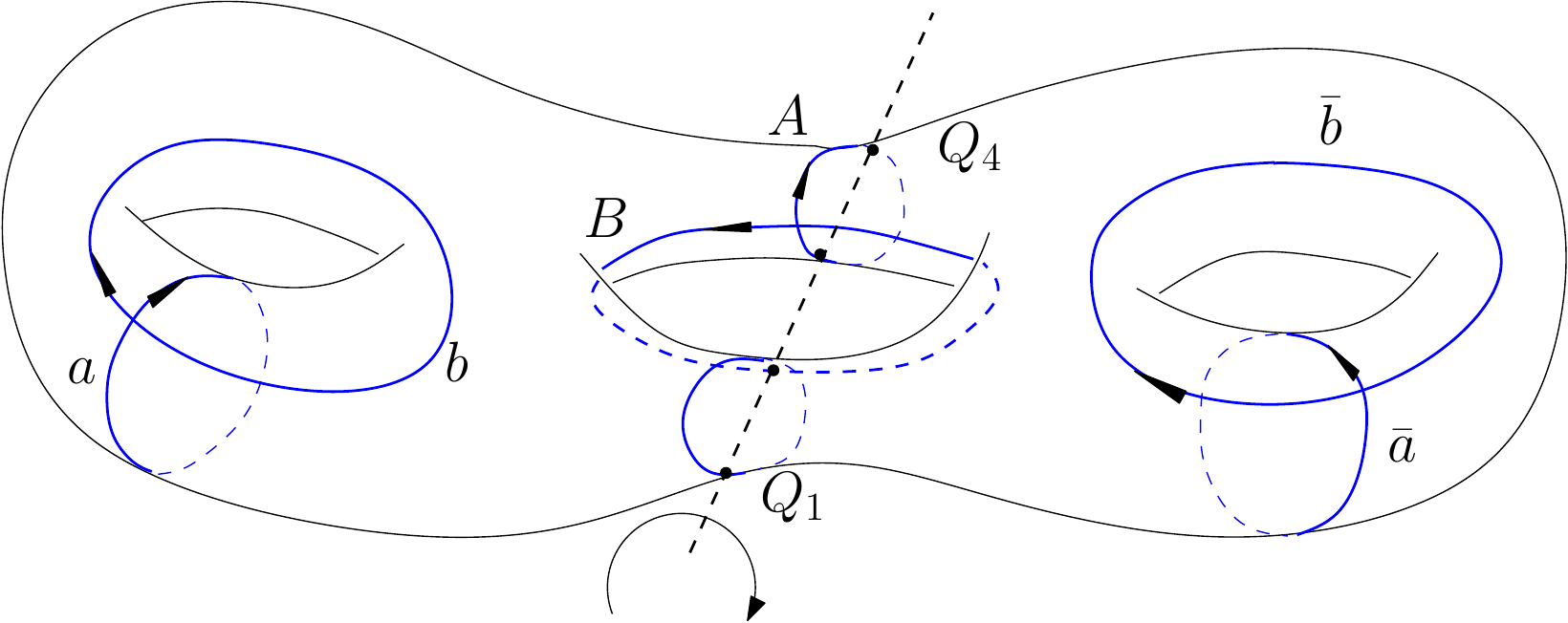}
\end{center} \caption{The genus 3 Riemann surface $C$:
the involution $\sigma$ acts on it as rotation by $\pi$ and has 4 fixed points $Q_1,\dots, Q_4$.}
\end{figure}

By definition, the Riemann matrix of ${\mathcal{C}}$ has the form
\[
\widehat{\tau}=\left(
\begin{array}{ccc}
\oint\limits_{\mathfrak{b}}u & \oint\limits_{\mathfrak{B}}u & \oint\limits_{\bar{\mathfrak{b}}}u \\
\oint\limits_{\mathfrak{b}}w & \oint\limits_{\mathfrak{B}}w & \oint\limits_{\bar{\mathfrak{b}}}w \\
\oint\limits_{\mathfrak{b}}\bar{u} & \oint\limits_{\mathfrak{B}}\bar{u} & \oint\limits_{\bar{\mathfrak{b}}}\bar{u}
\end{array}
\right).
\]
Note that the differential $\omega =u-\bar{u}$ is symmetric with respect to $\sigma$, and, therefore, is
reduced to the differential on $E$, whereas $W=u+\bar{u}$ is {\it anti-symmetric}. Then
$$
u=(W+\omega)/2,\quad \bar{u}=(W-\omega )/2,
$$
and we can write
\begin{align*}
\widehat{\tau}&=\left(
\begin{array}{ccc}
\oint\limits_{\mathfrak{b}}(W+\omega)/2 & \oint\limits_{\mathfrak{B}}(W+\omega )/2 & \oint\limits_{\bar{\mathfrak{b}}}(W+\omega )/2
\\
\oint\limits_{\mathfrak{b}}w & \oint\limits_{\mathfrak{B}}w & \oint\limits_{\bar{ \mathfrak{b}}}w \\
\oint\limits_{\mathfrak{b}}(W-\omega )/2 & \oint\limits_{\mathfrak{B}}(W-\omega )/2 & \oint\limits_{\bar{ \mathfrak{b}}}(W-\omega )/2%
\end{array}
\right)\\ &=\left(
\begin{array}{ccc}
\oint\limits_{\mathfrak{b}}(W+\omega)/2 & \oint\limits_{\mathfrak{B}}W/2 & \oint\limits_{\bar{ \mathfrak{b}}}(W+\omega )/2 \\
\oint\limits_{\mathfrak{b}}w & \oint\limits_{\mathfrak{B}}w & \oint\limits_{\bar{ \mathfrak{b}}}w \\
\oint\limits_{\mathfrak{b}}(W-\omega )/2 & \oint\limits_{\mathfrak{B}}W/2 & \oint\limits_{\bar{ \mathfrak{b}}}(W-\omega )/2
\end{array}
\right)
\end{align*}
Setting here
\[
\pi =\oint\limits_{\mathfrak{b}}W, \quad \Pi =\frac12\oint\limits_{\mathfrak{B}}W,\quad P=\frac12\oint\limits_{\mathfrak{B}}w, \quad p=\oint\limits_{\mathfrak{b}}w,
 \quad \tau =\oint\limits_{\mathfrak{b}}\omega
\]
and using the properties
$$
\oint\limits_{\mathfrak{B}}\omega =0,\quad
\oint\limits_{\mathfrak{b}} \omega = -\oint\limits_{\bar{ \mathfrak{b}}} \omega,\quad
 \oint\limits_{\mathfrak{b}} W =\oint\limits_{\bar{ \mathfrak{b}}}W, \quad  \Pi = p,
$$
we then have
\begin{equation} \label{RC}
\widehat{\tau}=\left(
\begin{array}{ccc}
\left(\pi +\tau \right) /2 & \Pi & \left( \pi -\tau \right) /2 \\
p & 2P & p \\
\left( \pi -\tau \right) /2 & \Pi & \left( \pi +\tau \right) /2
\end{array}
\right) . 
\end{equation}

The full $3\times 6$ period matrix of $\Jac(C)$ thus reads
$$
(1_3,\widehat{\tau}) \equiv(\boldsymbol{V}_1,\ldots, \boldsymbol{V}_6) \equiv
\left( \begin{array}{c}  \boldsymbol{R}_1\\ \boldsymbol{R}_2\\\boldsymbol{R}_3 \end{array}  \right) ,
$$
where $\boldsymbol{V}_i$, $\boldsymbol{R}_j$ are the corresponding columns and rows.
The columns of $(1_3,\widehat{\tau})$ defines a lattice in ${\mathbb C}^3=(z_1, z_2, z_3)$, the universal covering
of  $\Jac(C)$.

Now build another period matrix by taking linear combinations of
$\boldsymbol{V}_1,\dots, \boldsymbol{V}_6$ with integer coefficients
and applying, when necessary, linear changes of the coordinates $z_1, z_2, z_3$ (which
are equivalent to taking linear combinations of the rows $\boldsymbol{R}_1, \boldsymbol{R}_2, \boldsymbol{R}_3$)
as follows
\newpage

\begin{figure}
\begin{center}
\unitlength=1mm
\begin{picture}(120,50)(0,0)
\put(10,20){\makebox(0,0){$\left( 1_3;\widehat{\tau} \right)$}}
\put(35,25){\makebox(0,0){$\begin{array}{c}\boldsymbol{V}_4'=\boldsymbol{V}_4+\boldsymbol{V}_6 \\
\boldsymbol{V}_6'=\boldsymbol{V}_4-\boldsymbol{V}_6
\end{array} $}}
\put(20,20){\vector(1,0){30}}
\put(80,20){\makebox(0,0){$\left( \begin{array}{cccccc} 1&0&0   &\pi&\Pi&\tau\\   0&1&0 &2p&2P&0\\
 0&0&1&\pi&\Pi&-\tau   \end{array}\right)$}}
 \put(35,0){\makebox(0,0){$\begin{array}{c}\boldsymbol{R}_1'=\boldsymbol{R}_1+\boldsymbol{R}_3 \\
\boldsymbol{R}_3'=\boldsymbol{R}_1-\boldsymbol{R}_3
\end{array} $}}
 \put(20,-5){\vector(1,0){30}}
 \put(80,-5){\makebox(0,0){$\left( \begin{array}{cccccc} 1&0&1   &2\pi&2\Pi&0\\   0&1&0 &2p&2P&0\\
 1&0&-1&0&0&2\tau   \end{array}\right)$}}
 \put(35,-25){\makebox(0,0){$\begin{array}{c}\boldsymbol{V}_1'=\boldsymbol{V}_1+\boldsymbol{V}_3 \\
\boldsymbol{V}_3'=\boldsymbol{V}_1-\boldsymbol{V}_3
\end{array} $}}
 \put(20,-30){\vector(1,0){30}}
 \put(80,-30){\makebox(0,0){$\left( \begin{array}{cccccc} 2&0&0   &2\pi&2\Pi&0\\   0&1&0 &2p&2P&0\\
 0&0&2&0&0&2\tau   \end{array}\right)$}}
 \put(35,-45){\makebox(0,0){$\begin{array}{c}\boldsymbol{V}_3'=\boldsymbol{V}_3/2\\ \boldsymbol{V}_6'=\boldsymbol{V}_6/2
\end{array} $}}
 \put(20,-50){\vector(1,0){30}}
 \put(80,-50){\makebox(0,0){$\left( \begin{array}{cccccc} 2&0&0   &2\pi&2\Pi&0\\   0&1&0 &2p&2P&0\\
 0&0&1&0&0&\tau   \end{array}\right)$}}
\end{picture}
\end{center}
\end{figure}

\vskip 5cm

Thus, we arrived at the period matrix of a ''bigger'' (non-principally polarized)
Abelian variety $J \to \Jac(C)$, which is a direct product of the elliptic curve $E$
with the period matrix $(1\, \tau)$ and the Abelian variety with polarization
$(2,1)$, namely $\Prym (C,\sigma)$ with the period matrix
\begin{equation} \label{Prym_matrix}
\Lambda= \left( \begin{array}{cccc}
2 & 0 & 2\pi  & 2\Pi  \\
0 & 1 & 2p & 2P
\end{array}
\right) =\left(
\begin{array}{cccc}
2 & 0 & 2\oint\limits_{\mathfrak b} W & \oint\limits_{\mathfrak B} W \\
0 & 1 & 2\oint\limits_{\mathfrak b} w & \oint\limits_{\mathfrak B} w
\end{array}
\right),  \qquad \Pi =p.
\end{equation}
Thus, apart from $E$, the Jacobian of ${\mathcal{C}}$ contains
$\Prym ({\mathcal{C}}, \sigma)$ as an Abelian subtorus.

As shown in \cite{fay73},
the intersection $E \cap \Prym ({\mathcal{C}},\sigma)$ in $\Jac ({\mathcal{C}})$ consists of 4 points,
which are the half-periods of $E$ and some of the
half-periods of $\Prym ({\mathcal{C}},\sigma)$.

\section{Theta-functions and absolute invariants of a genus two curve}
Consider a hyperelliptic curve $\Gamma$ of genus two
\begin{align}
\begin{split}
w^2&=4 u_0 (z-e_1)(z-e_2)(z-e_3)(z-e_4)(z-e_5)(z-e_6)\\
&=u_0z^6+u_1z^5+u_2z^4+u_3z^3+u_4z^2+u_5z+u_6
\end{split} \label{sextic}
\end{align}
with branch points $e_i\neq e_j \in \mathbb{C}$.

Let $(\boldsymbol{\mathfrak{a}}; \boldsymbol{\mathfrak{b}})=
(\mathfrak{a}_1,\mathfrak{a}_2; \mathfrak{b}_1,\mathfrak{b}_2 ) $
 be canonic homology basis, $\mathfrak{a}_i\cap \mathfrak{b}_j=-\mathfrak{b}_i\cap \mathfrak{a}_j=\delta_{i,j}$,
 $\mathfrak{a}_i\cap \mathfrak{a}_j=\mathfrak{b}_i\cap \mathfrak{b}_j=\emptyset$, $i,j=1,2$, and
let
$$ 
\boldsymbol{\omega}=(\omega_1,\omega_2)^T, \quad
 \omega_1= \frac{\mathrm{d} z }{ w }, \quad \omega_2= \frac{z\mathrm{d} z }{ w }
$$ 
be canonic holomorphic differentials on $\Gamma$.
Introduce matrices of their $\mathfrak{a}$ and $\mathfrak{b}$-periods,
$$
\mathcal{A} = \left(\oint_{\mathfrak{a}_j}\omega_i\right)_{i,j=1,2},\quad
\mathcal{B} = \left(\oint_{\mathfrak{b}_j}\omega_i\right)_{i,j=1,2}
$$
and the Riemann period matrix
$$
\tau=\mathcal{A}^{-1}\mathcal{B},\quad   \tau^T=\tau, \quad \mathrm{Im} \;\tau >0 .
$$

Now let ${\mathbb C}^2=(z_1, z_2)$ be the universal covering of the Jacobian variety of $\Gamma$.
The Riemann theta-function $\theta[\varepsilon]( \boldsymbol{z}; \tau)$,
$\boldsymbol{z}=(z_1, z_2)$ with characteristics
$$
[\varepsilon] = [\boldsymbol{\varepsilon} \, \boldsymbol{\varepsilon'} ], \qquad
\boldsymbol{\varepsilon}=\left(  \begin{array}{c} \varepsilon_{1} \\  \varepsilon_{2}\end{array} \right),
\quad \boldsymbol{\varepsilon'}
=\left( \begin{array}{c} \varepsilon_{1}' \\  \varepsilon_{2}'\end{array}  \right), \quad
\varepsilon_i, \varepsilon_j' \in {\mathbb R}
$$
is definite on $\mathbb{C}^2\times \mathcal{S}_2$ by the Fourier series
\begin{equation}
\theta[\varepsilon](\boldsymbol{z};\tau)=
\sum_{\boldsymbol{n}\in \mathbb{Z}^2} \mathrm{exp}\left\{  \imath \pi ( \boldsymbol{n}+ \boldsymbol{\varepsilon'}  )^T\tau  ( \boldsymbol{n}+\boldsymbol{\varepsilon'}  ) +2\imath \pi ( \boldsymbol{n}+\boldsymbol{\varepsilon'}  )^T (\boldsymbol{z}+\boldsymbol{\varepsilon}  )\right\}.\label{theta}
\end{equation}
Many known periodic and modular properties of the theta-function are described in, e.g.,
\cite{fay73}, \cite{fk980}.
Non-vanishing values of this function with {\it half--integer} characteristics
$\varepsilon_i, \varepsilon_j'$  and their derivatives are called $\theta$-constants and denoted as
\begin{align*}
\theta[\varepsilon]   =\theta[\varepsilon](\boldsymbol{0};\tau), \qquad
\theta_{i}[\varepsilon]
=\left.\frac{\partial}{\partial z_i}\theta[\varepsilon](\boldsymbol{z};\tau)\right|_{\boldsymbol{z}=0},\quad i=1,2.
\end{align*}
Recall that a  half--integer characteristic $[\varepsilon]$ is odd if
$4 \boldsymbol{\varepsilon'}^T\boldsymbol{\varepsilon} = 1 \;\; (\text{mod} \; 2)$ and
even when
$4 \boldsymbol{\varepsilon'}^T\boldsymbol{\varepsilon} = 0 \; (\text{mod} \; 2)$.
There are 16 half-integer characteristics in total, and among them there are 6 odd and 10 even ones.

\paragraph{Absolute invariants of $\Gamma$ in terms of branch points and theta-constants.}
These invariants provide an effective tool which allow to check if two
genus 2 curves written in a hyperelliptic form \eqref{sextic} are birationally equivalent, i.e., are related
via a M\"obius transformation. There are 3 such invariants that can
be expressed either in terms of coefficients of the curves or in terms of theta-constants corresponding to the
period matrix $\tau$.

Namely, for a curve (\ref{sextic}) denote $(i,j)=e_i-e_j$. First, introduce
relative invariants with respect to the M\"obius transformations
\begin{align} \begin{split}
J_2(u)&=u_0^2\sum_{\rm{fifteen}}(12)^2(34)^2(56)^2,\\
J_4(u)&=u_0^4\sum_{\rm{ten}}(12)^2(23)^2(31)^2(45)^2(56)^2(64)^2,\\
J_6(u)&=u_0^6\sum_{\rm{sixty}}(12)^2(23)^2(31)^2(45)^2(56)^2(64)^2
(14)^2(25)^2(36)^2,\\
J_{10}(u)&=u_0^{10}\prod_{j<k}(j,k)^2. \end{split} \label{branchinv}
\end{align}
Here $J_{10}$ is the discriminant of the polynomial in the right hand side of \eqref{sextic}.
Direct computation leads to the following expressions for $J_{2}, J_{4}$ in
terms of coefficients of (\ref{sextic})
\begin{align*}
J_2&=2(20u_1u_5-8u_2u_4+3u_3^2-8u_1u_3-120u_6u_0+20u_0u_4 ),\\
J_4&=4(u_2^2u_4^2-3u_1u_3u_4^2-3u_2^2u_3u_5+9u_1u_2^2u_5+u_1u_2u_4u_5
-20u_1^2u_5^2+12u_2^3u_6\\&-45u_1u_2u_3u_6+75u_1^2u_4u_6
+12u_0u_4^3-135u_0u_1u_5u_6
-126u_0u_2u_4u_6+81u_0u_3^2u_6\\&+75u_0u_2u_5^2-45u_0u_3u_4u_5+405u_0^2u_6^2).
\end{align*}
These formulae, as well as similar expressions for $J_6,J_{10}$ (which are too long to be shown here),
can be found in \cite{sv04}.

On the other hand, given the curve \eqref{sextic} with the period matrix $\tau$,
the same relative invariants can be rewritten by in terms of the theta-constants
$ \theta[\varepsilon], \theta_i[\varepsilon]$ of half-integer theta-characteristics $\varepsilon$ by using formulae of
Bolza \cite{bolza86} and Rosenhain \cite{ro51}
as follows
\begin{align} \begin{split}
J_2&=48\, \pi^{12}\;\frac{\sum\limits_{15\;\rm{terms}
}\prod\limits_{k=1,\ldots,6,\;\sum{[\varepsilon_k]=0}}\;
\theta^4[\varepsilon_k]}
{\left(\prod\limits_{6 \;\rm{odd}\; [\delta]}
\theta_{1}[\delta]\right)^2 }\;,\\
J_4&=72\, \pi^{24}\;\frac{
\sum\limits_{10\;\rm{even}\; [\varepsilon]}\theta^8[\varepsilon_k]
\prod\limits_{10 \;\rm{even}\; [\varepsilon]}
\theta^4[\varepsilon] }
{\left(\prod\limits_{6 \;\rm{odd}\; [\delta]}
\theta_{1}[\delta]\right)^4 }\;,\\
J_6&=12\, \pi^{36}\;\frac{
\sum\limits_{60\;\rm{terms}}\theta[\varepsilon]^8
\sum\limits _{6\; \rm{terms}\;
[\varepsilon_k]\neq [\varepsilon],\;
\sum[\varepsilon_k]=0    } \prod\limits_{k=1}^6\;\theta^4[\varepsilon_k]
}
{\left(\prod\limits_{6 \;\rm{odd}\; [\delta]}
\theta_{1}[\delta]\right)^6 }\;,\\
J_{10}&=\pi^{60}\;\frac{
\prod\limits_{10 \;\rm{even}\; [\varepsilon]}
\theta^{12}[\varepsilon] }
{\left(\prod\limits_{6 \;\rm{odd}\; [\delta]}
\theta_{1}[\delta]\right)^{10} }\;. \end{split} \label{thetainv}
\end{align}

Alternative expressions for an another set of relative invariants in terms of theta-constants were
recently obtained in \cite{wen02}.

Following \cite{igusa62}, two genus 2 curves having the relative invariants $\{J_2,\dots, J_{10}\}$ and $\{J_2',\dots, J_{10}'\}$
are birationally equivalent (and period matrices of their Jacobians are symplectically equivalent)
if and only if $J'_k= r^k J_k$ with the same proportionality coefficient $r$.

Now, assuming that $J_2\neq 0$ and taking quotients of the relative invariants
 $J_2,\dots, J_{10}$ of the same order, we build three absolute invariants in the following form presented in \cite{sv04}:
\begin{align} \label{abs_invv}
i_1=144\frac{J_{4}}{J_{2}^2},\quad i_2=-1728\frac{(J_4J_2-3J_6)}{J_{2}^3},
\quad i_3=486\frac{J_{10}}{J_{2}^5} \, .
\end{align}
The latter are represented in terms of coefficients $u_0,\ldots, u_6$ of the sextic \eqref{sextic} or the
even $\theta$-constants given in \eqref{thetainv}. The importance of the above invariants is described by the following
theorem.

\begin{theorem}[\cite{sv04, ksv05}] \label{abs_inv_theta}
Two genus 2 curves are birationally equivalent (and period matrices of their Jacobians are symplectically equivalent)
if and only if they have the same triples of the absolute invariants.
\end{theorem}

We note that in some special cases $J_2$ can be zero, then
one can introduce other absolute invariants.

Below, in Appendix C.1,
we will use the above two representations of the absolute invariants to check the equivalence of
the Jacobians of the curves $\widetilde\Gamma_i, \Gamma_i$ obtained algebraically, by Theorem \ref{main_th}, and
analytically, by extracting their period matrices from that of the Prym variety $\Prym(C,\sigma)$.

\section{Numerical examples} The formulas for the genus 2 curves $\tilde \Gamma, \Gamma$ obtained in the main body of the
paper have been checked in various numerical examples: for the general case of the genus 3 curve $C$,
as well as for
the singular case (Section \ref{sec:sing_case}). Calculations given below use the
contents of the previous appendices and are made with {\tt MAPLE}, in particular the package {\tt algcurves}.

\subsection{Calculation of periods of the Prym variety and of the absolute invariants of the isogeneous Jacobians}
Consider the following example of the genus 3 curve \eqref{mastercurve}
\begin{equation} \label{num_cur}
{\mathcal{C}} \, : \; y^2 = x^2+3x+5+ \sqrt{4x^3-4x} , \quad \text{that is,} \quad
(y^2-x^2-3x-5)^2= 4x(x^2-1) ,
\end{equation}
which is a 2-fold covering of the elliptic curve $E=\{z^2=4x(x^2-1) \}$ whose canonical Weierstrass form is
$z^2 = 4x^3-g_2 x-g_3$ with the moduli $g_2=4, g_3=0$.

A basis of holomorphic differentials on ${\mathcal{C}}$ is given by
$$
 w_1= \frac{\mathrm{d}x}{y(y^2-x^2-3x-5) }, \quad w_2= \frac{x\, \mathrm{d}x}{y(y^2-x^2-3x-5) }, \quad
\omega=\frac{\mathrm{d}x}{y^2-x^2-3x-5 },
$$
or, equivalently,
$$
 w_1= \frac{\mathrm{d}x}{y\, \sqrt{4x^3-4x} }, \quad w_2= \frac{x\, \mathrm{d}x} {y\, \sqrt{4x^3-4x} }, \quad
\omega=\frac{\mathrm{d}x}{ \sqrt{4x^3-4x} },
$$
the first two differentials being {\it anti-invariant} with respect to the involution $\sigma : (x,y) \to (x,-y)$,
whereas the last one is invariant: it goes down to the differential $\mathrm{d}x /z$ on $E$.

The covering ${\mathcal{C}}\to E$ is ramified
at 4 points $Q_1,\dots, Q_4 \in E$, whose $x$-coordinates are the roots of $\psi(x)=(x^2+3x+5 )^2-4x(x^2-1)$.
Explicitly, one has
$$
x(Q_{2,1})=-0.97147712 \pm 0.69879610\,  \imath, \quad x(Q_{4,3}) = -0.028522874 \pm 4.17806957 \, \imath.
$$
The curve $\mathcal{C}$ can be depicted as  2 copies of $E$, and
each copy is realized as 2-fold covering of ${\mathbb P}=\{x \}$, as shown in Figure C.1.
Here the $x$-sheets 1,2 and 3,4 belong to the first and second copy of $E$, respectively. For any points $P_1$ and $P_2$
on sheets 1 and, respectively, 4, such that $x(P_1)=x(P_2)$, one has $y(P_1)=-y(P_2)$, the same holds for points on
sheets 2 and 3.
Applying the Maple command {\tt monodromy}, one observes that
the branch points $Q_1, Q_4$ connect the ''upper'' sheets 1 and 4, whereas
$Q_2, Q_3$ connect the ''lower'' sheets 2 and 3. On each copy of $E$, the points $Q_1, Q_2$ are joined
by cuts along which the copies are glued: when a point on $C$ cross such a cut,
it passes from one copy of $E$ to the other one, the same for the points $Q_3, Q_4$.

The remaining (horizontal) cuts join the points with $x=-1,0$ and
$x=1,\infty$, and realize connection between ''upper'' and ''lower'' sheets within the same copy of $E$.

The cycles \eqref{cycles} on $C$ are then drawn on the two copies of $E$ as shown on the same Figure C.1.

\begin{figure}[h,t]
\begin{center}
\includegraphics[width=1\textwidth]{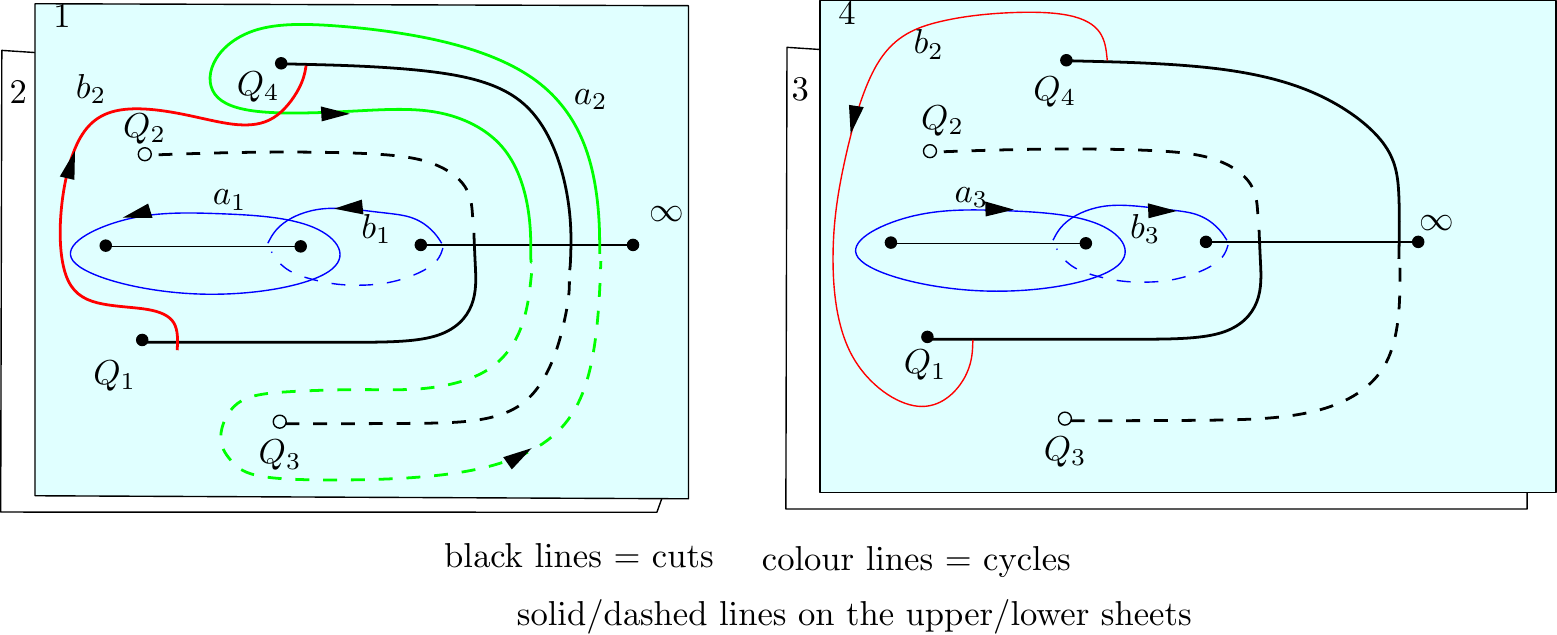}
\end{center} \caption{}
\end{figure}

To calculate the periods of $w_1,w_2, \omega$ along these cycles, one cannot use Maple command {\tt periodmatrix}:
it implements the algorithm of Trettkoff, which does not return the cycles with the required properties \eqref{s-cycles}.
Also, technically, it is quite difficult to calculate these periods in a manual
manner. For this reason, it is natural to use the elliptic parametrization
$$
x=\wp (u | g_2, g_3) , \quad \sqrt{4x^3-4x}= \wp'(u| g_2,g_3), \quad u\in {\mathbb C}, \qquad g_2=4,\; g_3=0 .
$$
Then, on ${\mathcal{C}}$ one has,
$$
y^2= \wp(u)^2+3\wp(u)+5+\wp'(u), \quad \mathrm{d}x=\wp'(u)\,du,
$$
and the above differentials take the form
$$
 w_1 = \frac{\mathrm{d}u}{\sqrt{\wp(u)^2+3\wp(u)+5+\wp'(u) } }, \quad
 w_2= \frac{\wp(u)\, \mathrm{d}u}{\sqrt{\wp(u)^2+3\wp(u)+5+\wp'(u) } }, \quad \omega=du\, .
$$
Note that the periods of $\omega$ along the cycles $\mathfrak{a}_{1},\mathfrak{b}_{1}$, as well as
$\mathfrak{a}_{3},\mathfrak{b}_{3}$, are, respectively
$$
2\varOmega_1 = 2\times 1.311028808, \quad 2\varOmega_3 =2\times 1.311028808 \, \imath,
$$
hence the normalized
period of $E$ is $\tau= \varOmega_3/\varOmega_1= \imath$. The periods of $\omega$ along $\mathfrak{a}_2,\mathfrak{b}_2$ are zero since
the projections of these cycles onto $E$ are homologically trivial.
The curve ${\mathcal{C}}$ now can be viewed as a 2-fold covering of the
parallelogram $\Sigma$ of periods of $E$ on the complex plane $u$ shown in Figure C.2 below. The (colored) quarters of the
parallelogram denoted as $1_+, 1_-$ (respectively as $2_+, 2_-$) correspond to the upper and lower half-planes of
the $x$-sheet 1 (respectively $x$-sheet 2) on Figure C.1.

The covering is ramified at
the $u$-images of the points $Q_1,\dots, Q_4\in E$, which (modulo the periods) are respectively
\begin{gather*}
q_1=2.279210426+1.770178281\,\imath, \quad q_2=2.279210426+0.851879\,\imath, \\
  q_3=0.3428471892+0.3451438934\,\imath, \quad  q_4=0.3428471892+2.276913703\,\imath
\end{gather*}
(One can check that
$$
y^2= \wp(q_i)^2+3\wp(q_i)+5+\wp'(q_i)=0, \qquad i=1,2,3,4,
$$
 as it should be, and that $q_1+\cdots+q_4=0$,
again modulo the periods of $E$.) Like $Q_1,\dots, Q_4$, the pairs $q_1, q_2$ and $q_3, q_4$ are connected by cuts along
which the 2 copies of the parallelogram $\Sigma$ are glued. The $u$-images of
$\mathfrak{a}_2,\mathfrak{b}_2$ embrace these cuts\footnote{Note that the images of the
cycles $\mathfrak{a}_1,\mathfrak{b}_1$ on $E$ in $\Sigma$ are just the horizontal an vertical straight line segments
passing through the half-period
$u=\varOmega_1+\varOmega_3=1.311028808 + 1.311028808 \,\imath$.
Then the cuts between $q_1, q_2$ and $q_3, q_4$ made within the parallelogram $\Sigma$ cross the above
segments, and, therefore, the corresponding images of $\mathfrak{a}_2,\mathfrak{b}_2$ would cross
$\mathfrak{a}_1,\mathfrak{b}_1$, which is not allowed because of canonicity of the cycles.
In order to avoid this, we replaced $q_1,q_2,q_4$ inside $\Sigma$ by their translations by full periods
of $E$.}.

It remains to integrate $w_1, w_2$ along the images of
$\mathfrak{a}_1,\mathfrak{b}_1,\mathfrak{a}_2,\mathfrak{b}_2$.
Note that, due to the symmetry/anti-symmetry of the differentials and the cycles, one has
$$
\oint\limits_{{\mathfrak a}_3} w_i = \oint\limits_{{\mathfrak a}_1} w_i, \quad
\oint\limits_{{\mathfrak a}_3} \omega = - \oint\limits_{{\mathfrak a}_1} \omega.
$$

\begin{figure}[h,t]
\begin{center}
\includegraphics[width=0.8\textwidth]{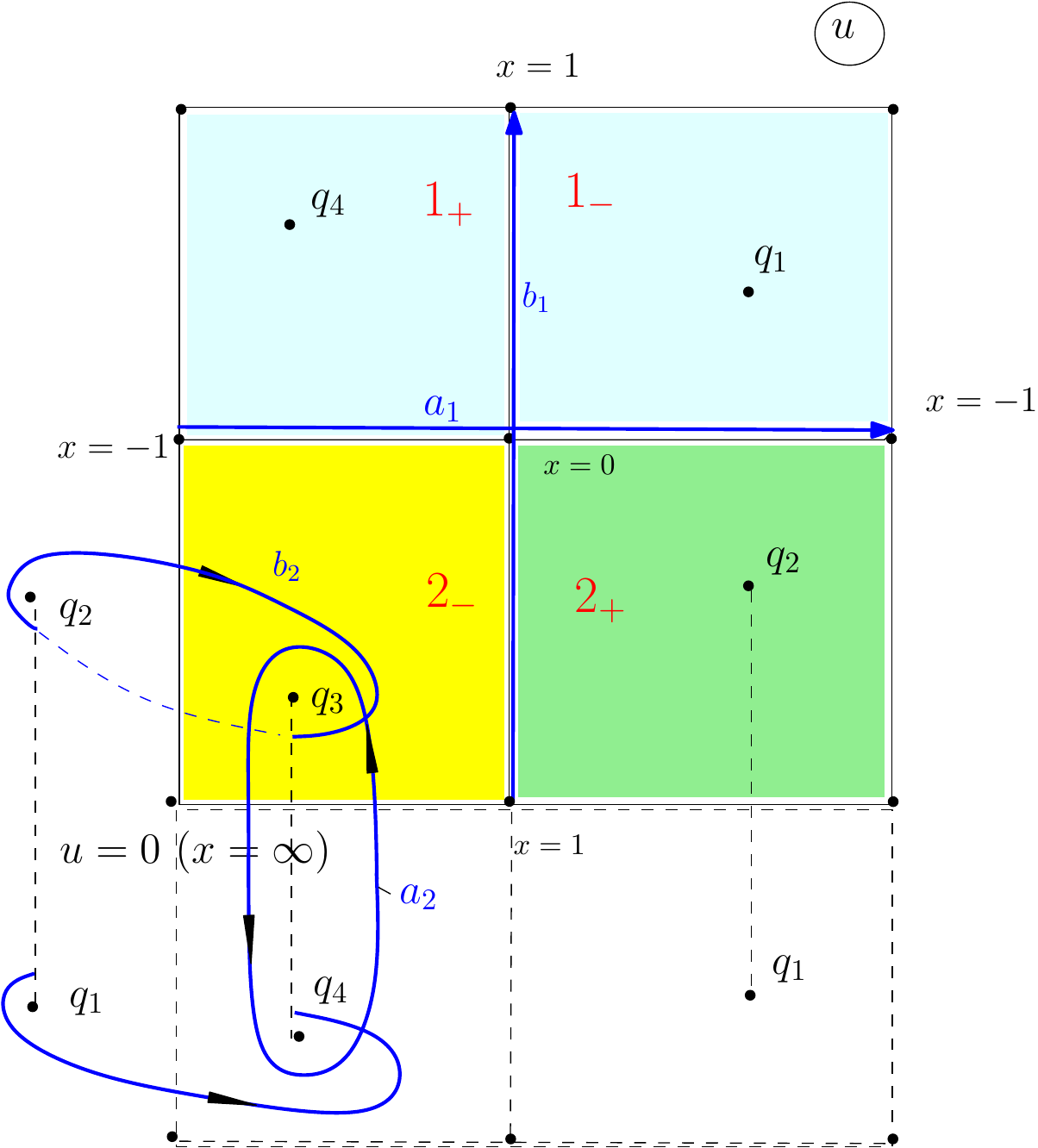}
\end{center} \caption{} \label{1.fig}
\end{figure}

Explicitly, for $j=1,2$, we have
\begin{align*}
  \oint\limits_{{\mathfrak a}_2} \frac{w_j}{4} &  = 2 \int_{q_4}^{q_3}
\frac{\wp(u)^{j-1}\, \mathrm{d}u}{4\sqrt{\wp(u)^2+3\wp(u)+5+\wp'(u) } }  = \left\{
\begin{aligned}
0.05372000635\,\imath \\ 0.6218457230\,\imath
\end{aligned}
  \right.  \\
   \oint\limits_{{\mathfrak b}_2}  \frac{w_j}{4}  & = 2 \int_{q_2}^{q_2}
\frac{\wp(u)^{j-1}\, \mathrm{d}u}{4\sqrt{\wp(u)^2+3\wp(u)+5+\wp'(u) } } = \left\{
\begin{aligned}
 0.2047172550  \\ -0.5928537135
\end{aligned}
  \right.  \\
\oint\limits_{{\mathfrak a}_1}  \frac{w_j}{4} & = \int_{\varOmega_3}^{\varOmega_3+2\varOmega_1}
\frac{\wp(u)^{j-1}\, \mathrm{d}u}{4\sqrt{\wp(u)^2+3\wp(u)+5+\wp'(u) } }
= \left\{
\begin{aligned}
0.340973237   \\ -0.167680659
\end{aligned}
  \right. \\
\oint\limits_{{\mathfrak b}_1}  \frac{w_j}{4}  & = \int_{\varOmega_1}^{\varOmega_1+2\varOmega_3}
\frac{\wp(u)^{j-1}\, \mathrm{d}u}{4\sqrt{\wp(u)^2+3\wp(u)+5+\wp'(u) } }  = \left\{
\begin{aligned}
 0.2557651545\, \imath  \\  0.107574277 \,\imath
\end{aligned}
  \right.
\end{align*}

Thus the matrices of periods of $w_1/4, w_2/4, \omega/4$ along
the cycles $(\mathfrak{a}_1,\mathfrak{a}_2,\mathfrak{a}_3)=(\mathfrak{a},\mathfrak{A},\bar{ \mathfrak{a}})$ and
$(\mathfrak{b}_1,\mathfrak{b}_2,\mathfrak{b}_3)=(\mathfrak{b},\mathfrak{B},\bar{ \mathfrak{b}})$ are
respectively
\begin{align*}
\widehat{{\mathcal A}} & =
\begin{pmatrix}
0.3409731370 & 0.05372000635\,\imath & 0.3409731370 \\
-0.1676806592 & 0.6218457230\,\imath & -0.1676806592 \\
0.655514404 & 0 & -0.655514404
\end{pmatrix} , \\
\widehat{{\mathcal   B}} & = \begin{pmatrix}
0.2557651545\,\imath & 0.2047172550 & 0.2557651545\,\imath \\
0.1075742777\,\imath & -0.5928537135 & 0.1075742777\,\imath \\
0.6555144040\,\imath & 0 & -0.6555144040\,\imath \end{pmatrix} .
\end{align*}

Then the numeric output for the  Riemann matrix \eqref{RC} of ${\mathcal{C}}$ is
$$
\widehat{\tau} =\widehat{\mathcal  A}^{-1}\widehat{\mathcal  B}= \begin{pmatrix}
0.846695699\,\imath & \,0.36000339 &  - 0.1533043007\, \imath \\
0.3599648533 & 0.759227705\,\imath & 0.359964853 \\
- 0.15330430078\,\imath & 0.36000339249 & 0.8466956992\,\imath \end{pmatrix} .
$$
Note that in view of the structure of the matrix \eqref{RC}, we have
$\tau=\widehat{\tau}_{1,1}-\widehat{\tau}_{1,3}= \imath$, i.e.,
the normalized period of the underlying elliptic curve $E$, as expected.
Comparing $\widehat{\tau}$ with \eqref{RC}, we then obtain the Prym period matrix \eqref{Prym_matrix} in the form
\begin{equation} \label{PrymN}
\Lambda = \begin{pmatrix} 2 & 0 & 1.386782796\imath & 0.7200067850 \\
         0 & 1 & 0.7199297066 & 0.75922770504 \imath  \end{pmatrix}
=  \left( \begin{array}{cccc}
2 & 0 & 2\pi  & 2\Pi  \\
0 & 1 & 2p & 2P
\end{array}
\right)  .
\end{equation}

Given $\Lambda$, we then evaluate the Riemann matrices of the three Jacobians
$\Jac(\tilde \Gamma_1)$, $\Jac(\tilde \Gamma_2)$, $\Jac(\tilde \Gamma_3)$ according to Diagram 1 in Section 3.
Next, using the formulas of
Theorem \ref{abs_inv_theta}, calculate the triple of absolute invariants
${\cal I}= \{\tilde i_1, \tilde i_2, \tilde i_3\}$ of each of the three curves, which are, respectively,
\begin{gather} \label{inv_num_theta}
\{ 2.04896, \; -2.57195, 0 \}, \quad \{-46.5226, \; 353.4564,\; 0.0348 \}, \quad
 \{0.007449, \; -0.000082, 0 \}.
\end{gather}

On the other hand, we calculate the branch points $d_1^2, d_2^2,d_3^3$ of the curves
$\tilde\Gamma_1, \tilde\Gamma_2, \tilde \Gamma_3$ in the hyperelliptic form \eqref{Gamma} by using expressions
\eqref{d_s}, \eqref{d_ss_a} of Theorem \ref{main_th}. From the data \eqref{num_cur} we have
\begin{equation} \label{dataN}
\begin{gathered}
 h_1 =3, \quad h_0=5, \quad h_3=1, \qquad c_1=-1, \; c_2=0, \; c_3=1, \\
s_{1,2}=-0.971477126\pm 0.6987961067\, \imath, \quad
s_{3,4}= -0.028522874 \pm 4.1780696 \,\imath .
\end{gathered}
\end{equation}
Then \eqref{d_s} yield the equations
\begin{gather*}
d_1^2+1/d_1^2= 2.06906849, \quad d_2^2+1/d_2^2= 5.90287078, \quad d_3^2+1/d_3^2= 16.00981635, \\
 d_1^2+1/d_1^2= 233.6540936, \quad d_2^2+1/d_2^2= 6.09954643 , \quad d_3^2+1/d_3^2=  3.142056369,  \\
 d_1^2+1/d_1^2= 1.93210387, \quad d_2^2+1/d_2^2=  0.02458074296, \quad d_3^2+1/d_3^2=  -1.111595606.
\end{gather*}
Take, for concreteness, the following solutions, respectively,
\begin{gather*}
\{ d_1^2=1.299602374, \quad d_2^2=5.728298886,\quad d_3^2= 15.94710906\}, \\
\{ d_1^2=233.6498137, \quad d_2^2=5.930939068, \quad d_3^2=.3593642226\} , \\
 \{d_1^2=0.966051936+0.2583479359\,\imath, \quad d_2^2=0.0122903715+0.99992447\,\imath, \\
\quad d_3^3=-0.555797803+0.83131751\,\imath \} .
\end{gather*}
and then calculate the algebraic absolute invariants of  $\tilde\Gamma_1, \tilde\Gamma_2, \tilde \Gamma_3$ by formulas
\eqref{branchinv} and \eqref{abs_invv} to get
\begin{gather*} 
\{ 2.04498067, \; -2.5634749, 0 \}, \quad \{-46.67437, \; 354.25472,\; 0.035048 \}, \\
 \{0.00749715, \; -0.000082, 0 \}.
\end{gather*}
Up to small errors of calculation, the above coincide with the invariants \eqref{inv_num_theta}, as expected.
\medskip

Next, substituting the numerical values \eqref{dataN} into the formulas of item 5 of Theorem \ref{main_th},
we calculate the Weierstrass equations \eqref{Gamma0} of the curves $\Gamma_1, \Gamma_2, \Gamma_3$ with the
following parameters, respectively,
\begin{gather*}
k_1^2=8.5992939, \quad k_2^2= 309.763893, \quad k_3^2=3.205312, \\
k_1^2= 12.4648624, \quad k_2^2= 4.14103076, \quad k_3^2=1.255478, \\
k_1^2= -0.253424+0.9673553 \,\imath, \quad k_2^2=-0.9743395+.2250834\,\imath, \\
 k_3^2=0.45+0.89302855\,\imath .
\end{gather*}
Then the expressions \eqref{branchinv}, \eqref{abs_invv} give the following algebraic absolute invariants
$\{i_1, i_2, i_3\}$ of $\Gamma_1, \Gamma_2, \Gamma_3$:
\begin{gather}
\{ -28.0500831,\; 254.6727861, \; 0.014329753 \}, \quad
\{6.382123, \; 4.05191,\; 14.54512 \}, \notag \\
 \{2.506817, \; 3.948671, 1.10004 \}.  \label{inv_num_alg}
\end{gather}

On the other hand, taking the expression \eqref{PrymN} of the period matrix of $\Prym({C},\sigma)$,
we calculate the period matrices
of $\Jac(\Gamma_1)$, $\Jac(\Gamma_2)$, $\Jac(\Gamma_3)$ by using the relations in Diagram 1. The triples of
absolute invariants of these three matrices, calculated via \eqref{thetainv}, \eqref{abs_invv}, coinside with the values
\eqref{inv_num_alg} up to small errors of order $10^{-4}$.

\subsection{Numerical example for the singular case}
Consider again the elliptic curve
$E$ in the canonical form $y^2 = 4 x(x^2-1)$, its parallelogram of periods is
$$
\{ 2\varOmega_1 {\mathbb Z}+ 2\varOmega_3 {\mathbb Z} \} \quad \text{with} \quad \varOmega_1= 1.31115187 \imath, \quad
\varOmega_3= 1.31115187
$$
Take the genus 3 curve $C$ in the form
\begin{gather}
w^2 = x^2-3x+4-2\sqrt{x(x^2-1)}, \label{ex_cur}
\end{gather}
Its coefficients satisfy the condition \eqref{cond_s1}, hence, according to Proposition \ref{Sing_alg_cond},
we have $S_{(12)\, (34)}^{(3)}=0$, and the Jacobian of $C$ must contain 3 elliptic curves.

The roots $s_1,\dots, s_4$ of $\psi(x)$ in \eqref{Psi0} are complex conjugated pairs
\begin{equation} \label{ssss}
s_{1,2}=0.2639320225 \pm 1.207561151\,\imath, \quad s_3=1.278004367, \quad s_4=8.194131588,
\end{equation}
and the Abel images $q_j\in {\mathbb C}$ of $Q_j=(s_j,2\sqrt{s_j^3-s_j})\in E$ such that $s_j=\wp(q_j)$ are
\begin{gather*}
q_1= -0.655514358+0.5487294304 \,\imath, \quad q_2= 1.966543166-0.5487294304\,\imath, \\
q_3= 0.9611650593, \quad q_4=0.3498637180 .
\end{gather*}
We observe that $q_1+q_2 = q_3+q_4 = \varOmega_3 = 1.31115187$, as expected and, again
$$
S_{(12)\, (34)}^{(3)} = (s_1-s_2)(s_3-s_4)(s_1 s_2+s_3 s_4-4 c_3- 8)=0
$$
The genus 2 curve $\widetilde\Gamma_1$ is singular and, according to item 2) of Theorem \ref{main_th},
\begin{gather*}
\text{for} \quad \tilde \Gamma_2 \; : \; d_3=1, \quad d_2^{2}+\frac{1}{d_2^2}=-1.01+3.897435\,\imath, \quad
d_1^{2}+\frac{1}{d_1^2}= 1.442+2.0378\,\imath  \\
\text{for} \quad \tilde \Gamma_3 \; : \; d_3=1, \quad
d_2^{1}+\frac{1}{d_1^2}= 1.01 - 3.897435\, \imath, \quad
d_3^{2}+\frac{1}{d_3^2}= 1.442+2.0378\, \imath  .
\end{gather*}
Solutions of the above equations give the following possible values of $\alpha, \beta$ in \eqref{abs}, \eqref{abs2}:
\begin{gather*}
\alpha_1 =\{-0.0575775-0.2278577\, \imath, \; -1.04242249+4.125298\, \imath \}, \\
\beta_1 =\{1.26620639+ 2.366331\, \imath, \; 0.1757936066 - 0.3285292\, \imath \},
\end{gather*}
and, respectively,
\begin{gather*}
\alpha_2 =\{-0.0575775+0.2278577\, \imath, \; -1.04242249-4.125298\, \imath \}, \\
  \beta_2 =\{1.26620639- 2.366331\, \imath, \; 0.1757936066 + 0.3285292\, \imath \}.
\end{gather*}
Then the moduli $C_\pm^{(j)}$ of the elliptic curves
${\mathcal W}_\pm^{(1)}\subset \Jac(\widetilde\Gamma_2)$ and
${\mathcal W}_\pm^{(2)}\subset \Jac(\widetilde\Gamma_3)$ in \eqref{elliptic_W} are
$$
C_-^{(1)} = 1.0580368, \quad C_+^{(1)} = 0.5+1.09996\, \imath, \qquad
 C_-^{(2)} = 1.0580368, \quad C_+^{(2)} = 0.5+1.100\, \imath .
$$
Hence, ${\mathcal W}_\pm^{(1)}$, ${\mathcal W}_\pm^{(2)}$  are pairwise isomorphic, as predicted by Proposition \ref{W12}.

The $J$-invariants of the curves are, respectively, $J_- = 81284.118$, $J_+ = -11.6898$. The same values are
obtained by the formula \eqref{Jpm}.
The normalized periods of the curves are
\begin{equation} \label{tauss}
\tau_1 = 0.5562073\, \imath, \quad \tau_2 = 1/2+0.9317386\, \imath.
\end{equation}

According to item 2) of Theorem \ref{3-curve}, in the considered case each of remaining regular
curves $\Gamma_1, \Gamma_2$ covers elliptic curves ${\cal U}_1, {\cal U}_2$. To obtain the equations
of $\Gamma_1, \Gamma_2$, we can either use the formulas of item 5) of Theorem \ref{main_th}
or consider the curve $\cal K$ dual to \eqref{ex_cur}, namely
$$
   w^2 = x^2-3x+4 + \sqrt{x^4-10 x^3-17 x^2-20 x+16} .
$$
Applying the projective transformation $X=\dfrac{x-s_4}{x-s_3}$, we take $\cal K$ to the canonical form
\eqref{mastercurve} with $h_2=1$:
$$
Y^2 = X^2 - 0.293375 X+ 25.87776 + 2\cdot 3.0362\, \sqrt{X(X^2-7.6411194\, X+ 25.8777647)}\,
$$
and then implement the same algorithm as above. Namely, let $s_1,\dots, s_4$ be the four values of $X$ for which $Y=0$.
Then
\begin{gather*}
 s_1=1,\quad s_2=4.0360465, \quad s_3=6.411661683, \quad
s_4=25.8777645 , \\
c_1=3.820559+3.358733\, \imath, \quad c_2=3.820559-3.358733\, \imath, \quad c_3=0, \\
h_1= - 0.293375 , \quad h_0=25.87776 , \quad h_3=3.0362 ,
\end{gather*}
and we again observe that
$$
S_{(14),(23)}^{(3)}= (s_{1}-s_{4})(s_{2}-s_{3}) ( (s_{1}s_{4}+s_{2}s_{3}) - 4 c_3 h_3^2-2h_0 )=0,
$$
as expected, so the curve $\Gamma_3$ is singular. For the other two curves we have the equations
\begin{gather*}
 \Gamma_1 \; : \; d_3=1, \quad {d_2}^{2}+\frac{1}{d_2^2}=6.666666787-3.711843044\, \imath, \quad
d_1^{2}+\frac{1}{d_1^2}= 6.66666+3.711843\, \imath ,  \\
 \Gamma_2 \; : \; d_3=1, \quad
d_2^{1}+\frac{1}{d_1^2}= -0.44-0.66813\, \imath , \quad
d_1^{2}+\frac{1}{d_1^2}= -0.44+0.66813\, \imath .
\end{gather*}
Following \eqref{abs}, we then choose $\alpha_1=6.552+3.7778\, \imath$, $\beta_1=6.552-3.7778\, \imath$
and, using an alalog of \eqref{elliptic_W}, obtain the following moduli of the corresponding elliptic curves
${\cal U}_1, {\cal U}_2$
$$
C_+ = -0.62597, \quad C_- =0.04484 .
$$
The normalized periods of ${\cal U}_1, {\cal U}_2$ are
$$
T_1 = 0.8988536\, \imath, \quad T_2= 1.8635126\, \imath .
$$
Comparing them with \eqref{tauss}, we notice that
$$
 T_1 = -\frac 12 \cdot \frac{1}{\tau_1}, \quad T_2 = 2\tau_2 -3 ,
$$
that is, up to unimodular transformations generated by $\tau \to \tau+1$, $\tau \to -1/\tau$,
we have $T_1=\tau_1/2$, $T_2= 2\tau_2$, as predicted by item 3) of Theorem \ref{3-curve}.

\section*{Acknowledgments} The authors are grateful to H. Braden for making independently
a series of hard calculations which confirmed the results of our numerical examples,
as well as for continuous stimulating discussions. 
We also grateful to A. Levin, J. C. Naranjo, and T. Shaska for valuable remarks.

We thank Department of Physics of the Oldenburg University and the School of Mathematics of the  university of Edinburgh for funding our research visits to these institutions, which allowed the completion of the present article.

The figures of the paper have been generated with a compact, but powerful and easy-to-use {\it IPE extensible drawing editor}.

Y.F acknowledges support of  the Spanish MINECO-FEDER Grants MTM2012-31714,  MTM2012-37070.
The work of V.E was supported by the School of Mathematics, university of Edinburgh, under the certificate of sponsorship C5E7V94128U.

\end{document}